\documentclass[11pt,a4paper]{article}

\usepackage[T1]{fontenc}
\usepackage[dvips]{graphicx}
\usepackage{color}
\usepackage{times}
\usepackage{hyperref}
\usepackage{amssymb}
\usepackage{rotating}
\usepackage[bf]{subfigure}

\title{Characterization of Complex Networks: \\A Survey of measurements}

\author{L.\ da F.\ Costa \and F.\ A.\ Rodrigues \and G.\ Travieso \and
  P.\ R.\ Villas Boas\\
  \small Instituto de F\'{\i}sica de S\~{a}o Carlos,
  Universidade de S\~{a}o Paulo\\
  \small Caixa Postal 369, 13560-970, S\~{a}o Carlos, SP, Brazil\\
  \texttt{luciano@ifsc.usp.br} }

\begin{document}

\maketitle

\begin{abstract}

  Each complex network (or class of networks) presents specific
  topological features which characterize its connectivity and highly
  influence the dynamics of processes executed on the network.  The
  analysis, discrimination, and synthesis of complex networks
  therefore rely on the use of measurements capable of expressing the
  most relevant topological features. This article presents a survey
  of such measurements.  It includes general considerations about
  complex network characterization, a brief review of the principal
  models, and the presentation of the main existing measurements.
  Important related issues covered in this work comprise the
  representation of the evolution of complex networks in terms of
  trajectories in several measurement spaces, the analysis of the
  correlations between some of the most traditional measurements,
  perturbation analysis, as well as the use of multivariate statistics
  for feature selection and network classification.  Depending on the
  network and the analysis task one has in mind, a specific set of
  features may be chosen.  It is hoped that the present survey will
  help the proper application and interpretation of measurements.

\end{abstract}

\vspace{1cm}

\pagebreak

\setcounter{tocdepth}{1}

\tableofcontents

\pagebreak

\pagebreak

\section{Introduction}\label{introduction}

Complex networks research can be conceptualized as lying at the
intersection between graph theory and statistical mechanics, which
endows it with a truly multidisciplinary nature.  While its origin can
be traced back to the pioneering works on percolation and random
graphs by Flory \cite{Flory}, Rapoport \cite{Rapoport:1951,
Rapoport:1953, Rapoport:1957}, and Erd\H{o}s and R\'enyi
\cite{Erdos-Renyi:1959,Erdos-Renyi:1960,Erdos-Renyi:1961}, research in
complex networks became a focus of attention only recently.  The main
reason for this was the discovery that real networks have
characteristics which are not explained by uniformly random
connectivity.  Instead, networks derived from real data may involve
community structure, power law degree distributions and hubs, among
other structural features.  Three particular developments have
contributed particularly for the ongoing related developments: Watts
and Strogatz's investigation of small-world networks \cite{Watts98},
Barab\'asi and Albert's characterization of scale-free models
\cite{Barabasi97}, and Girvan and Newman's identification of the
community structures present in many networks
(e.g. \cite{Girvan:2002}).

Although graph theory is a well-established and developed area in
mathematics and theoretical computer science (e.g.,
\cite{Bollobas98,West:2001}), many of the recent developments in
complex networks have taken place in areas such as sociology (e.g.,
\cite{Scott00,Newman:2003}), biology (e.g., \cite{Barabasi:2004})
and physics (e.g., \cite{Bornholdt03:book,Amaral04}). Current interest
has focused not only on applying the developed concepts to many real
data and situations, but also on studying the dynamical evolution of
network topology. Supported by the availability of high performance
computers and large data collections, results like the discovery of
the scale-free structure of the Internet \cite{Faloutsos99} and of the
WWW \cite{Albert99, Barabasi00} were of major importance for the
increased interest on the new area of complex networks, whose growing
relevance has been substantiated by a large number of recent related
publications. Reviews of such developments have been presented in four
excellent surveys \cite{Barabasi:survey, Dorogovtsev02,
Newman:2003:survey, Boccaletti05}, introductory papers
\cite{Barabasi:2003:SA, Hayes:2000a, Hayes:2000b, Amaral04} and
several books \cite{Bollobas:1985, Wasserman94, Scott00,
Barabasi:book, Dorogovtsev03, Bornholdt03:book, Newman06:book}.  For
additional information about the related areas of percolation,
disordered systems and fractals see \cite{Stauffer:1994,
Bunde_Havlin1, Bunde_Havlin2}; for complex systems, see
\cite{Boccara04:book, Bar-Yam92:book, Strogatz94:book}.

One of the main reasons behind complex networks popularity is their
flexibility and generality for representing virtually any natural
structure, including those undergoing dynamical changes of topology.
As a matter of fact, every discrete structure such as lists, trees, or
even lattices, can be suitably represented as special cases of
graphs. It is thus little surprising that several investigations in
complex network involve the \emph{representation} of the structure of
interest as a network, followed by an analysis of the topological
features of the obtained representation performed in terms of a set of
informative measurements.  Another interesting problem consists of
measuring the structural properties of evolving networks in order to
characterize how the connectivity of the investigated structures
changes along the process.  Both such activities can be understood as
directed to the \emph{topological characterization} of the studied
structures.  Another related application is to use the obtained
measurements in order to identify different categories of structures,
which is directly related to the area of \emph{pattern recognition}
\cite{Costa:01,Duda_Hart:01}.  Even when modeling networks, it is
often necessary to compare the realizations of the model with real
networks, which can be done in terms of the respective measurements.
Provided the measurements are comprehensive (ideally the
representation by the measurements should be one-to-one or
invertible), the fact that the simulated networks yield measurements
similar to those of the real counterparts supports the validity of the
model.

Particular attention has recently been focused on the relationship
between the structure and dynamics of complex networks, an issue
which has been covered in two excellent comprehensive reviews
\cite{Barabasi:survey, Newman:2003:survey}.  However, relatively
little attention has been given to the also important subject of
network measurements (e.g.~\cite{Ziv05:PRE016110}). Indeed, it is
only by obtaining informative quantitative features of the networks
topology that they can be characterized and analyzed and,
particularly, their structure can be fully related with the
respective dynamics.  The quantitative description of the networks
properties also provides fundamental subsidies for classifying
theoretical and real networks into major categories. The present
survey main objective is to provide a comprehensive and accessible
review of the main measurements which can be used in to quantify
important properties of complex networks.

\begin{figure}
 \centerline{\includegraphics[width=8cm]{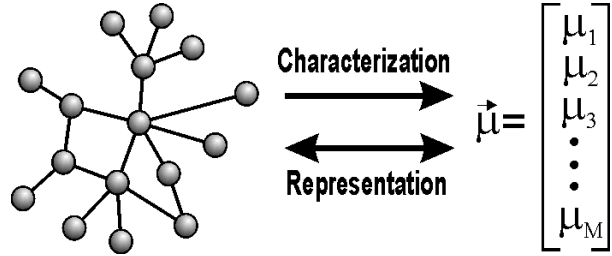}}
 \caption{The mapping from a complex network into a feature vector.
   Generic mappings can be used in order to obtain the
   characterization of the network in terms of a suitable set of
   measurements.  In case the mapping is invertible, we have a
   complete representation of the original structure.}
 \label{fig:char}
\end{figure}

\begin{figure}
 \centerline{\includegraphics[width=10cm]{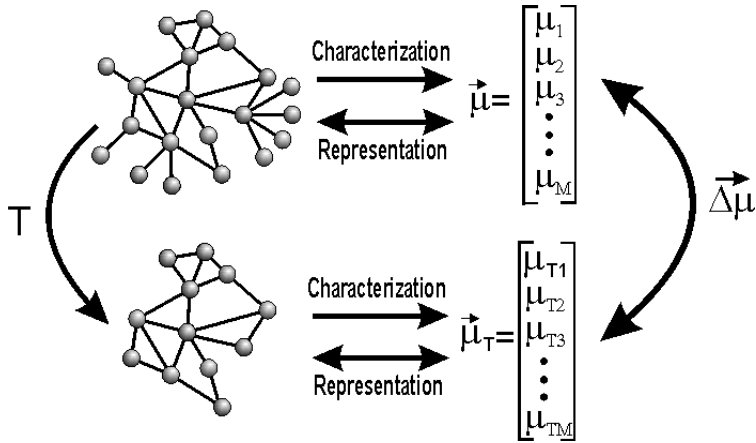}}
 \caption{Additional measurements of a complex network can be obtained
 by applying a transformation $T$ on it and obtaining a new feature
 vector $\vec{\mu}_T$.  The difference $\Delta\vec{\mu}$ between the
 original and transformed features vectors can also be considered in
 order to obtain additional insights about the properties of the
 original network.}
\label{fig:tchar}
\end{figure}

Network measurements are therefore essential as a direct or
subsidiary resource in many network investigations, including
representation, characterization, classification and modeling.
Figure~\ref{fig:char} shows the mapping of a generic complex network
into the feature vector $\vec{\mu}$, i.e.\ a vector of related
measurements such as average vertex degree, average clustering
coefficient, the network diameter, and so on.  In case the mapping
is invertible, in the sense that the network can be recovered from
the feature vector, the mapping is said to provide a
\emph{representation} of the network.  An example of invertible
mapping for networks with uniform vertices and edges is the
adjacency matrix (see \ref{basic_concepts}).  Note, however, that
the characterization and classification of networks does not
necessarily require invertible measurements.  An interesting
strategy which can be used to obtain additional information about
the structure of complex networks involves applying a transformation
to the original network and obtaining the measurements from the
resulting network, as illustrated in Figure~\ref{fig:tchar}.  In
this figure, a transformation $T$ (in this case, deletion of the
vertices adjacent to just one other vertex) is applied over the
original network to obtain a transformed structure from which new
measurements $\vec{\mu}_T$ are extracted.  In case the feature
vectors $\vec{\mu}$ and ${\vec{\mu}_T}$ correspond to the same set
of measurements, it is also possible to consider the difference
between these two vectors in order to obtain additional information
about the network under analysis.

Perturbations of networks, which can be understood as a special case
of the transformation framework outlined above, can also be used to
investigate the \emph{sensitivity} of the measurements. Informally
speaking, if the measurements considered in the feature vector are
such that small changes of the network topology (e.g., add/remove a
few edges or vertices) imply large changes in the measurements (large
values of $||\Delta\vec{\mu}||$), those measurements can be considered
as being highly sensitive or unstable.  One example of such an
unstable measurement is the average shortest path length between two
vertices (see Section~\ref{perturbation}).

\begin{figure}
 \centerline{\includegraphics[width=0.8\textwidth]{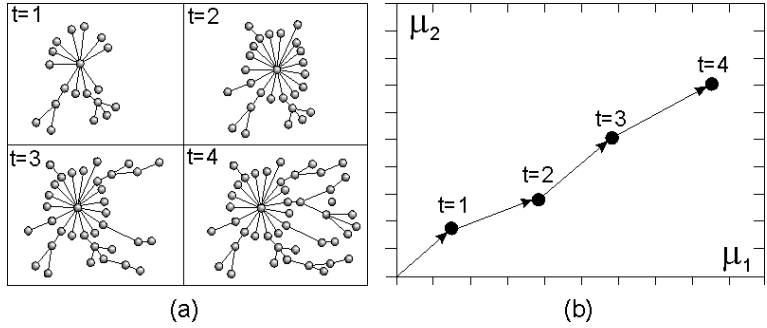}}
 \caption{Given a network undergoing some dynamical evolution (a) and
   a set of measurements (e.g., $\mu_1$ and $\mu_2$), trajectories can
   be defined in the feature space (b).}
\label{fig:dyn}
\end{figure}

Another possibility to obtain a richer set of measurements involves
the consideration of several instances along development/growth of the
network.  A feature vector $\vec{\mu}(t)$ is obtained at each ``time''
instant $t$ along the growth. Figure~\ref{fig:dyn} shows four
instances of an evolving network and the respective trajectory defined
in one of the possible feature (or phase) spaces involving two generic
measurements $\mu_1$ and $\mu_2$.  In such a way, the evolution of a
network can now be investigated in terms of a trajectory in a features
space.

Both the characterization and classification of natural and
human-made structures using complex networks imply the same
important question of \emph{how to choose the most appropriate
measurements}.  While such a choice should reflect the specific
interests and application, it is unfortunate that there is no
mathematical procedure for identifying the best measurements.  There
is an unlimited set of topological measurements, and they are often
correlated, implying redundancy. Statistical approaches to
decorrelation (e.g., principal component analysis and canonical
analysis) can help select and enhance measurements (see
Section~\ref{multivariate}), but are not guaranteed to produce
optimal results (e.g \cite{Costa:01}). Ultimately, one has to rely
on her/his knowledge of the problem and available measurements in
order to select a suitable set of features to be considered.  For
such reasons, it is of paramount importance to have a good knowledge
not only of the most representative measurements, but also of their
respective properties and interpretation.  Although a small number
of topological measurements, namely the average vertex degree,
clustering coefficient and average shortest path length, were
typically considered for complex network characterization during the
initial stages of this area, a series of new and more sophisticated
features have been proposed and used in the literature along the
last years. Actually, the fast pace of developments and new results
reported in this very dynamic area makes it particularly difficult
to follow and to organize the existing measurements.

This review starts by presenting the basic concepts and notation in
complex networks and follows by presenting several topological
measurements.  Illustrations of some of these measurements
respectively to Erd\H{o}s-R\'{e}nyi, Watts-Strogatz,
Barab\'asi-Albert, modular and geographical models are also included.
The measurements are presented in sections organized according to
their main types, including distance-based measurements, clustering
coefficients, assortativity, entropies, centrality, subgraphs,
spectral analysis, community-based measurements, hierarchical
measurements, and fractal dimensions.  A representative set of such
measurements is applied to the five considered models and the results
are presented and discussed in terms of their cross-correlations and
trajectories.  The important subjects of measurement selection and
assignment of categories to given complex networks are then covered
from the light of formal multivariate pattern recognition, including
the illustration of such a possibility by using canonical projections
and Bayesian decision theory.

\section{Basic Concepts}\label{basic_concepts}

\begin{figure}
  \centerline{\includegraphics[width=9cm]{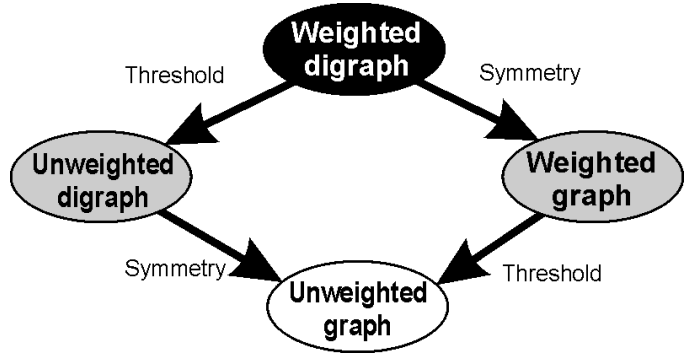}}
  \caption{The four main types of complex networks and their
    transformations. All network types can be derived from the
    weighted digraph through appropriate transformations.}
  \label{fig:types}
\end{figure}

Figure~\ref{fig:types} shows the four main types of complex networks,
which include weighted digraphs (directed graphs), unweighted
digraphs, weighted graphs and unweighted graphs.  The operation of
\emph{symmetry} can be used to transform a digraph into a graph, and
the operation of \emph{thresholding} can be applied to transform a
weighted graph into its unweighted counterpart. These types of graphs
and operations are defined more formally in the following, starting
from the concept of weighted digraph, from which all the other three
types can be derived.

A \emph{weighted directed graph}, $G$, is defined by a set
$\mathcal{N}(G)$ of $N$ \emph{vertices} (or \emph{nodes}), a set
$\mathcal{E}(G)$ of $M$ \emph{edges} (or \emph{links}), and a mapping
$\omega:\mathcal{E}(G)\mapsto\mathbb{R}$.  Each vertex can be
identified by an integer value $i = 1, 2, \ldots, N$; the edges are
identified by a pair $(i,j)$ that represents a connection going from
vertex $i$ to vertex $j$ to which a weight $\omega(i,j)$ is
associated. In the complex network literature, it is often assumed
that no self-connections or multiple connections exist; i.e.\ there
are no edges of the form $(i,i)$ and for each pair of edges
$(i_1,j_1)$ and $(i_2,j_2)$ it holds that $i_1 \neq i_2$ or $j_1 \neq
j_2$.  Graphs with self- or duplicate connections are sometimes called
\emph{multigraphs}, or \emph{degenerate} graphs. Only
non-degenerate graphs are considered henceforth. In an
\emph{unweighted digraph}, the edges have no weight, and the mapping
$\omega$ is not needed. For \emph{undirected graphs} (weighted or
unweighted), the edges have no directions; the presence of a edge
$(i,j)$ in $\mathcal{E}(G)$ thus means that a connection exist from
$i$ to $j$ and from $j$ to $i$.

A weighted digraph can be completely represented in terms of its
\emph{weight matrix} $W$, so that each element $w_{ij}=\omega(i,j)$
expresses the weight of the connection from vertex $i$ to vertex $j$.
The operation of \emph{thresholding} can be applied to a weighted
digraph to produce an unweighted counterpart.  This operation,
henceforth represented as $\delta_T(W)$, is applied to each element of
the matrix $W$, yielding the matrix $A=\delta_T(W)$.  The elements of
the matrix $A$ are computed comparing the corresponding elements of
$W$ with a specified threshold $T$; in case $|w_{ij}| > T$ we have
$a_{ij}=1$, otherwise $a_{ij}=0$.  The resulting matrix $A$ can be
understood as the \emph{adjacency matrix} of the unweighted digraph
obtained as a result of the thresholding operation. Any weighted
digraph can be transformed into a graph by using the \emph{symmetry}
operation $\sigma(W) = W+W^T$, where $W^T$ is the transpose of $W$.

For undirected graphs, two vertices $i$ and $j$ are said to be
\emph{adjacent} or \emph{neighbors}\label{neighbors} if $a_{ij} \neq
0$.  For directed graphs, the corresponding concepts are those of
\emph{predecessor} and \emph{successor}: if $a_{ij} \neq 0$ then $i$
is a predecessor of $j$ and $j$ is a successor of $i$.  The concept of
adjacency can also be used in digraphs by considering predecessors and
successors as adjacent vertices. The \emph{neighborhood} of a vertex
$i$, henceforth represented as $\nu(i)$, corresponds to the set of
vertices adjacent to $i$.

The \emph{degree of a vertex} $i$, hence $k_i$, is the number of edges
connected to that vertex, i.e.\ the cardinality of the set $\nu(i)$
(in the physics literature, this quantity is often called
``connectivity'' \cite{Dorogovtsev02}). For undirected networks it can
be computed as
\begin{equation}
  k_i = \sum_j a_{ij} = \sum_j a_{ji}.
\end{equation}
The \emph{average degree} of a network is the average of $k_i$ for all
vertices in the network,
\begin{equation}
  \langle k \rangle = \frac{1}{N} \sum_i k_i = \frac{1}{N} \sum_{ij}
  a_{ij} .
\end{equation}

In the case of directed networks, there are two kinds of degrees:
the \emph{out-degree}, $k_i^\mathrm{out}$, equal to the number of
outgoing edges (i.e.\ the cardinality of the set of successors), and
the \emph{in-degree}, $k_i^\mathrm{in}$, corresponding to the number
of incoming edges (i.e.\ the cardinality of the set of
predecessors),
\begin{eqnarray}
  k_i^\mathrm{out} & = & \sum_j a_{ij} , \\
  k_i^\mathrm{in}  & = & \sum_j a_{ji} . \\
\end{eqnarray}
Note that in this case the total degree is defined as $k_i =
k_i^\mathrm{in} + k_i^\mathrm{out}.$ The average in- and out-degrees
are the same (the network is supposed isolated)
\begin{equation}
  \langle k^\mathrm{out} \rangle =
  \langle k^\mathrm{in} \rangle =
  \frac{1}{N} \sum_{ij} a_{ij} .
\end{equation}
For weighted networks, the definitions of degree given above can be
used, but a quantity called \emph{strength} of $i$, $s_i$, defined as
the sum of the weights of the corresponding edges, is more generally
used \cite{Barthelemy04}:
\begin{eqnarray}
  s^\mathrm{out}_i & = & \sum_{j} w_{ij}\label{strength_out}, \\
  s^\mathrm{in}_i  & = & \sum_{j} w_{ji}\label{strength_in}.
\end{eqnarray}

In the general case, two vertices of a complex network are not
adjacent.  In fact, most of the networks of interest are sparse, in
the sense that only a small fraction of all possible edges are
present.  Nevertheless, two non-adjacent vertices $i$ and $j$ can be
connected through a sequence of $m$ edges
$(i,k_1),(k_1,k_2),\ldots,(k_{m-1},j)$; such set of edges is called a
\emph{walk} between $i$ and $j$, and $m$ is the \emph{length} of the
walk. We say that two vertices are \emph{connected} if there is at
least one walk connecting them.  Many measurements are based on the
length of these connecting walks (see Section~\ref{distance}).  A
\emph{loop} or \emph{cycle} is defined as a walk starting and terminating
in the same vertex $i$ and passing only once through each vertex
$k_n$.  In case all the vertices and edges along a walk are distinct,
the walk is a \emph{path}.

In undirected graphs, if vertices $i$ and $j$ are connected
and vertices $j$ and $k$ are connected, then $i$
and $k$ are also connected. This property can be used to partition the
vertices of a graph in non-overlapping subsets of connected vertices. These
subsets are called \emph{connected components} or \emph{clusters}.

If a network has too few edges, i.e.\ the average connectivity of its
vertices $\langle k\rangle$ is too small, there will be many isolated
vertices and clusters with a small number of vertices. As more edges
are added to the network, the small clusters are connected to larger
clusters; after some critical value of the connectivity, most of the
vertices are connected into a giant cluster, characterizing the
percolation \cite{Stauffer:1994} of the network. For the
Erd\H{o}s-R\'{e}nyi graph in the limit $N\rightarrow\infty$ this
happens at $\langle k\rangle=1$ \cite{Bollobas:1985}. Of special
interest is the distribution of sizes of the clusters in the
percolation point and the fraction of vertices in the giant
cluster. The critical density of edges (as well as average and
standard deviation) needed to achieve percolation can be used to
characterize network models or experimental phenomena.

Table~\ref{basic_symbols} lists the basic symbols used in the paper.

\begin{table}
  \caption{List of basic symbols used in the text.}
  \label{basic_symbols}
  \begin{center}
   \begin{tabular}{cl}
     \hline
     Symbol & Concept \\
     \hline
     $\mathcal{N}(G)$     & Set of vertices of graph $G$ \\
     $\mathcal{E}(G)$     & Set of edges of graph $G$ \\
     $|\mathcal{X}|$      & Cardinality of set $\mathcal{X}$\\
     $N$                  & Number of vertices, $|\mathcal{N}(G)|$ \\
     $M$                  & Number of edges, $|\mathcal{E}(G)|$ \\
     $W$                  & Weight matrix  \\
     $w_{ij}$             & Element of the weight matrix\\
     $A$                  & Adjacency matrix  \\
     $a_{ij}$             & Element of the adjacency matrix\\
     $k_i$                & Degree of a vertex $i$  \\
     $k_i^{\mathrm{out}}$ & Out-degree of a vertex $i$  \\
     $k_i^{\mathrm{in}}$  & In-degree of a vertex $i$  \\
     $s_i$                & Strength of a vertex $i$  \\
     $\nu(i)$             & Set of neighbors of vertex $i$\\
     $\| X \|$            & Sum of the elements of matrix $X$\\
     \hline
   \end{tabular}
  \end{center}
\end{table}

\section{Complex Network Models}\label{sec:models}

With the intent of studying the topological properties of real
networks, several network models have been proposed. Some of these
models have become subject of great interest, including random graphs,
the small-world model, the generalized random graph and
Barab\'{a}si-Albert networks.  Other models have been applied to the
study of the topology of networks with some specific features, as
geographical networks and networks with community structure. We do not
intend to cover a comprehensive a review of the various proposed
models.  Instead, the next subsections present some models used in the
discussion on network measurements (Sections~\ref{netdyn},
\ref{correlation}, and \ref{multivariate}).

\subsection{The Random Graph of Erd\H{o}s and R\'{e}nyi}

\begin{figure}
  \centerline{\includegraphics[width=10cm]{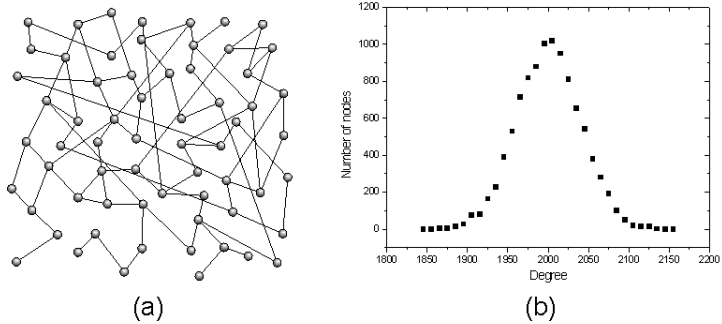}}
  \caption{The random graph of Erd\H{o}s and R\'{e}nyi: (a) an example
    of a random graph and (b) average degree distribution over $10$
    random networks formed by $10\,000$ vertices using a probability
    $p=0.2$.}
  \label{fig:random}
\end{figure}

The random graph developed by Rapoport \cite{Rapoport:1951,
Rapoport:1953, Rapoport:1957} and independently by Erd\H{o}s and
R\'{e}nyi \cite{Erdos-Renyi:1959, Erdos-Renyi:1960, Erdos-Renyi:1961}
can be considered the most basic model of complex networks. In their
$1959$ paper \cite{Erdos-Renyi:1959}, Erd\H{o}s and R\'{e}nyi
introduced a model to generate random graphs consisting of $N$
vertices and $M$ edges. Starting with $N$ disconnected vertices, the
network is constructed by the addition of $L$ edges at random,
avoiding multiple and self connections. Another similar model defines
$N$ vertices and a probability $p$ of connecting each pair of
vertices.  The latter model is widely known as Erd\H{o}s-R\'{e}nyi
(ER) model.  Figure~\ref{fig:random}(a) shows an example of this type
of network.

For the ER model, in the large network size limit ($N
\rightarrow\infty$), the average number of connections of each vertex
$\langle k \rangle$, given by
\begin{equation}
  \langle k \rangle = p(N-1),
  \label{kmed}
\end{equation}
diverges if $p$ is fixed.  Instead, $p$ is chosen as a function of
$N$ to keep $\langle k \rangle$ fixed: $p=\langle k \rangle/(N-1).$
For this model,  $P(k)$ (the degree distribution, see
Section~\ref{degrees}) is a Poisson distribution (see
Figure~\ref{fig:random}(b) and Table~\ref{models_tab}).

\subsection{The Small-World Model of Watts and Strogatz}

Many real world networks exhibit what is called the \emph{small world}
property, i.e.\ most vertices can be reached from the others through a
small number of edges. This characteristic is found, for example, in
social networks, where everyone in the world can be reached through a
short chain of social acquaintances \cite{Watts99,Watts03}. This
concept originated from the famous experiment made by Milgram in 1967
\cite{Milgram:1967}, who found that two US citizens chosen at random
were connected by an average of six acquaintances.

Another property of many networks is the presence of a large number
of loops of size three, i.e.\ if vertex $i$ is connected to vertices
$j$ and $k$, there is a high probability of vertices $j$ and $k$
being connected (the clustering coefficient,
Section~\ref{cc_section}, is high); for example, in a friendship
network, if B and C are friends of A, there is a high probability
that B and C are also friends. ER networks have the small world
property but a small average clustering coefficient; on the other
hand, regular networks with the second property are easy to
construct, but they have large average distances. The most popular
model of random networks with small world characteristics and an
abundance of short loops was developed by Watts and
Strogatz~\cite{Watts98} and is called the Watts-Strogatz (WS)
\emph{small-world model}.  They showed that small-world networks are
common in a variety of realms ranging from the \emph{C. elegans}
neuronal system to power grids. This model is situated between an
ordered finite lattice and a random graph presenting the small world
property and high clustering coefficient.

\begin{figure}
  \centerline{\includegraphics[width=9cm]{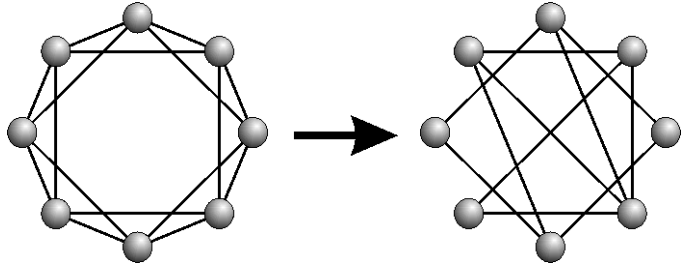}}
  \caption{The construction of a small-word network according to Watts
    and Strogatz: A regular network has its edges rewired with
    probability $p$. For $p\approx0$ the network is regular, with many
    triangles and large distances, for $p\approx1$, the network
    becomes a random network, with small distances and few triangles.}
  \label{small}
\end{figure}

\begin{figure}
  \centerline{\includegraphics[width=10cm]{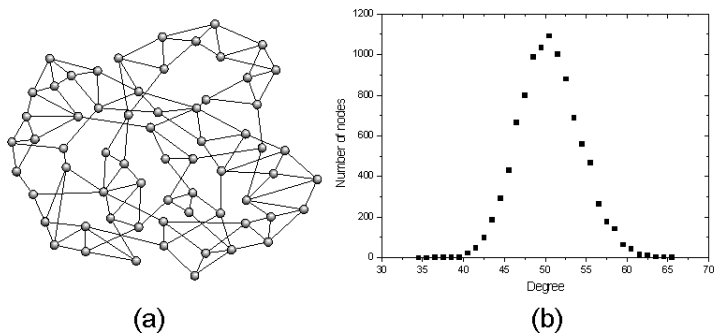}}
  \caption{The small-world model of Watts and Strogatz: (a) an example
    of a network with $N=64$ vertices, $\kappa=2$, $p=0.1$, and (b)
    average degree distribution over $10$ WS networks with $10\,000$
    vertices, $\kappa=25$ and $p=0.3$.}
  \label{fig:WS}
\end{figure}

To construct a small-word network, one starts with a regular lattice
of $N$ vertices (Figure~\ref{small}) in which each vertex is connected
to $\kappa$ nearest neighbors in each direction, totalizing $2\kappa$
connections, where $N\gg\kappa\gg\log(N)\gg 1$. Next, each edge is
randomly rewired with probability $p$. When $p = 0$ we have an ordered
lattice with high number of loops but large distances and when
$p\rightarrow 1$, the network becomes a random graph with short
distances but few loops. Watts and Strogatz have shown that, in an
intermediate regime, both short distances and a large number of loops
are present.  Figure~\ref{fig:WS}(a) shows an example of a
Watts-Strogatz network. Alternative procedures to generate small-world
networks based on addition of edges instead of rewiring have been
proposed \cite{Monasson99,Newman-Watts99}, but are not discussed here.

The degree distribution for small-world networks is similar to that of
random networks, with a peak at $\langle k \rangle = 2\kappa$ (see
also Table~\ref{models_tab}).

\subsection{Generalized Random Graphs}\label{genrandom}

A common way to study real networks is to compare their
characteristics with the values expected for similar random
networks. As the degrees of the vertices are important features of
the network, it is interesting to make the comparison with networks
with the same degree distribution. Models to generate networks with a
given degree distribution, while being random in other aspects, have
been proposed.

Bender and Canfield \cite{Bender78} first proposed a model to generate
random graphs with a pre-defined degree distribution called
configuration model. Later, Molloy and Reed
\cite{Molloy:1995,Molloy:1998} proposed a different method that
produces multigraphs (i.e.\ loops and multiple edges between the same
pair of vertices are allowed).

The common method used to generate this kind of random graph
involves selecting a degree sequence specified by a set $\{ k_i\}$
of degrees of the vertices drawn from the desired distribution
$P(k)$. Afterwards, to each vertex $i$ is associated a number $k_i$
of ``stubs'' or ``spokes'' (ends of edges emerging from a vertex)
according to the desired degree sequence. Next, pairs of such stubs
are selected uniformly and joined together to form an edge.  When
all stubs have been used up, a random graph that is a member of the
ensemble of graphs with that degree sequence is obtained
\cite{Newman:chapter,Newman:2002,Newman-Strogatz01}.

Another possibility, the \emph{rewiring method}, is to start with a
network (possibly a real network under study) that already has the
desired degree distribution, and then iteratively chose two edges and
interchange the corresponding attached vertices \cite{Milo:2003}.
This rewiring procedure is used in some results presented in
Section~\ref{perturbation}.

Due to its importance and amenability to analytical treatment, many
works deal with this model, including the papers of Newman
\cite{Newman:2003:survey}, Aiello \emph{et al.}\ \cite{Aiello00}, Chung and Lu
\cite{Chung02} and Cohen and Havlin \cite{Cohen03}.

\subsection{Scale-free Networks of Barab\'{a}si and Albert}

\begin{figure}
  \centerline{\includegraphics[width=10cm]{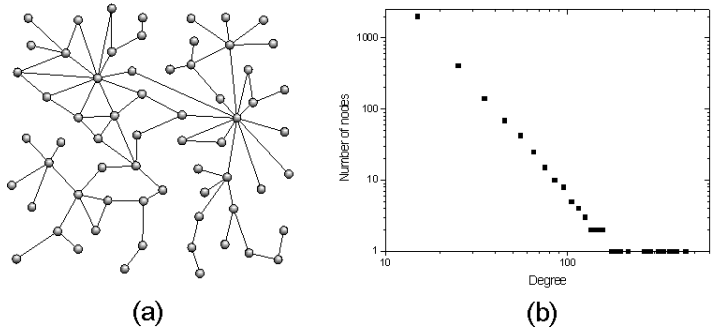}}
  \caption{The scale-free network of Barab\'{a}si and Albert. (a) an
    example of a scale-free network and (b) average degree
    distribution over $10$ Barab\'{a}si-Albert networks formed by
    $10,000$ vertices using $m=5$.  The degree distribution follows a
    power law, in contrast to that presented in
    Figure~\ref{fig:random}.}
  \label{fig:sfree}
\end{figure}

After Watts and Strogatz's model, Barab\'{a}si and Albert
\cite{Barabasi97} showed that the degree distribution of many real
systems is characterized by an uneven distribution. Instead of the
vertices of these networks having a random pattern of connections with
a characteristic degree, as with the ER and WS models (see
Figure~\ref{fig:random}(a)), some vertices are highly connected while
others have few connections, with the absence of a characteristic
degree.  More specifically, the degree distribution has been found to
follow a power law for large $k$,
\begin{equation}\label{scale_free_eq}
    P(k)\sim k^{-\gamma}
\end{equation}
(see Figure~\ref{fig:sfree}(b)). These networks are called
\emph{scale-free} networks.

A characteristic of this kind of network is the existence of
\emph{hubs}, i.e.\ vertices that are linked to a significant fraction
of the total number of edges of the network.

The Barab\'asi-Albert (BA) network model is based on two basic rules:
\emph{growth} and \emph{preferential attachment}. The network is
generated starting with a set of $m_0$ vertices; afterwards, at each
step of the construction the network \emph{grows} with the addition of
new vertices. For each new vertex, $m$ new edges are inserted between
the new vertex and some previous vertex. The vertices which receive
the new edges are chosen following a \emph{linear preferential
attachment} rule, i.e.\ the probability of the new vertex $i$ to
connect with an existing vertex $j$ is proportional to the degree of
$j$,
\begin{equation}
\mathcal{P}(i\rightarrow j) = \frac{k_j}{\sum_u k_u}.
\end{equation}
Thus, the most connected vertices have greater probability to
receive new vertices. This is known as ``the rich get richer''
paradigm.

Figure~\ref{fig:sfree}(a) shows an example of a Barab\'{a}si-Albert
network.

\begin{sidewaystable}
   \caption{Analytical result of some basic measurements for the
     Erd\H{o}s-R\'{e}nyi,Watts-Strogatz and Barab\'{a}si-Albert network
     models.}
   \label{models_tab}
   \begin{center}
   \begin{tabular}{|c|c|c|c|}
     \hline
     Measurement & Erd\H{o}s-R\'{e}nyi & Watts-Strogatz & Barab\'{a}si-Albert \\
     \hline
     Degree & & &\\
     distribution &$P(k) = \frac{e^{-\langle k\rangle}\langle k\rangle^{k}}{k!}$ &$P(k) = \sum_{i=1}^{min(k-\kappa,\kappa)} \left( \begin{array}{c} \kappa \\ i \\\end{array}\right) (1 -p)^ip^{\kappa-i}\frac{(p\kappa)^{k-\kappa-i}}{(k-\kappa-i)!}e^{-p\kappa}$&$  P(k)\sim k^{-3}$ \\    \hline
     Average & & & \\
     vertex degree & $\langle k\rangle = p(N-1)$ & $\langle k\rangle = 2\kappa^\star$ & $\langle k\rangle = 2m$ \\
     \hline
     Clustering & & &\\
     coefficient & $C = p$ & $C(p) \sim \frac{3(\kappa-1)}{2(2\kappa -1)}(1-p)^3$ & $C \sim N^{-0.75}$ \\
     \hline
     Average & & & \\ path length & $\ell \sim \frac{\ln N}{\ln\langle k \rangle }$ & $\ell(N,p) \sim p^\tau f(Np^\tau)^\ast$ &$\ell \sim \frac{\log N}{\log( \log N )}$ \\
     \hline
   \end{tabular}
   \end{center}
 \begin{small}
   {$\star$ In WS networks, the value $\kappa$ represents the number of
     neighbors of each vertex in the initial regular network (in
     Figure~\ref{small}, $\kappa = 4$). \\$\ast$ The function $f(u)=$
     constant if $u\ll 1$ or $f(u) = \ln(u)/u$ if $u\gg 1$.}
 \end{small}
 \end{sidewaystable}

\subsection{Networks with Community Structure}

Some real networks, such as social and biological networks, present
modular structure \cite{Girvan:2002}. These networks are formed by
sets or \emph{communities} of vertices such that most connections are
found between vertices inside the same community, while connections
between vertices of different communities are less common. A model to
generate networks with this property was proposed by Girvan and Newman
\cite{Girvan:2002}.  This model is a kind of random graph constructed
with different probabilities. Initially, a set of $N$ vertices is
classified into $c$ communities. At each following step, two vertices
are selected and linked with probability $p_\mathrm{in}$, if they are
in the same community, or $p_\mathrm{out}$, if they are in different
communities. The values of $p_\mathrm{in}$ and $p_\mathrm{out}$ should
be chosen so as to generate networks with the desired sharpness in the
distinction of the communities. When $p_\mathrm{out} \ll
p_\mathrm{in}$, the communities can be easily identified. On the other
hand, when $p_\mathrm{out} \approx p_\mathrm{in}$, the communities
become blurred.

\begin{figure}
  \centerline{\includegraphics[width=10cm]{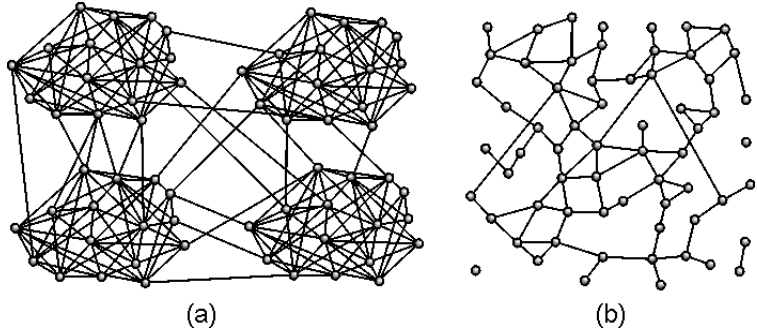}}
  \caption{(a) An example of a random network with community structure
    formed by $64$ vertices divides in $4$ communities. (b) An example
    of geographical network formed by $64$ vertices.}
  \label{com_geo}
\end{figure}

Figure~\ref{com_geo}(a) presents a network generated by using the
procedure above.

\subsection{Geographical Models}

Complex networks are generally considered as lying in an abstract
space, where the position of vertices has no particular meaning. In
the case of several kinds of networks, such as protein-protein
interaction networks or networks of movie actors, this consideration
is reasonable. However, there are many networks where the position of
vertices is particularly important as it influences the network
evolution.  This is the case of highway networks or the Internet, for
example, where the position of cities and routers can be localized in
a map and the edges between correspond to real physical entities, such
as roads and optical fibers \cite{Gastner06:EPJB}. This kind of
networks is called \emph{geographical} or \emph{spatial}
networks. Other important examples of geographical networks are power
grids \cite{Albert04,Kinney05}, airport networks
\cite{Barrat:2004,Guimera05,Hayashi05}, subway
\cite{Latora-Marchiori02} and neural networks \cite{Sporns02}.

In geographical networks, the existence of a direct connection
between vertices can depend on a lot of constraints such as the
distance between them, geographical accidents, available resources
to construct the network, territorial limitation and so on. The
models considered to represent these networks should consider these
constraints.

A simple way to generate geographical networks, used in the results
described in Sections~\ref{netdyn}, \ref{correlation}, and
\ref{multivariate}, is to distribute $N$ vertices at random in a
two-dimensional space $\Omega$ and link them with a given probability
which decays with the distance, for instance
\begin{equation}
\mathcal{P}(i\rightarrow j) \sim e^{-\lambda s_{ij}}\label{dist_geo}
\end{equation}
where $s_{ij}$ is the geographical distance of the vertices and
$\lambda$ fixes the length scale of the edges.  This model generates a
Poisson degree distribution as observed for random graphs and can be
used to model road networks (see Figure~\ref{com_geo}(b)).
Alternatively, the network development might start with few nodes
while new nodes and connections are added at each subsequent time step
(spatial growth). Such a model is able to generate a wide range of
network topologies including small-world and linear scale-free
networks~~\cite{Kaiser_Hilgetag:04}.

\section{Measurements Related with Distance}\label{distance}

For undirected, unweighted graphs, the number of edges in a path
connecting vertices $i$ and $j$ is called the \emph{length} of the
path.  A \emph{geodesic path} (or \emph{shortest path}), between
vertices $i$ and $j$, is one of the paths connecting these vertices
with minimum length (many geodesic paths may exist between two
vertices); the length of the geodesic paths is the \emph{geodesic
distance} $d_{ij}$ between vertices $i$ and $j$. If the graph is
weighted, the same definition can be used, but generally one is
interested in taking into account the edge weights. Two main
possibilities include: first, the edge weights may be proportionally
related to some physical distance, for example if the vertices
correspond to cities and the weights to distances between these
cities through given highways. In this case, one can compute the
distance along a path as the sum of the weights of the edges in the
path. Second, the edge weights may reflect the strength of
connection between the vertices, for example if the vertices are
Internet routers and the weights are the bandwidth of the edges, the
distance corresponding to each edge can be taken as the reciprocal
of the edge weight, and the path length is the sum of the reciprocal
of the weight of the edges along the path. If there are no paths
from vertex $i$ to vertex $j$, then $d_{ij} = \infty.$ For digraphs,
the same definitions can be used, but in general $d_{ij} \neq
d_{ji}$, as the paths from vertex $i$ to vertex $j$ are different
from the paths from $j$ to $i$.

Distance is an important characteristic that depends on the overall
network structure. The following describes some measurements based on
vertex distance.

\subsection{Average Distance}

We can define a network measurement by computing the mean value of
$d_{ij}$, known as \emph{average geodesic distance}:
\begin{equation}
 \ell = \frac{1}{N(N-1)}\sum_{i \ne j} d_{ij}.
 \label{meandist}
\end{equation}
A problem with this definition is that it diverges if there are
unconnected vertices in the network. To circumvent this problem, only
connected pairs of vertices are included in the sum. This avoids the
divergence, but introduces a distortion for networks with many
unconnected pairs of vertices, which will show a small value of
average distance, expected only for networks with a high number of
connections.  Latora and Marchiori \cite{Latora01} proposed a closely
related measurement that they called \emph{global efficiency}:
\begin{equation}
  E = \frac{1}{N(N-1)}\sum_{i \ne j} \frac{1}{d_{ij}},
  \label{netefficiency}
\end{equation}
where the sum takes all pairs of vertices into account. This measurement
quantifies the efficiency of the network in sending information
between vertices, assuming that the efficiency for sending information
between two vertices $i$ and $j$ is proportional to the reciprocal of
their distance. The reciprocal of the global efficiency is the
\emph{harmonic mean} of the geodesic distances:
\begin{equation}
  h = \frac{1}{E}.
  \label{hmeandist}
\end{equation}
As Eq.~(\ref{hmeandist}) does not present the divergence problem of
Eq.~(\ref{meandist}), it is therefore a more appropriate measurement
for graphs with more than one connected component.

The determination of shortest distances in a network is only possible
with global information on the structure of the network. This
information is not always available. When global information is
unavailable, navigation in a network must happen using limited, local
information and a specific algorithm. The effective distance between
two vertices is thus generally larger than the shortest distance, and
dependent on the algorithm used for navigation as well as network
structure \cite{Guimera:2002}.\label{navigation}

\subsection{Vulnerability}

In infrastructure networks (like WWW, the Internet, energy supply,
etc), it is important to know which components (vertices or edges)
are crucial to their best functioning. Intuitively, the critical
vertices of a network are their hubs (vertices with higher degree),
however there are situations in which they are not necessarily most
vital for the performance of the system which the network underlies.
For instance, all vertices of a network in the form of a binary tree
have equal degree, therefore there is no hub, but disconnection of
vertices closer to the root and the root itself have a greater
impact than of those near the leaves. This suggests that networks
have a hierarchical property, which means that the most crucial
components are those in higher positions in the hierarchy.

A way to find critical components of a network is by looking for the
most vulnerable vertices. If we associate the performance of a network
with its global efficiency, Eq.~(\ref{netefficiency}), the
\emph{vulnerability} of a vertex can be defined as the drop in
performance when the vertex and all its edges are removed from the
network \cite{Goldshtein04}
\begin{equation}
  V_i = \frac{E - E_i}{E}
  \label{nodevul}
\end{equation}
where $E$ is the global efficiency of the original network and $E_i$
is the global efficiency after the removal of the vertex $i$ and all
its edges. As suggested by Gol'dshtein \emph{et
al.} \cite{Goldshtein04}, the ordered distribution of vertices with
respect to their vulnerability $V_i$ is related to the network
hierarchy, thus the most vulnerable (critical) vertex occupies the
highest position in the network hierarchy.

A measurement of network vulnerability \cite{Latora04} is the maximum
vulnerability for all of its vertices:
\begin{equation}
  V = \max_{i} V_i.
  \label{netvulnerability}
\end{equation}

\section{Clustering and Cycles}\label{cc_section}

A characteristic of the Erd\H{o}s-R\'enyi model is that the local
structure of the network near a vertex tends to be a tree. More
precisely, the probability of loops involving a small number of
vertices goes to $0$ in the large network size limit. This is in
marked contrast with the profusion of short loops which appear in
many real-world networks. Some measurements proposed to study the
cyclic structure of networks and the tendency to form sets of
tightly connected vertices are described in the following.

\subsection{Clustering Coefficients}

One way to characterize the presence of loops of order three is
through the \emph{clustering coefficient}.

Two different clustering coefficients are frequently used. The first,
also known as \emph{transitivity} \cite{Newman00:PRE64}, is based on
the following definition for undirected unweighted networks:
\begin{equation}
  C = \frac{3 N_{\triangle}}{N_3},
  \label{clustcoeff}
\end{equation}
where $N_{\triangle}$ is the number of triangles in the network and
$N_3$ is the number of connected triples. The factor three accounts
for the fact that each triangle can be seen as consisting of three
different connected triples, one with each of the vertices as central
vertex, and assures that $0 \leq C \leq 1$. A triangle is a set of
three vertices with edges between each pair of vertices; a connected
triple is a set of three vertices where each vertex can be reached
from each other (directly or indirectly), i.e.\ two vertices must be
adjacent to another vertex (the central vertex).  Therefore we have
\begin{eqnarray}
  N_\triangle & = & \sum_{k>j>i} a_{ij}a_{ik}a_{jk},\\
  N_3         & = & \sum_{k>j>i} (a_{ij}a_{ik}+a_{ji}a_{jk}+a_{ki}a_{kj}),
  \label{N_3}
\end{eqnarray}
where the $a_{ij}$ are the elements of the adjacency matrix $A$ and
the sum is taken over all triples of distinct vertices $i$, $j$, and
$k$ only one time.

It is also possible to define the clustering coefficient of a given
vertex $i$ \cite{Watts98} as:
\begin{equation}
  C_i = \frac{N_{\triangle}(i)}{N_{3}(i)},
  \label{nodeclustcoeff}
\end{equation}
where $N_{\triangle}(i)$ is the number of triangles involving vertex $i$
and $N_{3}(i)$ is the number of connected triples having $i$ as the
central vertex:
\begin{eqnarray}
  N_\triangle(i) & = & \sum_{k>j} a_{ij}a_{ik}a_{jk},\\
  N_3(i)         & = & \sum_{k>j} a_{ij}a_{ik},
  \label{N_3_i}
\end{eqnarray}
If $k_i$ is the number of neighbors of vertex $i$, then
$N_3(i)=k_i(k_i-1)/2$.  $N_\triangle(i)$ counts the number of edges
between neighbors of $i$. Representing the number of edges between
neighbors of $i$ as $l_i$, Eq.~(\ref{nodeclustcoeff}) can be rewritten
as:
\begin{equation}
  C_i = \frac{2 l_i}{k_i(k_i-1)}.
  \label{nodeclustcoeff2}
\end{equation}

Using $C_i$, an alternative definition of the network clustering
coefficient (different from that in Eq.~(\ref{clustcoeff})) is
\begin{equation}
  \widetilde{C} = \frac{1}{N}\sum_{i}C_i.
  \label{clustercoeff2}
\end{equation}
The difference between the two definitions is that the average in
Eq.~(\ref{clustcoeff}) gives the same weight to each triangle in the
network, while Eq.~(\ref{clustercoeff2}) gives the same weight to each
vertex, resulting in different values because vertices of higher degree are
possibly involved in a larger number of triangles than vertices of
smaller degree.

For weighted graphs, Barth\'elemy \cite{Barthelemy04} introduced the
concept of \emph{weighted clustering coefficient} of a vertex,
\begin{equation}
  C^w_i = \frac{1}{s_i(k_i-1)}
  \sum_{(j,k)} \frac{w_{ij}+w_{ik}}{2} a_{ij}a_{ik}a_{jk},
  \label{nodeweightedcluster}
\end{equation}
where the normalizing factor $s_i(k_i-1)$ ($s_i$ is the strength of
the vertex, see Section~\ref{basic_concepts}) assures that $0 \leq
C^w_i \leq 1$. From this equation, a possible definition of clustering
coefficient for weighted networks is
\begin{equation}
  C^w = \frac{1}{N}\sum_i C^w_i.
  \label{weightedcluster}
\end{equation}
Another definition for clustering in weighted networks \cite{Onnela04}
is based on the intensity of the triangle subgraphs,
(see Section~\ref{weighted_subgraphs}),
\begin{equation}
  \widetilde{C}_i^w = \frac{2}{k_i(k_i-1)}\sum_{(j,k)}
  (\hat{w}_{ij}\hat{w}_{jk}\hat{w}_{ki})^{1/3},
  \label{weightedcluster2}
\end{equation}
where $\hat{w}_{ij}=w_{ij}/\max_{ij} w_{ij}$.

Given the clustering coefficients of the vertices, the clustering
coefficient can be expressed as a function of the degree of the
vertices:
\begin{equation}
  C(k) = \frac{\sum_i C_i\delta_{k_i k}}{\sum_i\delta_{k_i k}},
  \label{cxk}
\end{equation}
where $\delta_{ij}$ is the Kronecker delta. For some networks, this
function has the form $C(k)\sim k^{-\alpha}$.  This behavior has been
associated with a hierarchical structure of the network, with the
exponent $\alpha$ being called its \emph{hierarchical exponent}
\cite{Ravasz03}. Soffer and V\'{a}zquez \cite{Soffer04} found that
this dependence of the clustering coefficient with $k$ is to some
extent due to the degree correlations (Section~\ref{degrees}) of the
networks, with vertices of high degree connecting with vertices of low
degree. They suggested a new definition of clustering coefficient
without degree correlation bias:
\begin{equation}
  \hat{C}_i = \frac{l_i}{\omega_i},
  \label{cc_vazquez}
\end{equation}
where $l_i$ is the number of edges between neighbors of $i$ and
$\omega_i$ is the maximum number of edges possible between the
neighbors of vertex $i$, considering their vertex degrees and the fact
that they are necessarily connected with vertex $i$.

\subsection{Cyclic Coefficient}

Kim and Kim \cite{kim05} defined a cyclic coefficient in order to
measure how cyclic a network is. The \emph{local cyclic coefficient}
of a vertex $i$ is defined as the average of the inverse of the sizes
of the smallest cycles formed by vertex $i$ and its neighbors,
\begin{equation}
  \Theta_i = \frac{2}{k_i(k_i-1)}
        \sum_{(j,k)}{\frac{1}{S_{ijk}}a_{ij}a_{ik}},
  \label{eq:cyclicnode}
\end{equation}
where $S_{ijk}$ is the size of the smallest cycle which passes
through vertices $i$, $j$ and $k$. Note that if vertices $j$ and $k$
are connected, the smallest cycle is a triangle and $S_{ijk}=3$. If
there is no loop passing through $i$, $j$ and $k$, then these vertices
are tree-like connected and $S_{ijk}=\infty$. The cyclic coefficient
of a network is the average of the cyclic coefficient of all its
vertices:
\begin{equation}
  \Theta = \frac{1}{N}\sum_i \Theta_i.
  \label{eq:cyclicnet}
\end{equation}

\subsection{Rich-Club Coefficient}

In science, influential researchers of some areas tend to form
collaborative groups and publish papers together
\cite{Colliza06:NaturePhysics}. This tendency is observed in other
real networks and reflect the tendency of hubs to be well connected
with each other. This phenomenon, known as \emph{rich-club}, can be
measured by the \emph{rich-club coefficient}, introduced by Zhou and
Mondragon \cite{Zhou04:IEE}. The rich-club of degree $k$ of a network
$G$ is the set of vertices with degree greater than $k$,
$\mathcal{R}(k) = \{v\in\mathcal{N}(G)|k_v>k\}.$ The rich-club
coefficient of degree $k$ is given by
\begin{equation}
  \phi(k) = \frac{1}{|\mathcal{R}(k)|(|\mathcal{R}(k)| - 1)}
      \sum_{i,j\in\mathcal{R}(k)}a_{ij}
\end{equation}
(the sum corresponds to two times the number of edges between vertices
in the club). This measurement is similar to that defined before for
the clustering coefficient (see Eq.~(\ref{nodeclustcoeff2})), giving
the fraction of existing connections among vertices with degree higher
than $k$.

Colizza \emph{et al.}\ \cite{Colliza06:NaturePhysics} derived an
analytical expression of the rich-club coefficient, valid for
uncorrelated networks,
\begin{equation}
\phi_{\mathrm{unc}}(k) \sim \frac{k^2}{\langle k \rangle N}.
\end{equation}

The definition of the \emph{weighted rich-club coefficient} for
weighted networks is straightforward. If $\mathcal{R}^w(s)$ is the set
of vertices with strength greater than $s$, $\mathcal{R}^w(s) = \{v\in
G | s_v>s\},$
\begin{equation}
  \phi^{w}(s) = \frac
                  {\sum_{i,j\in\mathcal{R}^w(s)}w_{ij}}
                  {\sum_{i  \in\mathcal{R}^w(s)}s_i}
\end{equation}
(the sum in the numerator give two times the weight of the edges
between elements of the rich-club, the sum in the denominator gives
the total strength of the vertices in the club).

\section{Degree Distribution and Correlations}\label{degrees}

The degree is an important characteristic of a vertex
\cite{Dorogovtsev04}. Based on the degree of the vertices, it is
possible to derive many measurements for the network. One of the
simplest is the \emph{maximum degree}:
\begin{equation}
  k_{\mathrm{max}} = \max_{i} k_i.
  \label{maxdeg}
\end{equation}

Additional information is provided by the \emph{degree
distribution}, $P(k)$, which expresses the fraction of vertices in a
network with degree $k$. An important property of many real world
networks is their power law degree distribution \cite{Barabasi97}.
For directed networks there are an out-degree distribution
$P^\mathrm{out}(k^\mathrm{out})$, an in-degree distribution
$P^\mathrm{in}(k^\mathrm{in})$, and the joint in-degree and
out-degree distribution
$P^\mathrm{io}(k^\mathrm{in},k^\mathrm{out})$. The latter
distribution gives the probability of finding a vertex with
in-degree $k\mathrm{in}$ and out-degree $k_\mathrm{out}$. Similar
definitions considering the strength of the vertices can be used for
weighted networks.  An objective quantification of the level to
which a log-log distribution of points approach a power law can be
provided by the respective Pearson coefficient, which is henceforth
called \emph{straightness} and abbreviated as $\mathit{st}$.

It is often interesting to check for correlations between the degrees
of different vertices, which have been found to play an important role
in many structural and dynamical network properties
\cite{Maslov02}. The most natural approach is to consider the correlations
between two vertices connected by an edge. This correlation can be
expressed by the joint degree distribution $P(k,k')$, i.e.\ as the
probability that an arbitrary edge connects a vertex of degree $k$ to
a vertex of degree $k'$. Another way to express the dependence between
vertex degrees is in terms of the \emph{conditional probability} that
an arbitrary neighbor of a vertex of degree $k$ has degree $k'$
\cite{Boguna02:PRE,Boguna03:LNP},
\begin{equation}
  P(k'|k) = \frac{\langle k \rangle P(k,k')}{k P(k)}.
  \label{condkk}
\end{equation}
Notice that $\sum_{k'}P(k'|k) = 1$.  For undirected networks, $P(k,k')
= P(k',k)$ and $k'P(k|k')P(k') = kP(k'|k)P(k)$. For directed networks,
$k$ is the degree at the tail of the edge, $k'$ is the degree at the
head, both $k$ and $k'$ may be in-, out-, or total degrees, and in
general $P(k,k') \ne P(k,k')$. For weighted networks the strength $s$
can be used instead of $k$.

$P(k,k')$ and $P(k|k')$ characterize formally the vertex degree
correlations, but they are difficult to evaluate experimentally,
especially for fat-tailed distributions, as a consequence of the
finite network size and the resulting small sample of vertices with
high degree.  This problem can be addressed by computing the
\emph{average degree of the nearest neighbors} of vertices with a
given degree $k$ \cite{Pastor01}, which is given by
\begin{equation}
  k_{\mathrm{nn}}(k) = \sum_{k'} k' P(k'|k).
  \label{knn}
\end{equation}
If there are no correlations, $k_\mathrm{nn}(k)$ is independent of
$k$, $k_\mathrm{nn}(k) = \langle k^2 \rangle / \langle k
\rangle$. When $k_\mathrm{nn}(k)$ is an increasing function of $k$,
vertices of high degree tend to connect with vertices of high degree,
and the network is classified as \emph{assortative}, whereas whenever
$k_\mathrm{nn}(k)$ is a decreasing function of $k$, vertices of high
degree tend to connect with vertices of low degree, and the network is
called \emph{disassortative} \cite{Newman01}.

Another way to determine the degree correlation is by considering the
Pearson correlation coefficient of the degrees at both ends of the
edges \cite{Newman01}:
\begin{equation}
  r = \frac{
            \frac{1}{M} \sum_{j>i}k_i k_j a_{ij} -
            \left[ \frac{1}{M} \sum_{j>i}
                   \frac{1}{2} (k_i+k_j) a_{ij} \right]^2
           }
           {
            \frac{1}{M}\sum_{j>i}\frac{1}{2}(k_i^2+k_j^2) a_{ij} -
            \left[ \frac{1}{M}\sum_{j>i}
                   \frac{1}{2}(k_i+k_j) a_{ij}  \right]^2
           },
  \label{pearson}
\end{equation}
where $M$ is the total number of edges. If $r>0$ the network is
assortative; if $r<0$, the network is disassortative; for $r=0$
there are no correlation between vertex degrees.

Degree correlations can be used to characterize networks and to
validate the ability of network models to represent real network
topologies.  Newman \cite{Newman01} computed the Pearson correlation
coefficient for some real and model networks and discovered that,
although the models reproduce specific topological features such as
the power law degree distribution or the small-world property, most
of them (e.g., the Erd\H{o}-R\'{e}nyi and Barab\'{a}si-Albert
models) fail to reproduce the assortative mixing ($r = 0$ for the
Erd\H{o}-R\'{e}nyi and Barab\'{a}si-Albert models). Further, it was
found that the assortativity depends on the type of network. While
social networks tend to be assortative, biological and technological
networks are often disassortative. The latter property is
undesirable for practical purposes, because assortative networks are
known to be resilient to simple target attack, at the least. So, for
instance, in disease propagation, social networks would ideally be
vulnerable (i.e. the network is dismantled into connected
components, isolating the focus of disease) and technological and
biological networks should be resilient against attacks. The degree
correlations are related to the network evolution process and,
therefore, should be taken into account in the development of new
models as done, for instance, in the papers by Catanzaro \emph{et
al.}  \cite{Catanzaro04:PRE} on social networks, Park and Newman
\cite{Park03:PRE} on the Internet, and Berg \emph{et al.}
\cite{Berg04:BMC} on protein interaction networks.  Degree
correlations also have strong influence on dynamical processes like
instability \cite{Brede05}, synchronization \cite{Bernardo05,
Bernardo06} and spreading \cite{Boguna02:PRE, Madar04, Zhou06}. For
additional discussions about dynamical process as in networks see
Ref.~\cite{Boccaletti05}.

\section{Networks with Different Vertex Types}\label{assortativity}

Some networks include vertices of different types. For example, in a
sociological network where the vertices are people and the edges are a
social relation between two persons (e.g., friendship), one may be
interested in answering questions like: how probable is a friendship
relation between two persons of different economic classes?  In this
case, it is interesting to consider that the vertices are not
homogeneous, having different types.  In the following, measurements
associated with such kind of networks are discussed.

\subsection{Assortativity}

For networks with different types of vertices, a type mixing matrix $E$
can be defined, with elements $e_{st}$ such that $e_{st}$ is the
number of edges connecting vertices of type $s$ to vertices of type $t$ (or
the total strength of the edges connecting the two vertices of the given
types, for weighted networks).  It can be normalized as
\begin{equation}
  \hat{E} = \frac{E}{\| E \|},
  \label{enorm}
\end{equation}
where $\| X \|$ (cardinality) represents the sum of all elements of
matrix $X$.

The probability of a vertex of type $s$ having a neighbor of type $t$
therefore is
\begin{equation}
  P^\mathrm{(type)}(t|s) = \frac{\hat{e}_{st}}{\sum_u \hat{e}_{su}}.
  \label{pmix}
\end{equation}
Note that $\sum_t P^\mathrm{(type)}(t|s) = 1.$

$P^\mathrm{(type)}(s,t)$ and $\hat{E}$ can be used to quantify the
tendency in the network of vertices of some type to connect to
vertices of the same type, called \emph{assortativity}. We can define
an assortativity coefficient \cite{Gupta89,Newman:2003:survey} as:
\begin{equation}
  \widetilde{\mathbb{Q}} = \frac{\sum_s P^\mathrm{(type)}(s|s) - 1}{N_T - 1},
  \label{eq:guptaassor}
\end{equation}
where $N_T$ is the number of different vertex types in the network. It
can be seen that $0 \leq \tilde\mathbb{Q} \leq 1$, where
$\tilde\mathbb{Q}=1$ for a perfectly assortative network (only edges
between vertices of the same type) and $\tilde\mathbb{Q}=0$ for random
mixing. But each vertex type has the same weight in
$\tilde\mathbb{Q}$, regardless of the number of vertices of that type.
An alternative definition that avoids this problem \cite{Newman03b}
is:
\begin{equation}
  \mathbb{Q} = \frac{\mathrm{Tr}\,\hat{E} - \|\hat{E}^2\|}
           {1 - \|\hat{E}^2\|}.
  \label{eq:newmanassor}
\end{equation}

It is interesting to associate the vertex type to its degree.  The
Pearson correlation coefficient of vertex degrees,
Eq.~(\ref{pearson}), can be considered as an assortativity coefficient
for this case.

\subsection{Bipartivity Degree}\label{bipartite_section}

A special case of disassortativity is that of bipartite networks.  A
network is called \emph{bipartite} if its vertices can be separated
into two sets such that edges exist only between vertices of different
sets.  It is a known fact that a network is bipartite if and only if
it has no loops of odd length (e.g. \cite{Holme:2003}).  Although some
networks are bipartite by construction, others, like a network of
sexual contacts, are only approximately bipartite. A way to quantify
how much a network is bipartite is therefore needed. A possible
measurement is based on the number of edges between vertices of the
same subset in the best possible division \cite{Holme:2003},
\begin{equation}
  b = 1 - \frac{\sum_{ij}
  a_{ij}\delta_{\vartheta(i),\vartheta(j)}}{\sum_{ij} a_{ij}},
  \label{eq:bipartholme}
\end{equation}
where $\vartheta(i)$ maps a vertex $i$ to its type and $\delta$ is the
Kronecker delta. The smallest value of $b$ for all possible divisions
is the bipartivity of the network.  The problem with this measurement
is that its computation is NP-complete, due to the necessity of
evaluating $b$ for the best possible division.  A measurement that
approximates $b$ but is computationally easier was proposed in
\cite{Holme:2003}, based on a process of marking the minimum possible
number of edges as responsible for the creation of loops of odd
length.

Another approach is based on the subgraph centrality
\cite{Estrada:2005a} (Section~\ref{subgraphcent}). The subgraph
centrality of the network, Eq.~(\ref{eq:subcentnet}), is divided in a
part due to even closed walks and a part due to odd closed walks (a
closed walk is a walk, possibly with repetition of vertices, ending on
the starting vertex). As odd closed walks are not possible in bipartite
networks, the fraction of the subgraph centrality of the network due
to even closed walks can be used as the bipartivity degree
\cite{Estrada:2005a}:
\begin{equation}
  \beta = \frac{\mathit{SC}_\mathrm{even}}{\mathit{SC}} =
\frac{\sum_{j=1}^{N}\cosh\lambda_j}{\sum_{j=1}^{N}e^{\lambda_j}},
  \label{eq:bipartestrada}
\end{equation}
where $\mathit{SC}$ is the subgraph centrality of the network
(Section~\ref{subgraphcent}), $\mathit{SC}_\mathrm{even}$ is the
subgraph centrality due to the even closed walks and the $\lambda_j$
are the eigenvalues of the adjacency matrix of the network.

\section{Entropy}\label{entropy}

Entropy is a key concept in thermodynamics, statistical mechanics
\cite{Reif65} and information theory \cite{Brillouin04}.  It has
important physical implications related to the amount of
``disorder'' and information in a system \cite{Reichl98}. In
information theory, entropy describes how much randomness is present
in a signal or random event \cite{Shannon63:book}. This concept can
be usefully applied to complex networks.

\subsection{Entropy of the Degree Distribution}

The \emph{entropy of the degree distribution} provides an average
measurement of the heterogeneity of the network, which can be defined
as
\begin{equation}
  \label{eq:entropy}
  H = -\sum_{k} P(k) \log P(k).
\end{equation}
The maximum value of entropy is obtained for a uniform degree
distribution and the minimum value $H_{\mathrm{min}}=0$ is achieved
whenever all vertices have the same degree \cite{Wang05}. Network
entropy has been related to the robustness of networks, i.e.\ their
resilience to attacks \cite{Wang05}, and the contribution of vertices
to the network entropy is correlated with lethality in protein
interactions networks \cite{Demetrius04}.

Sol\'{e} and Valverde \cite{Sole04:LNP} suggested the use of the
remaining degree distribution to compute the entropy. The
\emph{remaining degree} of a vertex at one end of an edge is the
number of edges connected to that vertex not counting the original
edge. The remaining degree distribution can be computed as
\begin{equation}
q(k) = \frac{(k+1)P(k+1)}{\langle k \rangle}.
\end{equation}
The entropy of the remaining degree is given by
\begin{equation}
H^* = -\sum_{k} q(k) \log q(k).
\end{equation}

\subsection{Search Information, Target Entropy and Road Entropy}

The structure of a complex network is related to its reliability and
information propagation speed. The difficulty while searching
information in the network can be quantified through the information
entropy of the network \cite{Sneppen:2005,Rosvall06}.
Rosvall~\emph{et al.}\ \cite{Rosvall:2005} introduced measurements to
quantify the information associated to locating a specific target in a
network.  Let $p(i,b)$ be a shortest path starting at vertex $i$ and
ending at vertex $b$. The probability to follow this path in a random
walk is
\begin{equation}
  \mathcal{P}[p(i,b)] =
    \frac{1}{k_i}\prod_{j\in p(i,b)}\frac{1}{k_j - 1},
  \label{pathprob}
\end{equation}
where $k_j$ is the degree of vertex $j$ and the product includes all
vertices $j$ in the path $p(i,b)$ with the exclusion of $i$ and
$b$. The \emph{search information}, corresponding to the total
information needed to identify one of the shortest paths between $i$
and $b$, is given by
\begin{equation}
  \mathcal{S}(i,b) = -\log_2 \sum_{\{p(i,b)\}}\mathcal{P}[p(i,b)],
  \label{search_info_ij}
\end{equation}
where the sum is taken over all shortest paths $p(i,b)$ from $i$ to
$b$.

The average search information characterizes the ease or difficulty of
navigation in a network and is given by \cite{Rosvall:2005}
\begin{equation}
  \mathcal{S} = \frac{1}{N^2}\sum_{ij}\mathcal{S}(i,b).
  \label{searchinfo}
\end{equation}
This value depends on the structure of the network. As discussed by
Rosvall~\emph{et al.}\ \cite{Rosvall:2005}, city networks are more
difficult to be navigated than their random counterparts.

In order to measure how difficult it is to locate vertices in the
network starting from a given vertex $i$, the \emph{access
information} is used,
\begin{equation}\label{accessinfo}
  \mathcal{A}_i = \frac{1}{N} \sum_{b} \mathcal{S}(i,b),
\end{equation}
which measures the average number of ``questions'' needed to locate
another vertex starting from $i$. To quantify how difficult is to
find the vertex $b$ starting from the other vertices in the network,
the \emph{hide information} is used,
\begin{equation}
  \label{hideinfo}
  \mathcal{H}_b = \frac{1}{N} \sum_{i} \mathcal{S}(i,b).
\end{equation}

Note that the average value of $\mathcal{A}_i$ and $\mathcal{H}_b$
for a network is $\mathcal{S}$: $\sum_i \mathcal{A}_i =
\sum_b\mathcal{H}_b = \mathcal{S}N$.

Considering the exchange of messages in the network, it is possible to
define entropies in order to quantify the predictability of the
message flow.  Assuming that messages always flow through shortest
paths and all pairs of vertices exchange the same number of messages
at the same rate, the following entropies can be defined
\cite{Sneppen:2005}:
\begin{eqnarray}
  \mathcal{T}_i & = &
       -\sum_{ij} a_{ji} c_{ij} \log_2 c_{ij},
  \label{targetentropynode}\\
  \mathcal{R}_i & = &
       -\sum_{ij} a_{ji} b_{ij} \log_2 b_{ij},
  \label{roadentropynode}
\end{eqnarray}
where $a_{ji}$ is an element of the adjacency matrix, $c_{ij}$ is the
fraction of messages targeted at vertex $i$ that comes through vertex
$j$, and $b_{ij}$ is the fraction of messages that goes through vertex
$i$ coming from vertex $j$.  In addition, $\mathcal{T}_i$ is the
\emph{target entropy} of vertex $i$ and $\mathcal{R}_i$ is the
\emph{road entropy} of vertex $i$.  Low values of these entropies mean
that the vertex from where the next message originates (\emph{to}
vertex $i$ or passing \emph{through} vertex $i$) can be easily
predicted.

As a general measurement of the flows of messages, we can define
target and road entropies for the network as averages among all
vertices
\begin{eqnarray}
  \mathcal{T} & = & \frac{1}{N}\sum_i \mathcal{T}_i,
  \label{targetentropy}\\
  \mathcal{R} & = & \frac{1}{N}\sum_i \mathcal{R}_i.
  \label{roadentropy}
\end{eqnarray}
As shown in \cite{Sneppen:2005}, these quantities are related to the
organization of the network: a network with a low value of
$\mathcal{T}$ has a star structure and a low value of $\mathcal{R}$
means that the network is composed by hubs connected in a string.

Further works related to searchability in networks have been reported
by Trusina \emph{et al.}\ \cite{Trusina05:PRL}, who defined a search
information weighted by the traffic on the network, and
Rosvall \emph{et al.}\ \cite{Rosvall05:PRE} who studied networks with
higher order organization like modular or hierarchical structure.

\section{Centrality Measurements}\label{sec:central}

In networks, the greater the number of paths in which a vertex or edge
participates, the higher the importance of this vertex or edge for the
network. Thus, assuming that the interactions follow the shortest
paths between two vertices, it is possible to quantify the importance
of a vertex or a edge in terms of its
\emph{betweenness centrality} \cite{Freeman:1977} defined as:
\begin{equation}
  B_u = \sum_{ij} \frac{\sigma(i,u,j)}{\sigma(i,j)},
  \label{betweenness}
\end{equation}
where $\sigma(i,u,j)$ is the number of shortest paths between vertices
$i$ and $j$ that pass through \emph{vertex} or \emph{edge} $u$,
$\sigma(i,j)$ is the total number of shortest paths between $i$ and
$j$, and the sum is over all pairs $i,j$ of distinct vertices.

When one takes into account the fact that the shortest paths might not
be known and instead a search algorithm is used for navigation (see
Section~\ref{navigation}), the betweenness of a vertex or edge must be
defined in terms of the probability of it being visited by the search
algorithm.  This generalization, which was introduced by Arenas
\emph{et al.}\ \cite{Arenas:2002}, subsumes the betweenness centrality
based on random walks as proposed by Newman \cite{Newman:2005}.

The \emph{central point dominance} is defined as \cite{Freeman:1977}
\begin{equation}
  \mathit{CPD} = \frac{1}{N-1}\sum_{i}(B_{\mathrm{max}} - B_i),
  \label{eq:cpd}
\end{equation}
where $B_\mathrm{max}$ is the largest value of betweenness
centrality in the network. The central point dominance will be $0$
for a complete graph and $1$ for a star graph in which there is a
central vertex included in all paths. Other centrality measurements
can be found in the interesting survey by Kosch\"{u}tzki~\emph{et
al.} \cite{Koschutzki05:LNP}.

\section{Spectral Measurements}

The spectrum of a network corresponds to the set of eigenvalues
$\lambda_i$ ($i=1,2,\ldots,N$) of its adjacency matrix $A$. The
spectral density of the network is defined as \cite{Farkas,Goh01}:
\begin{equation}
  \rho(\lambda)=\frac{1}{N} \sum_{i} \delta(\lambda -
\lambda_i),
  \label{spectral}
\end{equation}
where $\delta(x)$ is the Dirac delta function and $\rho$ approaches a
continuous function as $N \rightarrow \infty$; e.g., for
Erd\H{o}s-R\'{e}nyi networks, if $p$ is constant as
$N\rightarrow\infty$, $\rho(\lambda)$ converges to a semicircle
\cite{Farkas}. Also, the eigenvalues can be used to compute the
$l$th-moments,
\begin{equation}
  M_l =
        \frac{1}{N}\sum_{i_1,i_2,\ldots,i_l}
           a_{i_1 i_2}a_{i_2 i_3}\cdots a_{i_l i_1} =
        \frac{1}{N}\sum_{i} (\lambda_i)^l.
  \label{moment}
\end{equation}

The eigenvalues and associated eigenvectors of a network are related
to the diameter, the number of cycles and connectivity properties of
the network \cite{Farkas,Goh01}. The quantity $D_l = N M_l$ is the
number of paths returning to the same vertex in the graph passing
through $l$ edges.  Note that these paths can contain already
visited vertices. Because in a tree-like graph a return walk is only
possible going back through the already visited edges, the presence
of odd moments is a sure sign of cycles in the graph; in particular,
as a walk can go through three edges and return to its starting
vertex only by following three different edges (if self-connections
are not allowed), $D_3$ is related with the number of triangles in
the network \cite{Goh01}.

In addition, spectral analysis allows the identification whether a
network is bipartite (if it does not contain any odd cycle
\cite{Estrada:2005a}, see section~\ref{bipartite_section}),
characterizing models of real networks
\cite{Rosato04:EPL,Mehta91:book}, and visualizing networks
\cite{Seary03}. In addition, spectral analysis of networks is
important to determine communities and subgraphs, as discussed in the
next section.

\section{Community Identification and Measurements}

\begin{figure}
\begin{center}
  \centerline{\includegraphics[width=5.5cm]{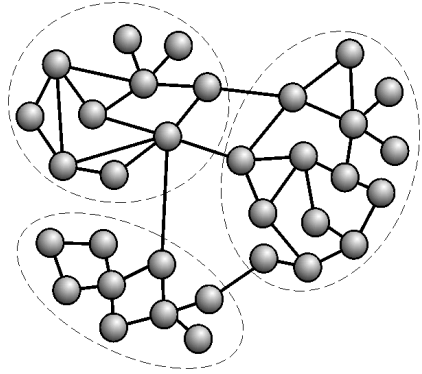}}
  \caption{A network with community structure represented by the
    dashed lines.  The communities are the groups of more intensely
    interconnected vertices.}
 \label{community}
\end{center}
\end{figure}

Many real networks present an inhomogeneous connecting structure
characterized by the presence of groups whose vertices are more
densely interconnected one another than with the rest of the
network. This modular structure has been found in many kinds of
networks such as social networks \cite{Arenas:2004,Gleiser:2003},
metabolic networks \cite{Guimera:2005} and in the worldwide flight
transportation network \cite{Guimera05}. Figure~\ref{community}
presents a network with a well-defined community structure.

Community identification in large networks is particularly useful
because vertices belonging to the same community are more likely to
share properties and dynamics.  In addition, the number and
characteristics of the existing communities provide subsidies for
identifying the category of a network as well as understanding its
dynamical evolution and organization.  In the case of the World Wide
Web, for instance, pages related to the same subject are typically
organized into communities, so that the identification of these
communities can help the task of seeking for information. Similarly,
in the case of the Internet, information about communities formed by
routers geographically close one another can be considered in order to
improve the flow of data.

Despite the importance of the concept of community, there is no
consensus about its definition. An intuitive definition was proposed
by Radichi \emph{et al.}\ \cite{Radicchi:2004} based on the
comparison of the edge density among vertices. Communities are
defined in a strong and a weak sense. In a \emph{strong sense}, a
subgraph is a community if all of its vertices have more connections
between them than with the rest of the network. In a \emph{weak
sense}, on the other hand, a subgraph is a community if the sum of
all vertex degrees inside the subgraph is larger than outside it.
Though these definitions are intuitive, one of their consequences is
that every union of communities is also a community. To overcome
this limitation a hierarchy among the communities can be assumed
\emph{a priori}, as discussed by Reichardt and Bornholdt
\cite{Reichardt0603718}, who defined community in networks as the
spin configuration that minimizes the energy of the spin glass by
mapping the community identification problem onto finding the ground
state of a infinite range Potts spin glass
\cite{Reichardt0606220,Reichardt04:PRL}.

Another fundamental related problem concerns how to best divide a
network into its constituent communities. In real networks, no
information is generally available about the number of existing
communities. In order to address this problem, a measurement of the
quality of a particular division of networks was proposed by Newman
and Girvan \cite{Newman:2004:PRE}, called \emph{modularity} and
typically represented by $Q$. If a particular network is split into
$c$ communities, $Q$ can be calculated from the symmetric $c\times
c$ \emph{mixing matrix}\label{mixmatrix} $E$ whose elements along
the main diagonal, $e_{ii}$, give the fraction of connections
between vertices in the same community $i$ while the other elements,
$e_{ij}$ ($i\ne j)$ identify the fraction of connections between
vertices in the different communities $i$ and $j$.  This is similar
to the definition used to compute assortativity,
Section~\ref{assortativity}. The calculation of $Q$ can then be
performed as follows:
\begin{equation}
  Q = \sum_i [ e_{ii} - ( \sum_j e_{ij} )^2 ]
       = \mathrm{Tr} E  - ||E^2||,
  \label{eq:modularity}
\end{equation}
The situation $Q = 1$ identifies networks formed by disconnected
modules. This quantity has been used in many community-finding
algorithms, as briefly reviewed in the following.

Though there are many ways to defined modularity, a generally accepted
definition of a module does not exist~\cite{Schlosser04}. The
definitions described above estimate the modularity in terms of a
given partitioning.  Ziv \emph{et al.}~\cite{Ziv05:PRE046117} proposed
the modularity to be defined in terms of information entropy (see
Section~\ref{entropy}). This algorithm, which has been called the \emph{Network
Information Bottleneck}, tends to allow performance better than the
algorithm based on betweenness centrality of Girvan and Newman.

It should be noted that this review of community finding methods
focus the subject of how specific network measurements have been
adopted to identify the communities. Since we do not attempt to
provide a comprehensive study of this important subject, the
interested reader should refer to recent papers by Newman
\cite{Newman:2004:EPJ} and Danon \emph{et al.}\ \cite{Danon:2005}
for further information and a more complete review of community
finding methods.  The following discussion has been organized into
subsections according to the nature of the adopted methodology.

\subsection{Spectral Methods}

Spectral methods are based on the analysis of the eigenvectors of
matrices derived from the networks \cite{Seary:1995}.  These methods
have been discussed in a recent survey by Newman \cite{Newman06}.  The
quantity measured corresponds to the eigenvalues of matrices
associated with the adjacency matrix. These matrices can be the
\emph{Laplacian matrix} (also known as Kirchhoff matrix),
\begin{equation}
  \label{laplacian}
  L = D - A,
\end{equation}
or the \emph{Normal matrix},
\begin{equation}
  \tilde{A} = D^{-1}A,
\end{equation}
where $D$ is the diagonal matrix of vertex degrees with elements
$d_{ii} = \sum_j a_{ij},$ $d_{ij}=0$ for $i\ne j$.

A particular method, called \emph{spectral bisection}
\cite{Fiedler:1973,Pothen:1990,Newman06}, is based on the
diagonalization of the Laplacian matrix. If the network is separated
into $c$ disconnected components, $L$ will be block diagonal and
have $c$ degenerated eigenvectors, all corresponding to eigenvalue
$0$. However, if the separation is not clear, the diagonalization of
$L$ will produce one eigenvector with eigenvalues $0$ and $c-1$
eigenvalues slightly different from $0$. The spectral bisection
considers the case when $c = 2$ and the division of the network is
obtained assigning positive components of the eigenvector associated
with the second eigenvalue (the positive eigenvalue most close to
$0$) to one community and the negative ones to another community.
Particularly, the second eigenvalue, called \emph{algebraic
connectivity}, is a measurement of how good the division is, with
small values corresponding to better divisions. Although spectral
bisection is easy to implement, it tends to be a poor approach for
detecting communities in real networks \cite{Newman06}. There are
many alternative methods based on spectral analysis
\cite{Capocci05}, to be found in \cite{Newman:2004:EPJ,Danon:2005}.

Recently, Newman \cite{Newman06:PNAS} proposed a method which
reformulates the modularity concept in terms of the eigenvectors of
a new characteristic matrix for the network, called \emph{modularity
matrix}. For each subgraph $g$, its modularity matrix $B^{(g)}$ has
elements
\begin{equation}
  b_{ij}^{(g)} = a_{ij} - \frac{k_i k_j}{2M} -
    \delta_{ij}\sum_{u\in\mathcal{N}(g)}
    \left[ a_{iu} - \frac{k_i k_u}{2M} \right],
\end{equation}
for vertices $i$ and $j$ in $g$. Thus, in order to split the network
in communities, first the modularity matrix is constructed and its
most positive eigenvalue and corresponding eigenvector are
determined. According to the signs of the elements of this vector, the
network is divided into two parts (vertices with positive elements are
assigned to a community and vertices with negative elements to
another). Next, the process is repeated recursively to each community
until a split which makes zero or a negative contribution to the total
modularity is reached. Following this idea, Newman proposed a new
definition of communities as indivisible subgraphs, i.e.\ subgraphs
whose division would increase the modularity. Currently, this method
is believed to be the most precise, as it is able to find a division
with the highest value of modularity for many networks
\cite{Newman06:PNAS}.

\subsection{Divisive Methods}

In a divisive method, the underlying idea is to find the edges which
connect different communities and remove them in a iterative form,
breaking the network into disconnected groups of vertices. The
computation of modularity can be used afterwards to determine the
best division of the network. Next we give a brief description of
the most known divisive methods according to the adopted measurement
used to choose the vertex to remove.

\subsubsection{Betweenness Centrality}

The most popular divisive method is the Girvan-Newman algorithm
\cite{Girvan:2002}. Because different communities are connected by
a small number of edges, this method considers that bottlenecks are
formed at the edges which connect communities, through which all
shortest paths should pass. In order to measure this traffic-related
property in networks, the algorithm uses the concept of \emph{edge
betweenness} \cite{Girvan:2002}, see Section~\ref{sec:central}. Edges
with high betweenness are progressively removed. After removing each
edge, the betweenness of each remaining edge must be calculated again.

Although this algorithm represents a powerful alternative to determine
communities (as shown in Figure~\ref{comp}), it has some
disadvantages. The main one is its high computational cost. As
discussed by Girvan and Newman \cite{Girvan:2002}, the entire
algorithm runs in worst-case time $O(M^2N)$ on networks with $M$ edges
and $N$ vertices. In order to overcome this limitation, some
improvements in the algorithm were proposed including the Tyler's
algorithm \cite{Tyler:2003}, which introduced a stochastic element to
the method, restricting the calculation of the betweenness only to a
partial set of edges and using statistics to estimate the real
betweenness.

\subsubsection{Edge Clustering Coefficient}

A different approach was proposed by Radicchi \emph{et al.}\
\cite{Radicchi:2004} (see also~\cite{Costa_perc:2004}), which is
based on counting short loops of order $l$ (triangles for $l = 3$) in
networks. The algorithm is similar to Girvan and Newman's method, but
instead of the betweenness centrality, it computes the \emph{edge
clustering coefficient}. This measurement is based on the fact that
edges which connect communities tend to exhibit a small value for this
coefficient. The clustering coefficient of edge $(i,j)$ is calculated
as
\begin{equation}\label{clustering}
  C_{ij} = \frac{Z_{ij} + 1}{\min(k_i-1,k_j-1)}
\end{equation}
where $Z_{ij}$ is the number of triangles to which $(i,j)$ belongs.
This method can be generalized to more complex loops, e.g.,
squares. Though this method is simple and fast ($O(M^4/N^2)$), it
fails whenever the network has a small average clustering coefficient,
because the value of $C_{ij}$ will be small for all edges. This
suggests that the method will work well only when applied to networks
with a high average clustering coefficient, such as social networks
\cite{Danon:2005}.

\subsection{Agglomerative Methods}

Some networks are characterized by the fact that the vertices
belonging to each community present similar features. So, it is in
principle possible to obtain the communities by considering such
similarities between vertices. In contrast to divisive methods,
agglomerative approaches start with all vertices disconnected and
then apply some similarity criterion to progressively join them and
obtain the communities.  It is interesting to note that this type of
method presents a direct relationship with pattern recognition and
clustering theory and algorithms (e.g.,
\cite{Anderberg73,JainDubes88,Romesburg90,Costa:01}), which have
been traditionally used in order to group individuals represented by
a vector of features into meaningful clusters.

\subsubsection{Similarity Measurements}

One important family of agglomerative methods is known as
\emph{hierarchical clustering}
\cite{Anderberg73,JainDubes88,Costa:01,Scott00}, which starts
with $N$ vertices and no edges.  Edges are added progressively to the
network in decreasing order of similarity, starting with the pair with
strongest similarity \cite{Hopcroft:2004,Newman:2004:EPJ}. To evaluate
the similarity associated with edge $(i,j)$, a possibility is to use
the so called \emph{Euclidian distance}, given by
\begin{equation}
  \label{euclidian}
  \sqrt{\sum_{k\neq i,j}(a_{ik}-a_{jk})^2},
\end{equation}
or the \emph{Pearson correlation} between vertices as represented in
the adjacency matrix, defined as
\begin{equation}
  \label{person}
  \frac{\frac{1}{N}\sum_k (a_{ik}-\mu_i)(a_{jk} -
\mu_j)}{\sigma_i \sigma_j},
\end{equation}
where $\mu_i = \frac{1}{N}\sum_j a_{ij}$ and $\sigma_i^2 =
\frac{1}{N-1}\sum_j(a_{ij}-\mu_{i})^2.$

Although this method is fast, the obtained division of the network is
not generally satisfactory for real networks, as discussed in
\cite{Newman:2004:EPJ}.

\subsection{Maximization of the Modularity}

Newman \cite{Newman:2004:PRE,Clauset:04PRE} proposed a method based on
joining communities in such a way as to maximize the modularity. In
this method, two communities $i$ and $j$ are joined according to a
measurement of affinity, given by the change of the modularity $Q$ of
the network (Eq.~\ref{eq:modularity}) when the communities are joined
\begin{equation}
  \Delta Q_{ij} = 2 \left(
        e_{ij} - \frac{\sum_j e_{ij} \sum_i e_{ij} }{2M}
                    \right).
  \label{eq:dmodularity}
\end{equation}

Thus, starting with each vertex disconnected and considering each of
them as a community, we repeatedly join communities together in
pairs, choosing at each step the joining that results in the
greatest increase (or smallest decrease) in the modularity $Q$. This
process can be repeated until the whole network is contained in one
community. Currently, as discussed by Danon \emph{et al.}\
\cite{Danon:2005}, the Newman's method is believed to be the fastest
one, running in $O(N\log^2 N)$. Also, this method is more precise
than the traditional method based on betweenness centrality
\cite{Newman:2004:PRE}. However, as discussed by Danon \emph{et
al.}\ \cite{Danon:06}, the fast Newman's method has a limitation
when the size of communities is not homogeneous, as a newly joined
community $i$ has the new values of $e_{ij}$ in
Eq.~\ref{eq:dmodularity} increased, and tends to be chosen for new
joining. In real networks the distribution of sizes of communities
tends to follow a power law. So, this approach fails in many real
networks. In order to overcome this limitation, it was proposed
\cite{Danon:06} to normalize the value of $\Delta Q$ by the number
of edges in community $i$,
\begin{equation}
\Delta \widehat{Q}_{ij} = \frac{\Delta Q_{ij}}{\sum_j e_{ij}} =
       \frac{2}{\sum_j e_{ij}} \left( e_{ij} -
          \frac{\sum_j e_{ij} \sum_i e_{ij}}{2M} \right).
\end{equation}

This alteration on the local modularity makes the method more
precise while not affecting its execution time.

\subsubsection{Extremal Optimization}

The extremal optimization method proposed by Duch and Arenas
\cite{Duch:2005} is a heuristic search for optimizing the value of
the modularity $Q$. The \emph{local modularity} represents the
contribution of individual vertex $i$ to the modularity $Q$. If
$c_i$ is the community of vertex $i$, the local modularity is given
by
\begin{equation}
  q_i = \sum_j a_{ij}\delta_{c_i,c_j} - k_i \sum_{c_k} e_{c_i c_k}
\end{equation}
where $e_{ij}$ are the elements of the community mixing matrix
(page~\pageref{mixmatrix}) and $\delta$ is the Kronecker delta. In
order to keep the value of this contribution in the interval
$[-1,1]$ and independent of vertex degree, it should be normalized
by the degree of the vertex, i.e.\ $\hat{q}_i = q_i/k_i$. The value
of $\hat{q}_i$ is used as the \emph{fitness} for the extremal
optimization algorithm. A heuristic search is performed to obtain
the maximum value of the modularity. Initially, the network is split
into two random partitions with the same number of vertices. After
each step, the system self-organizes by moving the vertex with
lowest fitness from one partition to another. The process stops when
the maximum value of $Q$ is reached.

\begin{figure}
\begin{center}
  \centerline{\includegraphics[width=9cm]{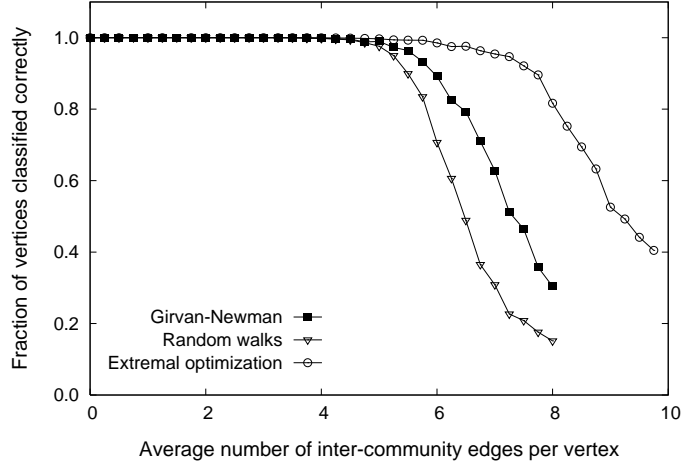}}
  \caption{Comparison of precision between the methods developed by
  Girvan and Newman \cite{Girvan:2002} (based on removing edges with
  highest betweenness centrality value), the same method based on
  random walks \cite{Newman:2003:PRE}, and the method developed by
  Duch and Arenas, based on extremal optimization \cite{Duch:2005}.
  Each point in this graph is an average of 100 realizations of
  networks with 128 vertices organized into 4 communities.}
 \label{comp}
\end{center}
\end{figure}

Although this method is not particularly fast, scaling as $O(N^2\log
N)$, it can achieve high modularity values \cite{Duch:2005}.  By
comparing the precision of some methods as presented in
Figure~\ref{comp}, we can see that the extremal optimization method is
more precise than the methods based on removing edges with highest
betweenness centrality value. Moreover, it is clear that the
computation of betweenness centrality by counting the number of
shortest paths passing through each edge is more precise than
calculating this coefficient by random walks \cite{Newman:2004:PRE}.

\subsection{Local Methods}

More recently, some methods have been developed to detect the local
community of a vertex based only on local information about the
network topology. One such method was proposed by Bagrow and Bolt
\cite{Bagrow05}, which is based on the change of the hierarchical
degree between two consecutive distances (see
Section~\ref{sec:hier}). Starting from a vertex $v_0$, the vertices of
successive hierarchical rings are added to the community, as long as
the relation between the successive hierarchical degrees is greater
than a specified threshold $\alpha$
\begin{equation}
\frac{k_d(v_0)}{k_{d-1}(v_0)} > \alpha.
\end{equation}
When the expansion reaches a distance $d$ for which the above
condition fails, the community stops growing.

Despite its favorable speed, this approach has an important
limitation: the division is precise only when $v_0$ is equidistant
from all parts of its enclosing community's boundary
\cite{Clauset:2005}. In order to overcome this drawback, it has been
suggested \cite{Bagrow05} that the algorithm be executed $N$ times
starting from each vertex and then achieve a consensus about the
detected communities. However, this approach increases the execution
time of the algorithm.

Another local method was proposed by Clauset \cite{Clauset:2005}
which is based on computing the \emph{local modularity}. The idea is
that of a step-by-step growth of the community together with the
exploration of the network. The community $\mathcal{C}$ starts with
only the original vertex $v_0$. When a vertex is explored, a list of
its neighbors is known. The set $\mathcal{U}$ is a list of all
vertices that are not in $\mathcal{C}$ but are adjacent to some of
its vertices; the set $\mathcal{B}$ (the \emph{boundary} of
$\mathcal{C}$) is the subset of vertices in $\mathcal{C}$ that are
adjacent of at least one vertex in $\mathcal{U}.$ The local
modularity is defined as the ratio of the number of edges with one
end point in $\mathcal{B}$ and neither end point in $\mathcal{U}$ to
the number of edges with end points in $\mathcal{B}.$ Considering
undirected networks, this can be written as
\begin{equation}
  R = \frac
      {\sum_{i\in\mathcal{B},j\in\mathcal{C}}a_{ij}}
      {\sum_{i\in\mathcal{B},j}a_{ij}}.
  \label{eq:local}
\end{equation}
The algorithm consists in choosing iteratively from the set
$\mathcal{U}$ the vertex that would result in the largest increase
(or smallest decrease) in the value of $R$ when added to
$\mathcal{C}$. The iteration stops when a pre-defined number of
vertices was included in the community.

\subsection{Method Selection}

Despite the many interesting alternative methods, including those
briefly reviewed above, it should be noted that the problem of
community finding remains a challenge because no single method is
fast and sensitive enough to ensure ideal results for general, large
networks, a problem which is compounded by the lack of a clear
definition of communities.  If communities are to be identified with
high precision, the spectral method proposed by Newman
\cite{Newman06:PNAS} is a good choice. However, if priority is
assigned to speed, methods such as those using greedy algorithms
(runs in $O(N\log^2 N)$) should be considered \cite{Clauset:04PRE}.
In brief, the choice of the best method to be used depends on the
configuration of the problem and the kind of desired results
\cite{Danon:2005}.

One fact that should have become clear from our brief review of
community finding approaches is the essential importance of the
choice of the \emph{measurements} adopted to express the separation
of the communities.  As a matter of fact, such measurements
ultimately represent an objective definition of communities.
Therefore, an interesting perspective for further research would be
to consider the possible adaptation and combination of some of the
measurements reported in this survey with the specific objective of
community characterization.

\subsection{Roles of Vertices}

After community identification, it is possible to determine the role
of vertices \cite{Guimera:2005} by using the \emph{z-score of the
within-module degree}, $z_i$, and the \emph{participation
coefficient}, $P_i$.  The z-score measures how ``well-connected''
vertex $i$ is to the other vertices in the community, being defined by
\begin{equation}\label{zscore}
z_i = \frac{q_i - \overline{q}_{s_i}}{\sigma_{q_{s_i}}},
\end{equation}
where $q_i$ is the number of connections $i$ makes with other vertices
in its own community $s_i$, $\overline{q}_{s_i}$ is the average of $q$
over all vertex in $s_i$, and $\sigma_{q_{s_i}}$ is the standard
deviation of $q$ in $s_i$.

The participation coefficient measures how ``well-distributed'' the
edges of vertex $i$ are among different communities,
\begin{equation}\label{pcoeff}
P_i = 1 - \sum_{s=1}^{N_M}\left(\frac{q_{is}}{k_i}\right)^2,
\end{equation}
where $q_{is}$ is the number of edges from vertex $i$ to community $s$
and $k_i$ is the degree of vertex $i$.  This value is zero if all
edges are within its own community and it is close to one if its edges
are uniformly distributed among all communities. Based on these two
index, a $zP$ parameter-space can be constructed, allowing the
classification of vertices into different roles (see e.g.,
\cite{Guimera:2005}).

\section{Subgraphs}

\begin{figure}
  \centerline{\includegraphics[width=7.5cm]{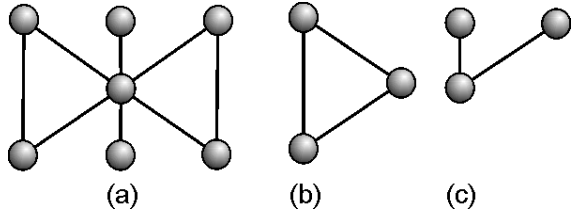}}
  \caption{A network such as that in (a) includes several subgraphs,
  such as cycles (b) and trees (c).}
 \label{subgraphs}
\end{figure}

A graph $g$ is a \emph{subgraph} of the graph $G$ if $\mathcal{N}(g)
\subseteq \mathcal{N}(G)$ and $\mathcal{E}(g)\subseteq
\mathcal{E}(G)$, with the edges in $\mathcal{E}(g)$ extending over
vertices in $\mathcal{N}(g)$. If $g$ contains all edges of $G$ that
connect vertices in $\mathcal{N}(g)$, the subgraph $g$ is said to be
\emph{implied} by $\mathcal{N}(g)$.  Important subgraphs include
loops, trees (connected graphs without loops) and complete subnetworks
(cliques). Figure~\ref{subgraphs} shows a network and some
subnetworks. The probability distribution of subgraphs in random
graphs has been studied for some time \cite{Bollobas:1985}, but
interest has increased recently as a consequence of the discovery of
network motifs as discussed below.

\subsection{Network Motifs}

\begin{figure}
  \centerline{\includegraphics[width=7.5cm]{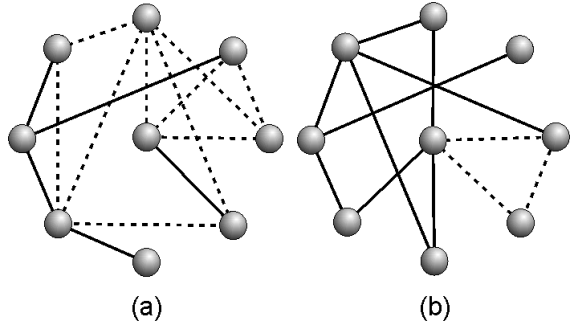}}
  \caption{In a real network (a), the number of motifs (represented
    here by three vertices linked by dashed lines) is greater than in an
    equivalent random network (b).}
 \label{motifs}
\end{figure}

\begin{figure}
  \centerline{\includegraphics[width=8.5cm]{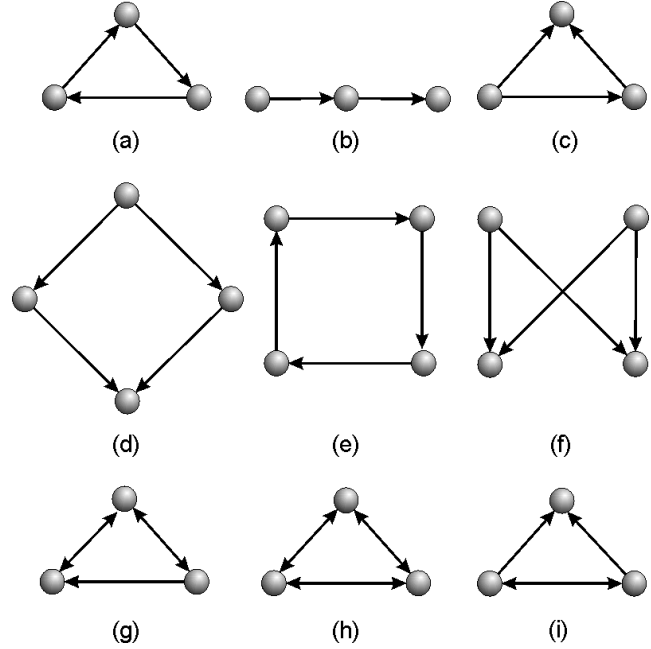}}
  \caption{Some types of motifs: (a) three-vertex feedback, (b) three
    chain, (c) feed-forward loop, (d) bi-parallel, (e) four-vertex
    feedback, (f) bi-fan, (g) feedback with two mutual dyads, (h)
    fully connected triad and (i) uplinked mutual dyad.}
 \label{type_motifs}
\end{figure}

Network motifs are subgraphs that appear more frequently in a real
network than could be statistically expected
\cite{ShenOrr:2002,Milo:2002,Middendorf05:PNAS} (see Figure~\ref{motifs}).
Figure~\ref{type_motifs} shows some possible motifs of directed
networks and their conventional names. To find the motifs in a real
network, the number of occurrences of subgraphs in the network is
compared with the expected number in the ensemble of networks with the
same degree for each vertex. A large number of randomized networks
from this ensemble is generated in order to compute the statistics of
occurrence of each subgraph of interest. If the probability of a given
subgraph to appear at least the same number of times as in the real
network is smaller than a given threshold (usually $0.01$), the
subgraph is considered a motif of the network.

In order to quantify the significance of a given motif, its
\emph{$Z$-score} can be computed. If $N^{(\mathrm{real})}_i$ is the
number of times that a motif $i$ appears in the real network,
$\langle N^{(\mathrm{rand})}_i \rangle$ the ensemble average of its
number of occurrences, and $\sigma^{(\mathrm{rand})}_i$ the standard
deviation of the number of occurrences, then:
\begin{equation}
  Z_i =
\frac{N^{(\mathrm{real})}_i-\langle N^{(\mathrm{rand})}_i
\rangle}{\sigma^{ (\mathrm{rand})}_i}.
  \label{eq:zscore}
\end{equation}

It is also possible to categorize different networks by the $Z$-scores
of their motifs: networks that show emphasis on the same motifs can be
considered as part of the same family \cite{Milo:2004}. For this
purpose, the \emph{significance profile} of the network can be
computed. The significance profile is a vector that, for each motif of
interest $i$, is used to compute the importance of this motif with
respect to other motifs in the network:
\begin{equation}\label{SP}
  \mathit{SP}_i = \frac{Z_i}{\sum_j Z_j^2}.
\end{equation}

It is interesting to note that motifs are related to network
evolution. As described by Milo \emph{et al.}\ \cite{Milo:2002},
different kinds of networks present different types of motifs e.g.,
for transcription networks of \emph{Saccharomyces cerevisiae} and
\emph{Escherichia coli} two main motifs are identified: feed-forward
loop and bi-fan; for neurons: feed-forward loop, bi-fan and
bi-parallel; for food-webs: three chain and bi-parallel; for
electronic circuits: feed-forward loop, bi-fan, bi-parallel, four-node
feedback and three node feedback loop; and for the WWW: feedback with
two mutual dyads, fully connected triad and uplinked mutual
dyad. Thus, motifs can be considered as building blocks of complex
networks and many papers have been published investigating the
functions and evolution of motifs in networks \cite{Barabasi:2004}.

\subsection{Subgraphs and Motifs  in Weighted Networks}\label{weighted_subgraphs}

In weighted networks, a subgraph may be present with different values
for the weights of the edges. Onnela \emph{et al.}\ \cite{Onnela04}
suggested a definition for the \emph{intensity} of a subgraph based on
the geometric mean of its weights on the network.  Given a subgraph
$g$, its intensity is defined by
\begin{equation}
  I(g) = \left( \prod_{(i,j)\in \mathcal{E}(g)} w_{ij} \right)^{1/n_g},
  \label{subgraph_intensity}
\end{equation}
where $n_g = |\mathcal{E}(g)|$ is the number of edges of subgraph
$g$.

In order to verify whether the intensity of a subgraph is small
because all its edges have small weight values or just one of the
weights is too small, the \emph{coherence} of the subgraph $\Psi$,
defined as the ratio between geometric and arithmetic mean of its
weights, can be used:
\begin{equation}
  \Psi(g) = \frac{I(g) n_g}{\sum_{(i,j)\in \mathcal{E}(g)} w_{ij}}.
  \label{subgraph_coherence}
\end{equation}

All possible subgraphs of the weighted graph can be categorized into
sets of \emph{topologically equivalent} subgraphs.\footnote{Two
subgraphs are topologically equivalent if their only difference is on
the weight of the existing edges.} Let $M$ be one such set of
topologically equivalent subgraphs. The intensity of $M$ is given by
$I_M = \sum_{g \in M}I(g)$ and its coherence by $\Psi_M = \sum_{g\in
M}\Psi(g)$. An intensity score $ZI_M$ can be accordingly defined by
\begin{equation}
ZI_M = \frac{I_M - \langle
I_M^{\mathrm{(rand)}}\rangle}{\sigma_{I_M}^{\mathrm{(rand)}}}
\end{equation}
and the coherence score,
\begin{equation}
Z\Psi_M = \frac{\Psi_M - \langle \Psi_M^{\mathrm{(rand)}}
\rangle}{\sigma^{\mathrm{(rand)}}_{\Psi_M}},
\end{equation}
where $\langle I_M^{\mathrm{(rand)}} \rangle$ and
$\sigma_{I_M}^{\mathrm{(rand)}}$ are the mean and the standard
deviation of the intensities in a randomized graph ensemble;
$\langle \Psi_M^{\mathrm{(rand)}} \rangle$ and
$\sigma_{\Psi_M}^{\mathrm{(rand)}}$ are the average and the standard
deviation of the coherence in the randomized ensemble. When the
network is transformed to its unweighted version, $ZI_M$ and
$Z\Psi_M$ tend to $Z$ (see Eq.~(\ref{eq:zscore})).

\subsection{Subgraph Centrality}\label{subgraphcent}

A way to quantify the centrality of a vertex based on the number of
subgraphs in which the vertex takes part has been proposed
\cite{Estrada:2005}.  The respective measurement, called
\emph{subgraph centrality}, considers the number of subgraphs that
constitute a closed walk starting and ending at a given vertex $i$,
with higher weights given to smaller subgraphs. This measurement is
related to the moments of the adjacency matrix, Eq.~(\ref{moment}):
\begin{equation}
  \mathit{SC}_i = \sum_{k=0}^{\infty} \frac{(A^k)_{ii}}{k!},
  \label{eq:subcent}
\end{equation}
where $(A^k)_{ii}$ is the $i$th diagonal element of the $k$th power of
the adjacency matrix $A$, and the factor $k!$ assures that the sum
converges and that smaller subgraphs have more weight in the sum.
Subgraph centrality can be easily computed \cite{Estrada:2005} from
the spectral decomposition of the adjacency matrix,
\begin{equation}
  \mathit{SC}_i = \sum_{j=1}^{N} v_j(i)^2 e^{\lambda_j},
  \label{eq:subcentspec}
\end{equation}
where $\lambda_j$ is the $j$th eigenvalue and $v_j(i)$ is the $i$th
element of the associated eigenvector. This set of eigenvectors should
be orthogonalized. The subgraph centrality of a graph is given by
\cite{Estrada:2005a}:
\begin{equation}
\mathit{SC} = \frac{1}{N}\sum_{i=1}^{N}{\mathit{SC}_i} =
\frac{1}{N}\sum_{i=1}^{N}{e^{\lambda_i}}.
  \label{eq:subcentnet}
\end{equation}

\section{Hierarchical Measurements}\label{sec:hier}

\begin{figure}
  \centerline{\includegraphics[width=6cm]{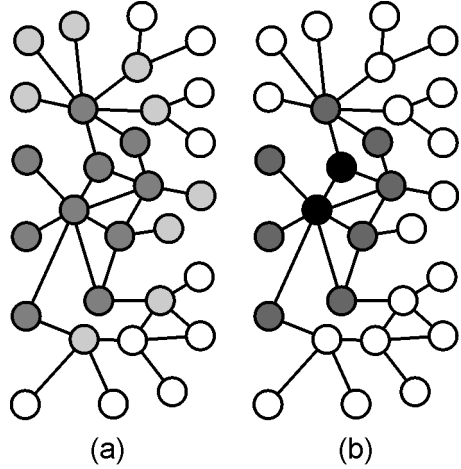}}
  \caption{Example of morphological operations: (a) Dilation: the
    dilation of the initial subnetwork (dark gray vertices)
    corresponds to the dark and light gray vertices; (b) Erosion: the
    erosion of the original subnetwork, given by the dark gray
    vertices in (a), results in the subnetwork represented by the
    black vertices in (b).}
 \label{fig:dilation}
\end{figure}

Using concepts of mathematical morphology
\cite{Vincent:1989,Dougherty:2003,Heijmans:1990,Costa:2005c}, it is
possible to extend some of the traditional network measurements and
develop new ones \cite{CostaPRL:2004,Costa:2004,Costa:2005a}. Two
fundamental operations of mathematical morphology are
\emph{dilation} and \emph{erosion} (see Figure~\ref{fig:dilation}).
Given a subgraph $g$ of a graph $G$, the complement of $g$, denoted
$\overline{g}$, is the subgraph implied by the set of vertices in
$G$ that are not in $g$, $$ \mathcal{N}(\overline{g}) =
\mathcal{N}(G)\setminus \mathcal{N}(g) $$ ($\setminus$ is the
operator of set difference).  The dilation of $g$ is the subgraph
$\delta(g)$ implied by the vertices in $g$ plus the vertices
directly connected to a vertex in $g$. The erosion of $g$, denoted
$\varepsilon(g)$, is defined as the complement of the dilation of
the complement of $g$: $$ \varepsilon(g) =
\overline{\delta(\overline{g})}.  $$ These operations can be applied
repeatedly to generate the $d$-dilations and $d$-erosions:
\begin{eqnarray}
  \delta_d(g) & = & \underbrace{\delta(\delta( ... (g) ... ))}_d,
  \label{eq:dilation} \\
  \varepsilon_d(g) & = &
  \underbrace{\varepsilon(\varepsilon(...(g)...))}_d.
  \label{eq:erosion}
\end{eqnarray}
The first operation converges to the entire network $G$ and the
second converges to an empty network.

\begin{figure}
  \centerline{\includegraphics[width=9cm]{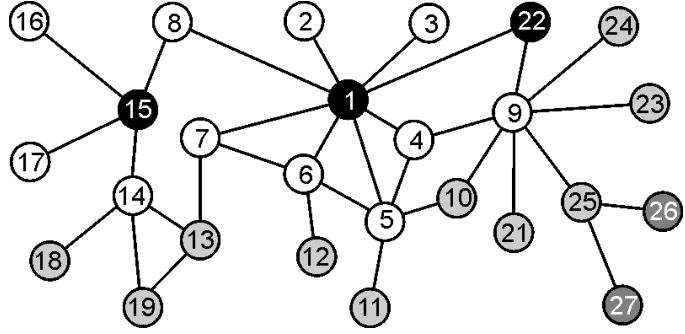}}
  \caption{The subgraph of interest is defined by black vertices, $g =
    \{1,15,22\}$.  The first hierarchical level of $g$ is given by the
    first dilation around $g$, represented by the white vertices; the
    second hierarchical level is obtained dilating the subnetwork
    again, represented by the gray vertices. The hierarchical degree
    of the first level is given by the number of edges from white to
    light gray vertices, $k_1(g) = 12$, and the hierarchical degree of
    the second level is the number of edges from light gray to dark
    gray vertices, $k_2(g) = 2$.}
 \label{hdegree}
\end{figure}

The $d$-ring of subgraph $g$, denoted $R_d(g)$, is the subgraph
implied by the set of vertices
$$\mathcal{N}(\delta_d(g))\setminus\mathcal{N}(\delta_{d-1}(g));$$ the
$rs$-ring of $g$, denoted $R_{rs}(g)$, is the subgraph implied by $$
\mathcal{N}(\delta_s(g))\setminus\mathcal{N}(\delta_{r-1}(g)).$$ Note
that $R_d(g) = R_{dd}(g).$ The same definitions can be extended to a
single vertex considering the subgraph implied by that vertex, and to
an edge considering the subgraph formed by the edge and the two
vertices that it connects. In the case of a single vertex $i$ the
abbreviations $R_d(i)$ and $R_{rs}(i)$ are used. For example, in
Figure~\ref{hdegree}, $R_1(15)$ includes the vertices
$\{8,14,16,17\}$; $R_2(15)$ includes $\{1,13,18,19\}$; for the graph
$g$ implied by the vertices $\{1,15,22\}$ (in black), $R_1(g)$
includes the vertices in white: $\{2,3,4,5,6,7,8,9,14,16,17\}$.

The \emph{hierarchical degree} of a subgraph $g$ at distance $d$,
henceforth represented as $k_d(g)$, can be defined as the number of
edges connecting rings $R_d(g)$ to $R_{d+1}(g)$. Note that $k_0(i)$ is
equal to $k_i$.

Another measurement which can be hierarchically extended is the
clustering coefficient. The \emph{$rs$-clustering coefficient} of
$g$, $C_{rs}(g)$, can be defined as the number of edges in the
respective $rs$-ring $n_{rs}$, divided by the total of possible
edges between the vertices in that ring, i.e.\ for undirected
networks
\begin{equation}
  C_{rs}(g) = \frac{2 n_{rs}(g)}
  {|\mathcal{N}(R_{rs}(g))|(|\mathcal{N}(R_{rs}(g))|-1)}.
  \label{eq:hcc}
\end{equation}

Other possible hierarchical measurements are briefly described in the
following. The \emph{convergence ratio} at distance $d$ of subgraph
$g$, $\mathit{cv}_d(g)$, corresponds to the ratio between the
hierarchical subgraph degree at distance $d-1$ and the number of
vertices in the ring at distance $d$; it can be understood as the
average number of edges received by each vertex in the hierarchical
level $d$ from the previous level,
\begin{equation}
  \mathit{cv}_d(g) = \frac{k_{d-1}(g)}{|\mathcal{N}(R_d(g))|}.
  \label{cratio}
\end{equation}
It is also possible to define the \emph{divergence ratio}, which
corresponds to the reciprocal of the convergence ratio
\begin{equation}
  \mathit{dv}_d(g) = \frac{|\mathcal{N}(R_d(g))|}{k_{d-1}(g)}.
  \label{dratio}
\end{equation}

The \emph{intra-ring degree} is obtained by taking the average among
the degrees of vertices in the subnetwork $R_d(g)$; note that only
internal ring edges are considered. On the other hand, the
\emph{inter-ring degree} is defined by the average number of
connections between vertices in ring $R_d(g)$ and those in
$R_{d+1}(g)$.

\section{Fractal Dimensionality}

Fractals are objects or quantities that display self-similarity (or
self-affinity) in all scales. For complex networks, the concept of
self-similarity under a length-scale transformation was not expected
because of the small world property, implying the average shortest
path length of a network increases logarithmically with the number of
vertices. However, Song \emph{et al.}\ \cite{Song:2005} analyzed
complex networks by using fractal methodologies and verified that real
complex networks may consist of self-repeating patterns on all length
scales.

In order to measure the fractal dimension of complex networks, a
\emph{box counting method} and a \emph{cluster growing method} has
been proposed \cite{Song:2005}. In the former, the network is
covered with $N_B$ boxes, where all vertices in each of them are
connected by a minimum distance smaller than $l_B$. $N_B$ and $l_B$
are found to be related by
\begin{equation}
N_B\sim l_B^{-d_B},
\end{equation}
where $d_B$ is the \emph{fractal box dimension} of the network.

For the cluster growing method, a seed vertex is chosen at random
and a cluster is formed by vertices distant at most $l$ from the
seed. This process is repeated many times and the average mass of
resulting clusters is calculated as a function of $l$, resulting in
the relation
\begin{equation}
\langle M_c \rangle \sim l^{d_f},
\end{equation}
where the average mass $\langle M_c \rangle$ is defined as the number
of vertices in the cluster and $d_f$ is the \emph{fractal cluster
dimension}.

For a network whose vertices have a typical number of connections,
both exponents are the same, but this is not the case for scale-free
networks.

Another scaling relation is found with a renormalization procedure
based on the box counting method \cite{Song:2005}. A renormalized
network is created with each box of the original network transformed
into a vertex and two new vertices are connected if at least one edge
exists between vertices of the corresponding boxes in the original
network.  By considering the degree $k'$ of each vertex of the
renormalized network versus the maximum degree $k$ in each box of the
original network we have that:
\begin{equation}
k' \approx l_B^{-d_k}k,
\end{equation}

The exponents $\gamma$ (of the power law of the degree
distribution), $d_B$ and $d_k$ are related by \cite{Song:2005}:
\begin{equation}
\gamma = 1 + d_B/d_k.
\end{equation}
Thus, scale-free networks, characterized by the exponent $\gamma$,
can also be described by the two length invariant exponents $d_B$
and $d_k$.

\section{Other measurements}\label{sec:other}

This section describes additional, complementary measurements related
to network complexity, edge reciprocity and matching index.

\subsection{Network Complexity}

It might be of interest to quantify the `complexity' of a
network. Lattices and other regular structures, as well as purely
random graphs, should have small values of complexity. Some recent
proposals are briefly presented below.

Machta and Machta \cite{Machta05} proposed the use of the
computational complexity of a parallel algorithm \cite{Codenotti93}
for the generation of a network as a complexity measurement of the
network model. If there is a known parallel algorithm for the
generation of the network of order $\mathcal{O}(f(N))$, with $f(x)$ a
given function, then the complexity of the network model is defined as
$\mathcal{O}(f(N))$. For example, Barab\'{a}si-Albert networks can be
generated in $\mathcal{O}(\log \log N)$ parallel steps
\cite{Machta05}.

Meyer-Ortmanns \cite{Meyer03} associated the complexity of the
network with the number of topologically non-equivalent graphs
generated by splitting vertices and partitioning the edges of the
original vertex among the new vertices, the transformations being
restricted by some constraints to guarantee the generation of valid
graphs.

The \emph{off-diagonal complexity}, proposed by Claussen
\cite{Claussen04} is defined as an entropy of a specially defined
vertex-vertex edge correlation matrix. An element with indexes $(k,l)$
of this matrix has contributions from all edges that connect a vertex
of degree $k$ to a vertex of degree $l$ (only values $k>l$ are used).

\subsection{Edge Reciprocity}\label{recip}

For directed networks, it is often interesting to know if their
edges are reciprocal, i.e.\ if vertex $i$ is linked to vertex $j$,
is vertex $j$ also linked to vertex $i$? Such information helps to
obtain a better characterization of the network, can be used to test
network models against real networks and gives indication of how
much information is lost when the direction of the edges is
discarded (e.g. for the computation of some measurements that only
apply to undirected networks).

A standard way to obtain information about reciprocity is to compute
the fraction of bilateral edges:
\begin{equation}
  \varrho = \frac{\sum_{ij} a_{ij} a_{ji}}{M},
  \label{eq:recipnormal}
\end{equation}
where $M$ is the total number of edges.

A problem with this measurement is that its value is only relevant
with respect to a random version of the network, as networks with
higher connectivity tend to have a higher number of reciprocal edges
due exclusively to random factors.  Garlaschelli and Loffredo
\cite{garlaschelli04} proposed the use of the correlation
coefficient of the adjacency matrix:
\begin{equation}
  \rho = \frac{ \sum_{ij} (a_{ij}-\bar{a}) (a_{ji}-\bar{a}) }
              { \sum_{ij} (a_{ij}-\bar{a})^2 },
  \label{eq:reccorr}
\end{equation}
where $\bar{a}$ is the mean value of the elements of the adjacency
matrix. This expression, known as \emph{edge reciprocity},
simplifies to
\begin{equation}
  \rho = \frac{\varrho-\bar{a}}{1-\bar{a}}.
  \label{eq:recgarla}
\end{equation}
This value is an absolute quantity, in the sense that values of $\rho$
greater than zero imply larger reciprocity than the random version
(\emph{reciprocal} networks), while values below zero imply smaller
reciprocity than a random network (\emph{antireciprocal}
networks). This concept can be easily extended to weighted networks by
substituting $a_{ij}$ for $w_{ij}$ in the above expressions.

\subsection{Matching Index}

A \emph{matching index} can be assigned to each edge in a network in
order to quantify the similarity between the connectivity of the two
vertices adjacent to that edge \cite{Kaiser04}. A low value of the
matching index identifies an edge that connects two dissimilar regions
of the network, thus possibly playing an important role as a shortcut
between distant network regions \cite{Kaiser04}. The matching index of
edge $(i,j)$ is computed as the number of matching connections of
vertices $i$ and $j$ (i.e.\ connections to the same other vertex $k$),
divided by the total number of connections of both vertices (excluding
connections between $i$ and $j$),
\begin{equation}
  \mu_{ij} = \frac{\sum_{k\ne i,j} a_{ik}a_{jk}}
                  {\sum_{k\ne j}a_{ik}+\sum_{k\ne i}a_{jk}}.
  \label{eq:matching}
\end{equation}
For directed networks, matching connections are only those in the
same direction, and incoming and outgoing connections of vertices
$i$ and $j$ should be considered separately.  The matching index
has also been adapted to apply to consider all the immediate
neighbors of a node, instead of a single edge~\cite{Costa0607272}.

\section{Measurements of Network Dynamics and Perturbation}\label{netdyn}

This section covers two important related issues, namely the use of
trajectories to characterize the dynamical evolution of complex
networks connectivity and a brief discussion about the 
sensitivity of measurements to perturbations.

\subsection{Trajectories}\label{trajectories}

\begin{sidewaysfigure}
  \begin{center}
    \includegraphics[width=\linewidth]{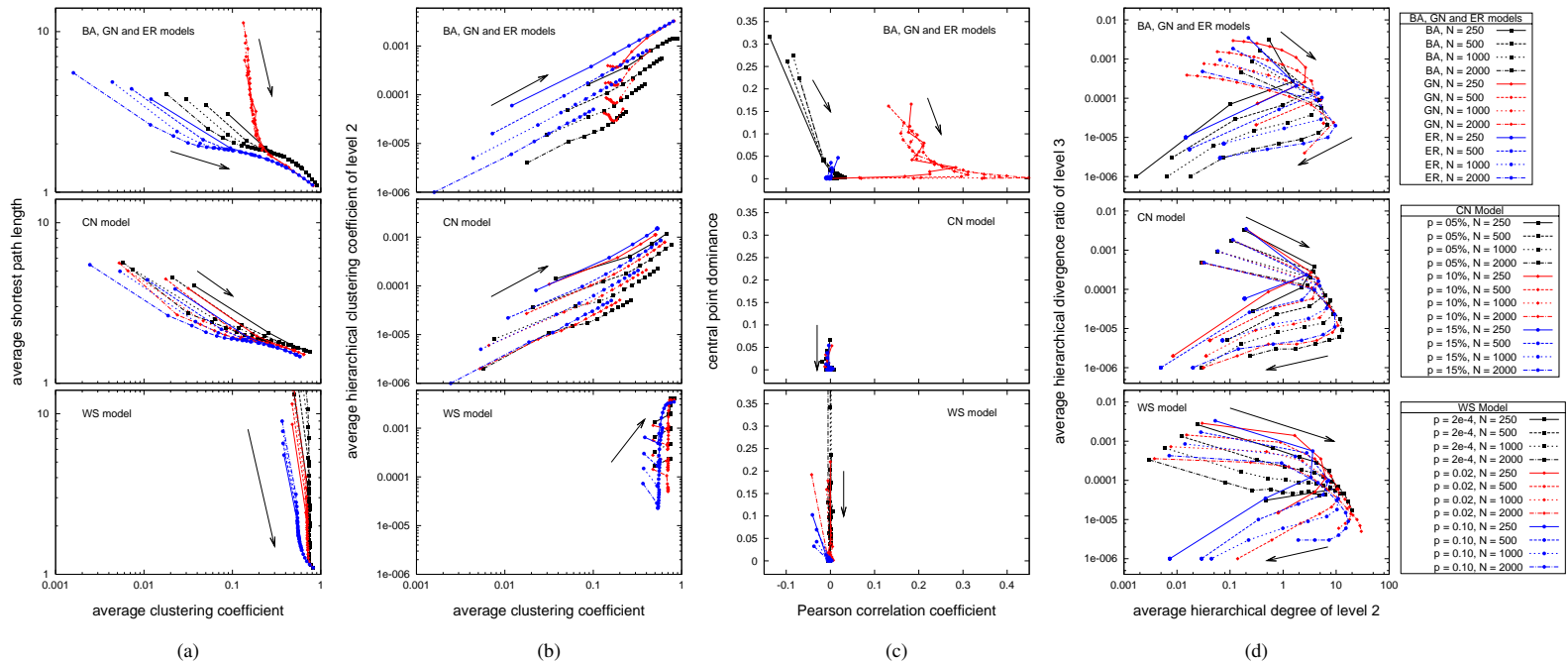}
  \end{center}
  \caption{Trajectories defined by pairs of measurements. Each
  point corresponds to $10$ network model realizations. Network sizes
  used are $250, 500, 1000,$ and $2000$; average degrees vary from $4$
  to $204$ in steps of $20$; for the community model,
  $p_\mathrm{out}/p_\mathrm{in}$ is $5\%, 10\%,$ and $15\%$; for the
  WS model the rewiring probability values are $0.0002, 0.02,$ and
  $0.1$.}
  \label{fig:traject}
\end{sidewaysfigure}

As motivated in the Introduction of this work, in the following we
analyze the behavior of trajectories (see Figure~\ref{fig:traject})
defined by tuples of measurements as the analyzed network undergoes
progressive modification, such as during their growth. The network
models considered for the illustration of trajectories include:
Erd\H{o}s-R\'{e}nyi random graphs (ER), random networks with community
structure (CN), Watts-Strogatz small-worlds (WS), Geographical
Networks (GN), and Barab\'{a}si and Albert scale-free networks (BA)
(see section~\ref{sec:models} for a description of these models).  The
number of vertices considered was $250, 500, 1000$ and $2000$, and the
number of edges varies so that the average vertex degree ranges from
$4$ to $204$, increasing by steps of $20$.  In the case of the GN
model, the vertices were randomly distributed through a square box of
unit size.  The $\lambda$ parameter in Eq.~\ref{dist_geo} was adjusted
in order to guarantee the desired average degree. The CN networks
include four communities interconnected by using
$p_\mathrm{out}/p_\mathrm{in}$ (see Section~\ref{sec:models}) equal to
$5\%$, $10\%$ and $15\%$. In the WS model, the probability $p$ of
rewiring the edges was $0.0002$, $0.02$ and $0.1$. For the sake of
better visualization, the trajectories of the WS and CN models are
drawn separately from the other cases.  The direction of evolution of
the trajectories as more edges are included is indicated by arrows in
Figures~\ref{fig:traject}.  These results are discussed subsequently
with respect to several pairs of measurements.

\subsubsection{Average Clustering Coefficient and Average Shortest Path Length}

By inspecting the trajectories associated to this pair of measurements
(see Figure~\ref{fig:traject}(a)), two distinct behaviors can be
identified. First, the average clustering coefficient $\widetilde{C}$
exhibits a high variation while the average shortest path length
$\ell$ remains almost constant with addition of edges for the ER, CN
and BA models. Second, an opposite effect is observed for GN and WS
models. In the latter case, the $\ell$ value undergoes a steep
decrease, while staying almost constant for the other network
models. This effect is related to the fact that GN and WS models are
formed by vertices that tend to link to closer neighbors. Hence, with
the addition of edges some long-range connections may decrease $\ell$
while $\widetilde{C}$ remains almost unchanged. Furthermore, $\ell$
decreases faster for WS model than for GN while $\widetilde{C}$ for
the former model remains larger than in the other cases. This can be
explained by the fact that the WS model is more regular than the GN
and has larger $\widetilde{C}$.

In the case of ER, CN and BA models, the values of $\ell$ and
$\widetilde{C}$ are smaller than for the other models. For $\ell$ the
connections are not limited by proximity, adjacency or geography, and
for $\widetilde{C}$ loops of order three appear when new edges are
added to them, increasing this measurement.

Another interesting fact observed from Figure~\ref{fig:traject}(a) is
that all curves converge to the same point, corresponding to fully
connected graphs, as the networks become denser. Therefore, $\ell$ and
$\widetilde{C}$ tend in this stage to a unit value.

\subsubsection{Average Clustering Coefficient and Average Hierarchical
Clustering Coefficient of Second Level}

The combination\footnote{Note that $\widetilde{C}$ is the same as the
average hierarchical clustering coefficient of first level.}  of
$\widetilde{C}$ and the average hierarchical clustering coefficient of
second level $\langle C_2(i) \rangle$, where the average is taken over
all vertices in the network, see Figure~\ref{fig:traject}(b), tend to
follow a power law for all trajectories except for the GN and WS
models, whose curves have a minimum value for $\langle C_2(i)
\rangle$. Nonetheless, the highest growth rate is observed for the
trajectories of WS model after their minimum value of $\langle C_2(i)
\rangle$ is reached.

Another interesting characteristic of this combination of
measurements is that $\widetilde{C}$ is greater than $\langle C_2(i)
\rangle$. This can be explained by the fact of $\langle C_2(i)
\rangle$ is related to the presence of loops of order five without
additional connections between their vertices \cite{Costa:2005a}.
Since loops of higher orders are less likely to appear in the
considered networks, $\widetilde{C}$ is larger than $\langle C_2(i)
\rangle$.

\subsubsection{Pearson Correlation Coefficient and Central Point Dominance}

For all network models, except for the GN, the $r$ value is close to
zero even with the addition of new edges, as can be seen in
Figure~\ref{fig:traject}(c), which shows the trajectories defined by
the pair of measurements Pearson correlation of the degrees $r$
(Section~\ref{degrees}) and central point dominance $\mathit{CPD}$
(Section~\ref{sec:central}) as the average degree increases. This
property can be explained by the fact that in ER, CN and WS models
edges are placed without regard to vertex degree, while the BA model
is based on growth \cite{Newman01}, which leads to non-assortative
mixing (i.e.\ no correlation between vertex degrees). The $r$ value
for the GN model is greater than zero in almost all cases because its
growing dynamics is based on the geographic proximity of vertices. As
the position of vertices is randomly chosen, some regions may result
highly populated, implying the respective vertices to have a high
probability of becoming highly interconnected. On the other hand,
vertices belonging to the regions barely populated have a small
likelihood to become ``hubs'' while still having a good chance of
being connected. These two opposite behaviors lead to a $r$ value
greater than zero.

The central point dominance is a measurement of the maximum
betweenness of any point in the network \cite{Freeman:1977} (see
Section~\ref{sec:central} for further details). By observing
Figure~\ref{fig:traject}(c), one can see that most network models
exhibit average values of this measurement close to zero, except for
the BA, GN and WS cases. In BA networks, values significantly larger
than zero only occur in the beginning of the growth process (i.e.\
with few edges).

For WS models, the way in which they are normally constructed (see
Section~\ref{sec:models}) directly contributes to producing a network
with modular structure, hence a high $\mathit{CPD}$
value. Nevertheless, when new edges are added, the network gets denser
and the value of this measurement goes to zero.  In CN models, the
$\mathit{CPD}$ coefficient depends on the relation between the average
vertex degrees inside and outside communities, i.e.\ when the network
is highly modular, the $\mathit{CPD}$ value tends to become larger.

\subsubsection{Average Hierarchical Degree of Second Level and
  Average Hierarchical Divergence Ratio of Third Level}

As shown in Figure~\ref{fig:traject}(d), all curves obtained for the
average hierarchical degree of second level\footnote{Notice that
$\langle k_2(i) \rangle$ (average taken over all vertices $i$ in the
network) depends on the network connectivity.} $\langle k_2(i)
\rangle$ and the average hierarchical divergence ratio of level
three $\langle \mathit{dv}_3(i) \rangle$ have similar behavior. When
the networks are sparse and new edges are added to them increasing
the average vertex degree, the average hierarchical counterpart
increases until a maximum value.  Afterwards, since the networks
have a finite size, further increase of the connectivity tends to
reduce the number of hierarchical levels in the networks and, as
consequence, the average hierarchical vertex degree of levels higher
than one tends to decrease.  The hierarchical divergence ratio of
level three decreases with larger average vertex degree.

\subsubsection{Discussion}

As presented in Figure~\ref{fig:traject}, each measurement is
specifically sensitive to the effects of addition of new edges to a
network.  Interestingly, the sensitivity also depend strongly on the
network model.  Some trajectories resulted closer one another for
specific network models as a consequence of inherent structural
similarities.  This effect is particularly pronounced in trajectories
defined by the average clustering coefficient and average shortest
path length, where two classes of trajectories appear, one for ER, BA
and CN, and another for GN and WS.

The analysis of network dynamics provides insights about model
similarities. If network trajectories evolve in a similar fashion,
it is possible to infer that these networks have similar structure
concerning the respective pair of measurements. However, for other
measurements, this similarity may be weaker or non-existent.  For
instance, in the space defined by the average clustering coefficient
and average shortest path length, the curves obtained for RA and BA
evolve in similar fashion. This behavior is not observed in the
space defined by the central point dominance and Pearson correlation
coefficient. Also, by inspecting the trajectories, it is possible to
determine the correlation between measurements during the network
evolution. For instance, the dynamics of the average clustering
coefficient and the average hierarchical clustering coefficient of
second level present correlation for ER and CN.

The trajectory-based study described here can be immediately extended
to real network analysis and modeling. In the case of the WWW, for
instance, by inspecting its evolution in the measurements space it is
possible to develop more precise models to represent and characterize
its structure. For citation networks, it is possible to characterize
the networks generated for different knowledge areas and obtain
insights about their structure and evolution.  All in all,
trajectories provide a visually clear and accessible interpretation
and dynamics about the evolution of complex networks connectivity.

\subsection{Perturbation Analysis}\label{perturbation}

Another important property of a given measurement relates to how much
it changes when the networks undergo small \emph{perturbations} (e.g.,
rewiring, edge or vertex attack, weight changes, etc.).  For instance,
the shortest path length provides a good example of a particularly
sensitive measurement, in the sense that the modification of a single
connection may have great impact in its value.  The quantification of
the sensitivity of measurements to different types of perturbations
and networks therefore provides valuable information to be considered
for characterization, analysis and classification of complex networks.

\begin{figure}
  \begin{center}
    \subfigure[]{\includegraphics[width=0.48\linewidth]{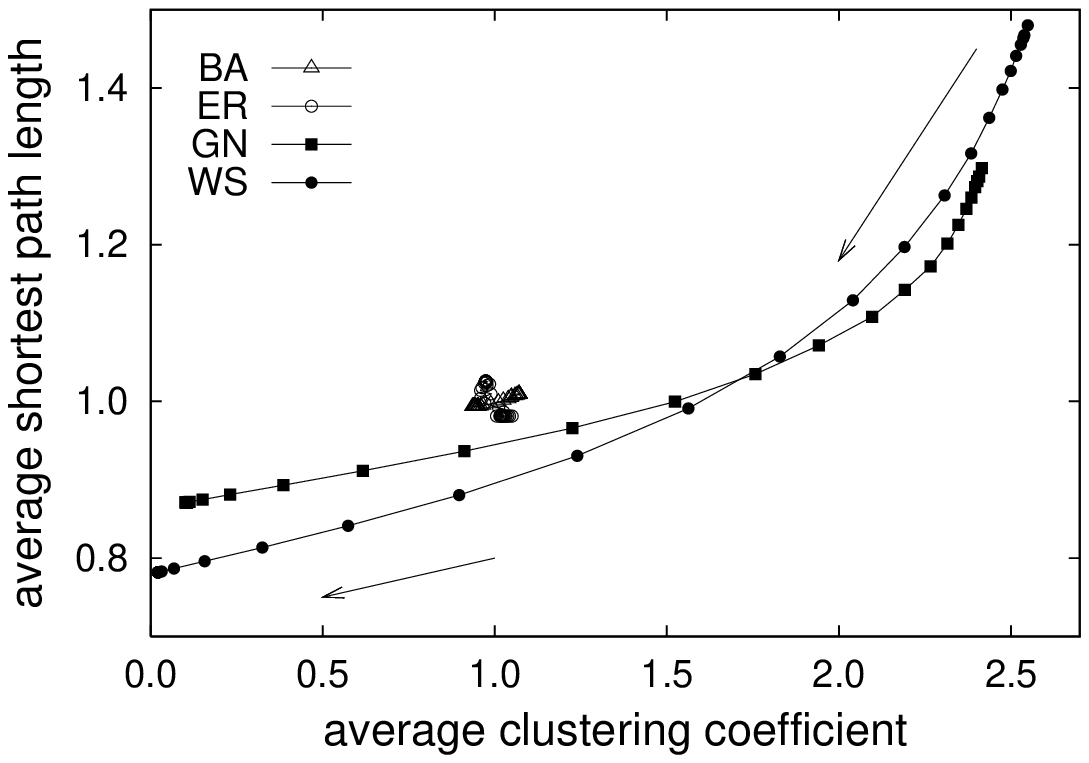}}
    \, \,
    \subfigure[]{\includegraphics[width=0.48\linewidth]{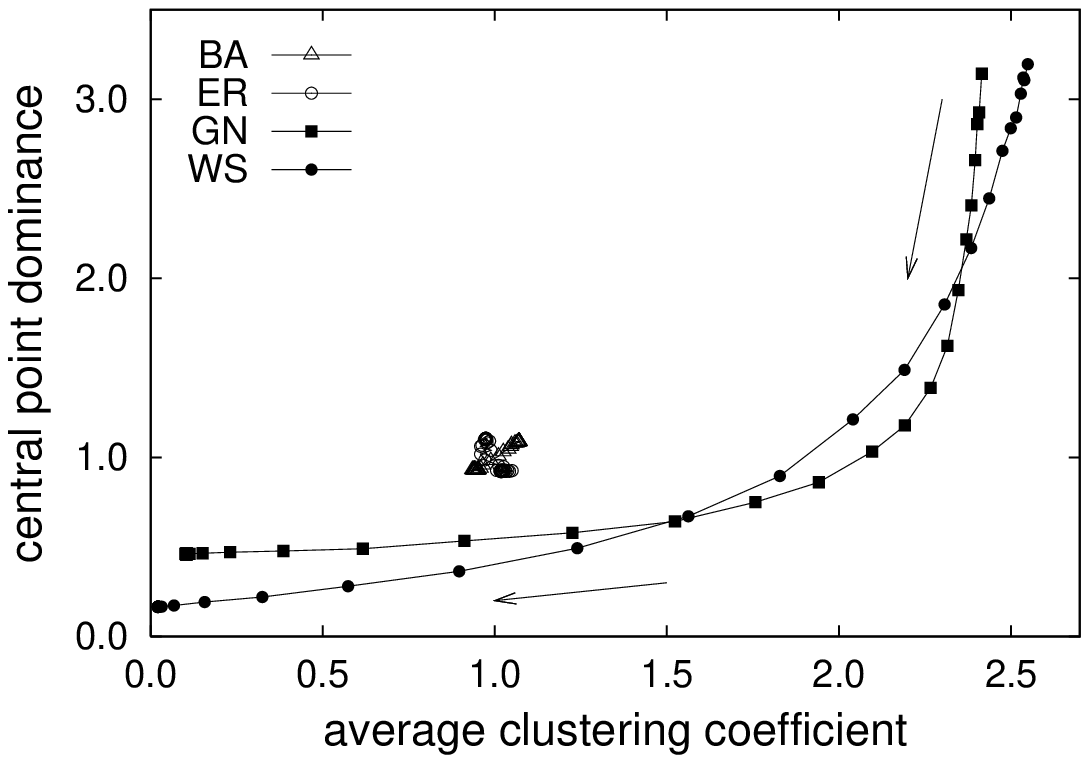}}
  \end{center}
  \caption{Example of perturbations. Each point corresponds to $100$
  realizations of networks with $N=1000$ and $\langle k \rangle=6$.}
  \label{fig:pert}
\end{figure}

Interesting insights can be obtained as far as this subject is
concerned by performing progressive perturbations in a specific
network and observing the respective relative variations in the
measurement of interest.  Figure~\ref{fig:pert} shows the trajectories
obtained in the $\{\widetilde{C},\ell \}$ and $\{\widetilde{C},
\mathit{CPD}\}$ feature spaces shown in Figure~\ref{fig:pert}(a) and
(b), respectively, while considering edge randomization
\cite{Milo:2003} (see also Section~\ref{genrandom}) progressively
performed for the BA, ER, GN and WS models ($N=1000$, $\langle k
\rangle = 6$).  The values shown in this figure were normalized
through division by the respective averages of the measurements in
order to provide suitable visual comparison.

Each successive point along the trajectories, which are indicated by
arrows in Figure~\ref{fig:pert}, was obtained after a number of
rewirings which are successive integer powers of two (for the sake of
obtaining more uniform visualization). It is clear from the results
that the sensitivity of the two pairs of measurements varies
substantially with respect to the type of network under
consideration. More specifically, much wider variations were observed
for the GN and WS models.  The stable trajectories obtained for the
shortest path length in the case of the BA model is a direct
consequence of the fact that these networks are inherently
characterized by low overall average shortest path length.  A similar
situation is verified for the clustering coefficient, which tends to
be small in those two types of networks. The evolution of the
trajectory regarding the clustering coefficient and average shortest
path length for the GN and WS networks is a direct consequence of the
fact that the progressive edge rewiring tend to strongly reduce those
two measurements.  The marked difference of sensitivity of
measurements to perturbations depending on the type of network model
suggests that quantifications of the sensitivity (e.g. the standard
deviation or entropy) can be potentially useful as measurements for
network identification.

\section{Correlations Analysis}\label{correlation}

In order to investigate possible linear relationships between
several of the measurements considered, we calculated them for the
BA, ER and GN models ($N = 1000$ and $\left< k \right> = 4$) and
estimated the respective pairwise correlations.
Table~\ref{tab:correl} shows the values obtained with respect to
each model and also considering all models together (`All').

\begin{table}
  \caption{Correlation between measurements for the BA, ER and GN
    models and ``All'' jointly. Values estimated from $1000$
    realizations (for each model) of networks with $N=1000$ and
    $\langle k \rangle = 4$.}
  \label{tab:correl}
  {\footnotesize
  \begin{tabular}{|c|c|c|c|c|c|c|c|c|c|}
    \hline
    & & $\mathit{st}$ & $r$ & $\widetilde{C}$ & $\ell$ & $\mathit{CPD}$ & $\langle k_2(i) \rangle$ & $\langle C_2(i) \rangle$ & $\langle \mathit{dv}_3(i) \rangle$ \\
    \hline
 & BA & ~1.00 &  &  &  &  &  &  &  \\
 $\mathit{st}$ & ER & ~1.00 &  &  &  &  &  &  &  \\
 & GN & ~1.00 &  &  &  &  &  &  &  \\
 & All & ~1.00 &  &  &  &  &  &  &  \\
\hline
 & BA & -0.22 & ~1.00 &  &  &  &  &  &  \\
 $r$ & ER & -0.01 & ~1.00 &  &  &  &  &  &  \\
 & GN & -0.13 & ~1.00 &  &  &  &  &  &  \\
 & All & ~0.71 & ~1.00 &  &  &  &  &  &  \\
\hline
 & BA & ~0.06 & -0.29 & ~1.00 &  &  &  &  &  \\
 $\widetilde{C}$ & ER & -0.01 & ~0.07 & ~1.00 &  &  &  &  &  \\
 & GN & ~0.04 & -0.00 & ~1.00 &  &  &  &  &  \\
 & All & ~0.31 & ~0.82 & ~1.00 &  &  &  &  &  \\
\hline
 & BA & -0.01 & ~0.38 & -0.63 & ~1.00 &  &  &  &  \\
 $\ell$ & ER & -0.06 & ~0.04 & -0.08 & ~1.00 &  &  &  &  \\
 & GN & -0.10 & ~0.02 & ~0.03 & ~1.00 &  &  &  &  \\
 & All & ~0.69 & ~0.96 & ~0.88 & ~1.00 &  &  &  &  \\
\hline
 & BA & -0.09 & ~0.23 & ~0.39 & -0.58 & ~1.00 &  &  &  \\
 $\mathit{CPD}$ & ER & -0.61 & ~0.10 & ~0.03 & ~0.07 & ~1.00 &  &  &  \\
 & GN & -0.05 & -0.02 & ~0.03 & ~0.23 & ~1.00 &  &  &  \\
 & All & -0.87 & -0.44 & ~0.02 & -0.41 & ~1.00 &  &  &  \\
\hline
 & BA & ~0.01 & -0.30 & ~0.63 & -0.99 & ~0.60 & ~1.00 &  &  \\
 $\langle k_2(i) \rangle$ & ER & ~0.04 & ~0.03 & ~0.08 & -0.90 & -0.06 & ~1.00 &  &  \\
 & GN & ~0.08 & ~0.28 & -0.02 & -0.65 & -0.13 & ~1.00 &  &  \\
 & All & -0.96 & -0.80 & -0.43 & -0.79 & ~0.85 & ~1.00 &  &  \\
\hline
 & BA & ~0.02 & ~0.02 & ~0.58 & -0.74 & ~0.59 & ~0.76 & ~1.00 &  \\
 $\langle C_2(i) \rangle$ & ER & -0.03 & ~0.04 & ~0.45 & -0.16 & ~0.02 & ~0.19 & ~1.00 &  \\
 & GN & -0.00 & ~0.09 & ~0.59 & ~0.18 & ~0.07 & -0.11 & ~1.00 &  \\
 & All & ~0.37 & ~0.86 & ~0.99 & ~0.91 & -0.05 & -0.49 & ~1.00 &  \\
\hline
 & BA & ~0.01 & ~0.26 & -0.57 & ~0.91 & -0.52 & -0.94 & -0.69 & ~1.00 \\
 $\langle \mathit{dv}_3(i) \rangle$ & ER & ~0.03 & -0.10 & -0.01 & -0.25 & -0.01 & -0.16 & -0.04 & ~1.00 \\
 & GN & -0.02 & -0.28 & -0.09 & -0.03 & -0.00 & -0.50 & -0.21 & ~1.00 \\
 & All & -0.14 & -0.74 & -0.97 & -0.79 & -0.18 & ~0.27 & -0.96 & ~1.00 \\
\hline
  \end{tabular}
}
\end{table}

\begin{figure}
  \begin{center}
    \includegraphics[width=0.6\linewidth]{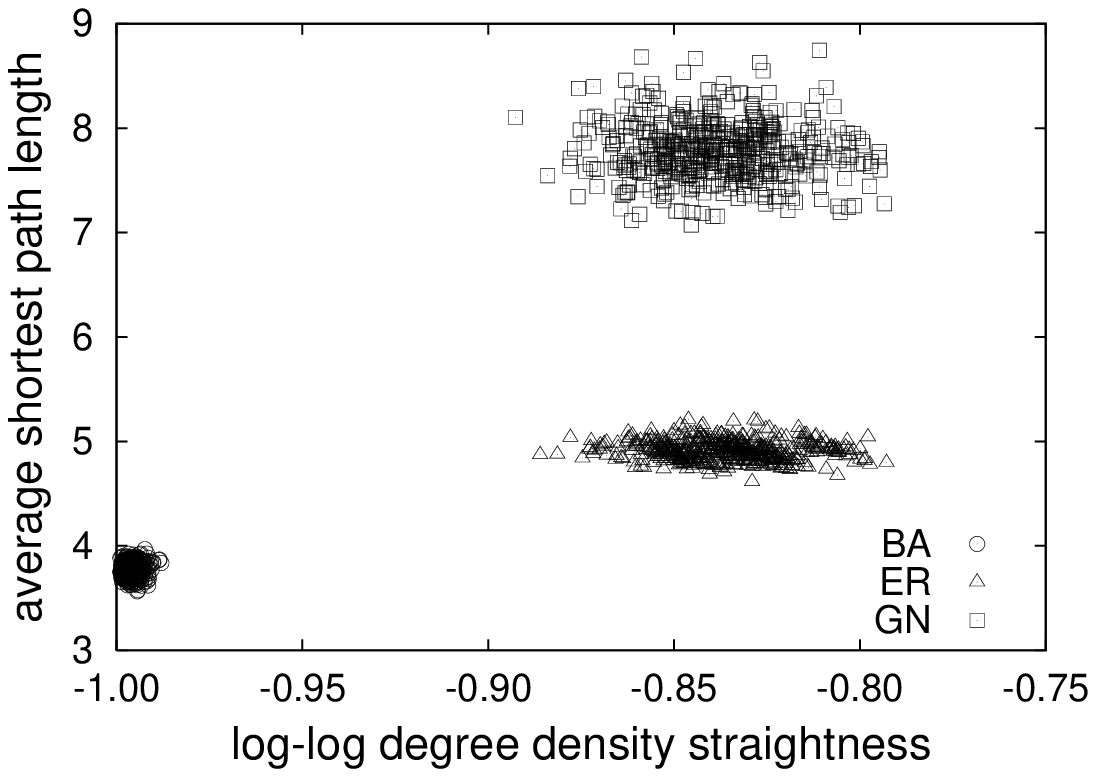}
  \end{center}
  \caption{Example of scatterplot showing the low correlation between
  log-log degree density straightness and the average shortest path
  length for all the individual models and the high correlation for
  all models together. The networks have $N=1000$ and $\langle k
  \rangle=4$; $500$ realizations of each model were used.}
  \label{fig:correl}
\end{figure}

Several interesting facts can be inferred from this table.  First,
particularly high absolute values of correlations have been obtained
for the BA model, with low absolute values observed for the ER and GN
cases.  This seems to represent a particularly interesting property of
the BA networks.  Another interesting finding regards the fact that
the correlations obtained for specific network models not necessarily
agree with that obtained when the three models are considered
together.  This is the case, for instance, of the low correlation
observed between the measurement average shortest path length $\ell$
and log-log degree distribution straightness $\mathit{st}$ for each of
the three individual models and high correlation otherwise obtained
when these three models are considered jointly.  This interesting
behavior can be immediately understood from Figure~\ref{fig:correl},
which indicates that the three low correlations groups obtained for
the individual models tend to align globally, therefore implying the
relatively strong negative correlation.  Such situations indicate that
the individual and global correlations provide information about
different types of relationships and should be treated accordingly.

It is also clear from the results in Table~\ref{tab:correl} that
particularly high correlations were obtained between the average
shortest path length $\ell$ and the vertex degree at the second
hierarchical level $\langle k_2(i) \rangle$.  This fact suggests
that this specific hierarchical vertex degree may be considered, at
least for the three types of networks, as an estimation of the
average shortest path length, allowing substantial computational
saving. Another interesting result is that the highest correlations
were obtained for the BA model, a possible consequence of the
respective hubs. For instance, the correlation between the average
shortest path length and the average clustering coefficient was
equal to -0.63 for the BA models. This is a consequence of the fact
that additional links tend to be established with the hubs and
therefore contribute to higher clustering and shortest paths.

\section{Multivariate Statistical Methods for Measurement Selection
  and Network Classification}\label{multivariate}

\begin{figure}
  \centerline{\includegraphics[width=0.65\linewidth]{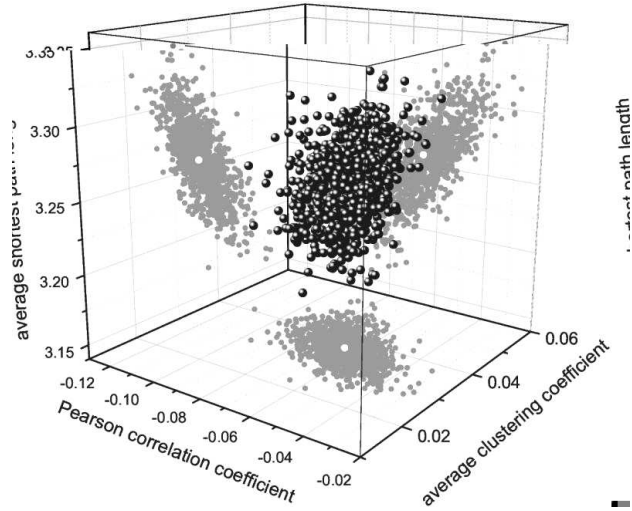}}
  \caption{The spatial distribution in the $(r,\widetilde{C},\ell)$
    phase space of $1000$ different realizations of the BA model with
    $N=1000$ and $m=3$. The distribution has been projected into the
    three main planes (gray shadows) for the purpose of better
    visualization. The \emph{white circles} in the middle of the gray
    shadows represent the mean projected into these planes.}
  \label{fig:exmult}
\end{figure}

The intrinsic statistical variability of the connectivity of real
and simulated complex networks, even when produced by the same
process or belonging to the same class, implies that sound
characterization, comparison and classification of networks should
take into account not only the average measurements, but also
additional information about their variability including higher
statistical moments (e.g., variance, kurtosis, etc.)\ as well as
multivariate statistical distribution of the measurements.  For
example, realizations obtained by using the Barab\'asi-Albert (BA)
model with fixed parameters will produce networks which, though not
identical, will have equivalent statistical distribution of their
properties. Figures~\ref{fig:exmult} shows a scatterplot obtained by
considering 1000 realizations of the BA model with $N=1000$ and
$m=3$ with respect to the measurements $(r,\widetilde{C},\ell)$,
where $r$ is the Pearson correlation coefficient of vertex degrees;
$\widetilde{C}$, the average clustering coefficient; and $\ell$, the
average shortest path length. Although the obtained points form a
well-defined cluster around the average point $(- 0.0653, \, 0.0365,
\, 3.255)$, there is a significant dispersion of cases around this
center, implying that additional measurements other than the mean
need to be used for proper characterization of the network under
analysis.  Therefore, any objective attempt at characterizing,
comparing or classifying complex networks needs to take into account
distributions in phase spaces such as that in
Figure~\ref{fig:exmult}.  Such an important task can be effectively
accomplished by using traditional and well-established concepts and
methods from \emph{Multivariate Statistics} (e.g.,
\cite{Costa:01,Duda_Hart:01},McLachlan:04) and 
\emph{Pattern Recognition} (e.g., \cite{Costa:01,Fukunaga:90,Duda_Hart:01}).

As far as the choice and interpretation of network measurements are
concerned, two multivariate methods stand out as being particularly
useful, namely \emph{Principal Component Analysis} --- PCA
(e.g. \cite{Costa:01,Duda_Hart:01}) and \emph{Canonical Variable
Analysis} (e.g., \cite{McLachlan:04}).  While the former procedure
allows the reduction of the dimensionality of the measurement phase
space, obtained in terms of projections so as to concentrate the
variation of the data along the first axes (i.e.\ those associated to
the highest covariance matrix eigenvalues), the second method
implements such projections so as to achieve best separation, in
terms of inter and intra-class distances (see below), between the
involved classes of networks under analysis.

Another situation in multivariate statistics which is particularly
important for complex network research concerns network
identification.  Indeed, it is often a critical issue to decide to
which of several models a given theoretical or experimentally
obtained network belongs.  This important problem can be approached
in a sound way by using Bayesian decision theory
\cite{Duda_Hart:01}, a well-established methodology which, provided
we have good probabilistic models of the properties of the networks,
allows near-optimal classification performance~\footnote{Optimal
performance is guaranteed provided the involved mass and conditional
properties are perfectly known (see Section~\ref{sec:bayes} and
\cite{Costa:01,Duda_Hart:01}).}.

The current section presents and illustrates in a self-contained and
accessible fashion these two families of methods from multivariate
statistics, i.e.\ dimensionality reduction (PCA and canonical
analysis) and classification (Bayesian decision theory).  The
possibility to apply Bayesian decision theory on phase spaces obtained
through canonical variable analysis projections of multidimensional
measurement spaces is also illustrated.  The potential for
applications of these methods is illustrated with respect to three
reference complex network models --- namely Erd\H{o}s and R\'enyi
random graph (ER), Barab\'asi-Albert (BA) and Geographical Network
model (GN), against which some real-world networks are classified.

It should be observed that many other methods from multivariate
statistical analysis, including hierarchical clustering and structural
equation modeling, can also be valuable for investigations in complex
network research.  Though the potential of hierarchical clustering for
suggesting relationships between the classes is briefly illustrated in
the following, further information about such methods can be found in
textbooks such as
\cite{Fukunaga:90,Duda_Hart:01,Costa:01,Hair_etal:98,Johnson_Wichern:02,McLachlan:04}.

\subsection{Principal Component Analysis}

Let the connectivity properties of a set of $R$ complex networks,
irrespective of their type or origin, be described in terms of $P$
scalar measurements $x_i$, $i = 1, 2, \ldots, P$, organized as the
feature vector $\vec{x} = (x_1, x_2, \ldots, x_P)^T$.  The covariance
matrix $K$ can be estimated as
\begin{equation}
  K = \frac{(\vec{x}-\vec{\left< x \right>})(\vec{x}-\vec{\left< x
  \right>})^T}{R},
\end{equation}
where $\vec{\left< x \right>}$ is the average feature vector, each
element of which corresponds to the average of the respective
measurement. As $K$ is a real and symmetric $P \times P$ matrix, a
set of $P$ decreasing eigenvalues $\lambda_i$ and respectively
associated eigenvectors $\vec{v_i}$ can be obtained.  Moreover, if
all eigenvalues are distinct, the eigenvectors will be
orthogonal~\footnote{Otherwise, orthogonal eigenvectors can still be
assigned to repeated eigenvalues.}.  These eigenvectors can be
stacked to obtain the transformation matrix $T$, i.e.\
\begin{equation}
  T = \left[
    \begin{array}{ c }
      \leftarrow \vec{v_1} \rightarrow \\
      \leftarrow \vec{v_2} \rightarrow \\
      \ldots \\
      \leftarrow \vec{v_P} \rightarrow
    \end{array} \right].
\end{equation}

The original feature vectors $\vec{x}$ can now be transformed into a
new coordinates reference through the following linear
transformation corresponding to axes rotation:
\begin{equation}
  \vec{X} = T \vec{x}
\end{equation}
which defines the \emph{principal component projections}.

It can be shown \cite{Johnson_Wichern:02} that the distribution of
points in the new phase space obtained by the above transformation is
such that the largest variance is observed along the first axis,
followed by decreasing variances along the subsequent axes, with the
initial axes called \emph{principal}. Such an important property
allows, by considering only the principal eigenvalues, the original
cloud of points to be projected along phase spaces of a smaller
dimension $p$.  In order to do so, the transformation matrix is
constructed while taking into account only the first $T_p$
eigenvectors associated to the largest eigenvalues, i.e.\
\begin{equation}
 T_p = \left[
  \begin{array}{ c }
     \leftarrow \vec{v_1} \rightarrow \\
     \leftarrow \vec{v_2} \rightarrow \\
     \ldots \\
     \leftarrow \vec{v_p} \rightarrow
  \end{array} \right].
\end{equation}

\begin{figure}
  \centerline{\includegraphics[width=0.6\linewidth]{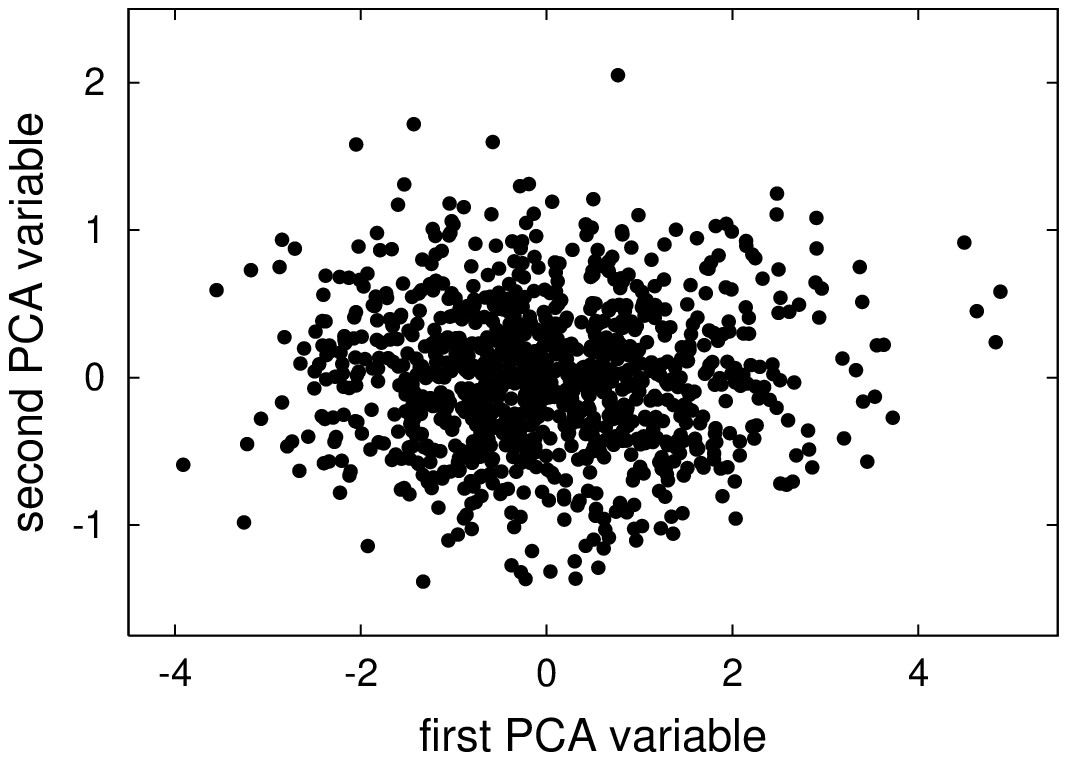}}
  \caption{The principal component projection of the distribution of
    measurements in Figure~\ref{fig:exmult}. Measurement values were
    first normalized by subtracting the corresponding mean value and
    dividing by the standard deviation, to avoid distortions due to
    the different absolute values. The first and second PCA variable
    have projecting vectors $(-0.005, 0.707, -0.707)$ and $(0.006, 0.707,
    0.707)$ in the space defined by $(r,\widetilde{C},\ell)$, respectively. }
  \label{fig:expca}
\end{figure}

Figure~\ref{fig:expca} shows the effect of projecting the cloud of
points in Figure~\ref{fig:exmult} onto the two main axes where the
variance of the samples is maximized.  Although useful for
implementing dimensionality reduction --- which favors
visualization, redundancy reduction, and computational savings ---
the principal component analysis method is limited as it does not
explicitly consider the category of each individual.  This
limitation is overcome by the canonical variable analysis described
below.  Note that the obtained projections are little influenced by
the measurement $r$, which is compatible with the fact that degree
correlations are almost absent in BA networks~\cite{Boccaletti05}.

\subsection{Canonical Variable Analysis}

The method known as \emph{canonical variable analysis} provides a
powerful extension of principal component analysis by performing
projections which optimize the separation between the known categories
of objects.  Before presenting the method, we introduce a series of
scatter measurements considering each category separately and also all
categories, from which the overall criterion for class separation is
defined.

Let us consider that the $R$ complex networks of interest can be
divided into $N_c$ classes, each one with $N_i$ objects and
identified as $C_i$, $i = 1, 2, \ldots, N_c$, and that each object
$\xi$ is represented by its respective feature vector $\vec{x_\xi} =
(x_1, x_2, \ldots, x_P)^T$ (see the previous section). The
\emph{total scatter matrix}, $S$, expressing the overall dispersion
of the measurements \cite{Costa:01} is defined as follows
\begin{equation}
  S = \sum_{\xi \, =1}^{R} \left ( \vec{x_\xi} - \vec{\left< x \right>} \right
  ) \left ( \vec{x_\xi} - \vec{\left< x \right>} \right )^T.
\end{equation}

The \emph{scatter matrix for each class $C_i$} is given as
\begin{equation}
  S_i = \sum_{\xi \in \; C_i} \left ( \vec{x_\xi} - \vec{\left< x \right>}_i
  \right ) \left ( \vec{x_\xi} - \vec{\left< x \right>}_i \right
  )^T,
\end{equation}
where $\vec{\left< x \right>}_i$ is the average feature vector of the
class $C_i$.

The \emph{intraclass scatter matrix}, providing the dispersion
inside each of the classes, is defined as
\begin{equation}
  S_{\mathrm{intra}} = \sum_{i=1}^{N_c} S_i .
\end{equation}

Finally, the \emph{interclass scatter matrix}, characterizing the
dispersion between each pair of classes, is given as
\begin{equation}
  S_{\mathrm{inter}} = \sum_{i=1}^{N_c} N_i \left ( \vec{\left< x
    \right>}_i - \vec{\left< x \right>} \right ) \left ( \vec{\left< x
    \right>}_i - \vec{\left< x \right>} \right )^T.
\end{equation}

It can be verified that
\begin{equation}
  S = S_{\mathrm{intra}} + S_{\mathrm{inter}}.
\end{equation}

The objective of the canonical analysis method is to maximize the
interclass dispersion while minimizing the intraclass scattering
(e.g., \cite{McLachlan:04}).  This can be achieved through the following
linear transformation
\begin{equation}
  \vec{X_\xi} = \Gamma \vec{x_\xi}
\end{equation}
where $\Gamma = [ \vec{\gamma_1}, \vec{\gamma_2}, \ldots,
\vec{\gamma_p}]^T$ is chosen so that $\vec{\gamma_1}$ maximizes the
ratio
\begin{equation}
  \frac{\vec{\gamma_1}^T S_{\mathrm{inter}}
  \vec{\gamma_1}}{\vec{\gamma_1}^T S_{\mathrm{intra}}
  \vec{\gamma_1}},
\end{equation}
and $\vec{\gamma_\xi}$, $\xi=2, 3, \ldots, p$, maximizes a similar
ratio and
\begin{equation}
  \vec{\gamma_\xi}^T S_{\mathrm{intra}} \vec{\gamma_\xi} = 0.
\end{equation}
It can be shown that the vectors $\vec{\gamma_1}, \vec{\gamma_2},
\ldots,\vec{\gamma_\xi}$ correspond to the eigenvectors of the
matrix $S_{\mathrm{intra}}^{-1} S_{\mathrm{inter}}$.

\begin{figure}
  \begin{center}
    \subfigure[]{\includegraphics[width=0.4\linewidth]{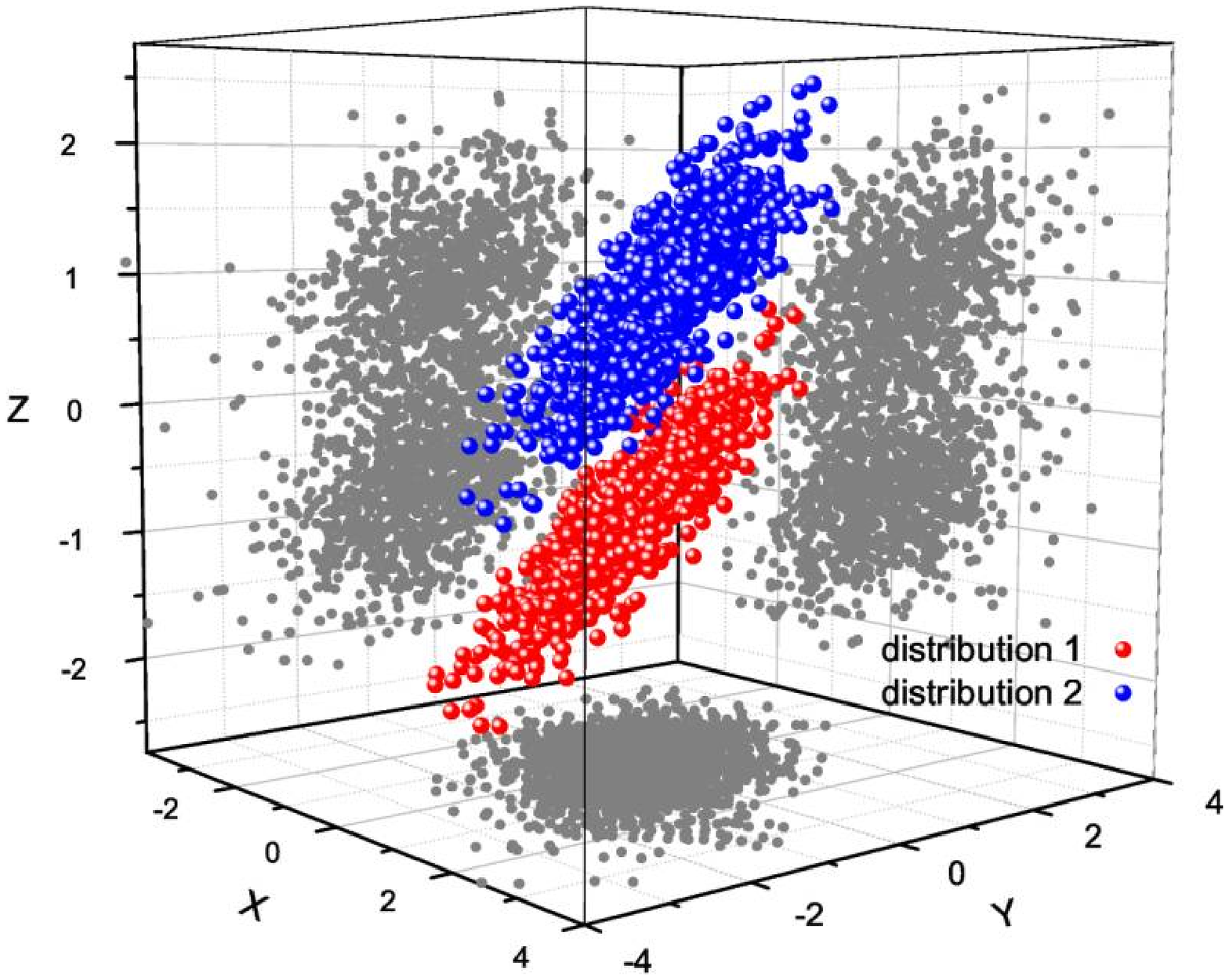}}
    \linebreak
    \subfigure[]{\includegraphics[width=0.4\linewidth]{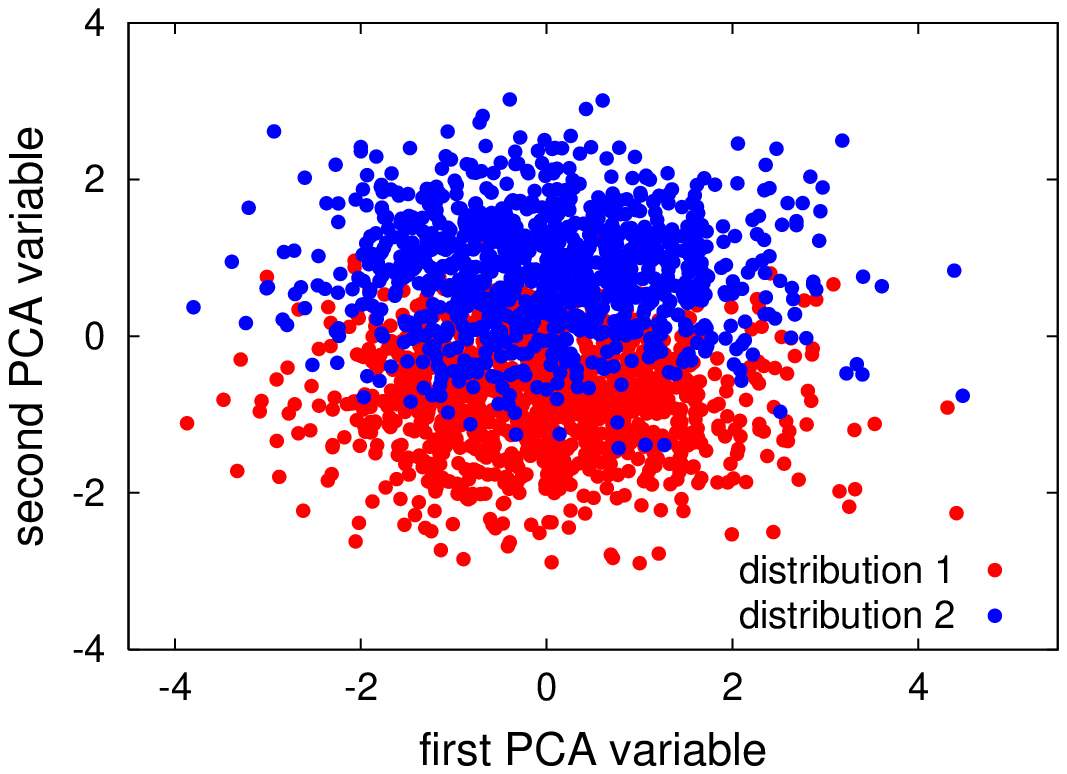}}
    \, \,
    \subfigure[]{\includegraphics[width=0.4\linewidth]{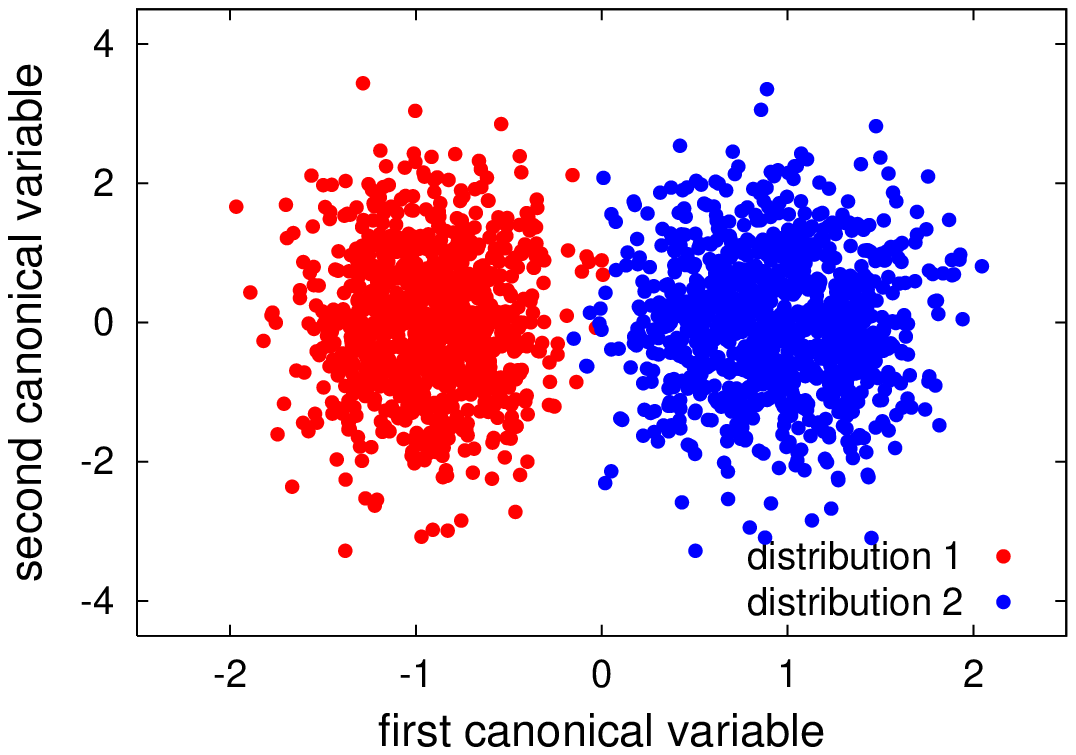}}
  \end{center}
  \caption{A phase space (scatterplot) containing two distributions of
    points (a) and respective PCA (b) and canonical (c)
    projections. Note that neither the projections into the three main
    planes (a) nor the PCA projection (b) separate the distributions,
    in contrast to the canonical projection (c).}
  \label{fig:excan}
\end{figure}

Figure~\ref{fig:excan} illustrates a phase space (a) containing two
distributions of observations, as well as the respective PCA (b) and
canonical analysis (c) projections considering two dimensions.  The
potential of the canonical approach for obtaining separated clusters is
evident from this example.

\subsection{Bayesian Decision Theory}\label{sec:bayes}

The elegant and sound methodology known as \emph{Bayesian decision
theory} provides an intuitive and effective means for classifying
objects into a given set of categories.  In principle, it is assumed
that the mass probabilities $P_i$, as well as the conditional
probability densities, $p(\vec{x_\xi} | C_i)$, are all given or can be
properly estimated (e.g., by using parametric or non-parametric
methods, see
\cite{Duda_Hart:01,Fukunaga:90,Costa:01}). The mass probability
$P_i$ corresponds to the probability that an object, irrespective of
its properties, belongs to class $C_i$, and therefore can be
estimated from the respective relative frequency.  The conditional
probabilities $p(\vec{x_\xi} | C_i)$ provide a statistical model of
how the measurements in the feature vectors are distributed inside
each category.  Given an object with unknown classification, the
most likely category $c$ to be assigned to it is the one for which
the respectively observed feature vector $\vec{x}$ produces the
highest value of $P_\xi p(\vec{x} | C_\xi)$.  In case the
probability functions are not available, it is still possible to use
an approximate classification method, known as \emph{$k$-nearest
neighbors} (e.g.~\cite{Duda_Hart:01}), which consists in identifying the set 
of the $k$ individuals which are closer (i.e.\ smaller distance between 
feature vectors) to the sample to be classified, and take as the resulting
category that corresponding to the most frequent class among the
nearest neighbors.

\begin{figure}
  \begin{center}
    \subfigure[]{\includegraphics[width=0.5\linewidth]{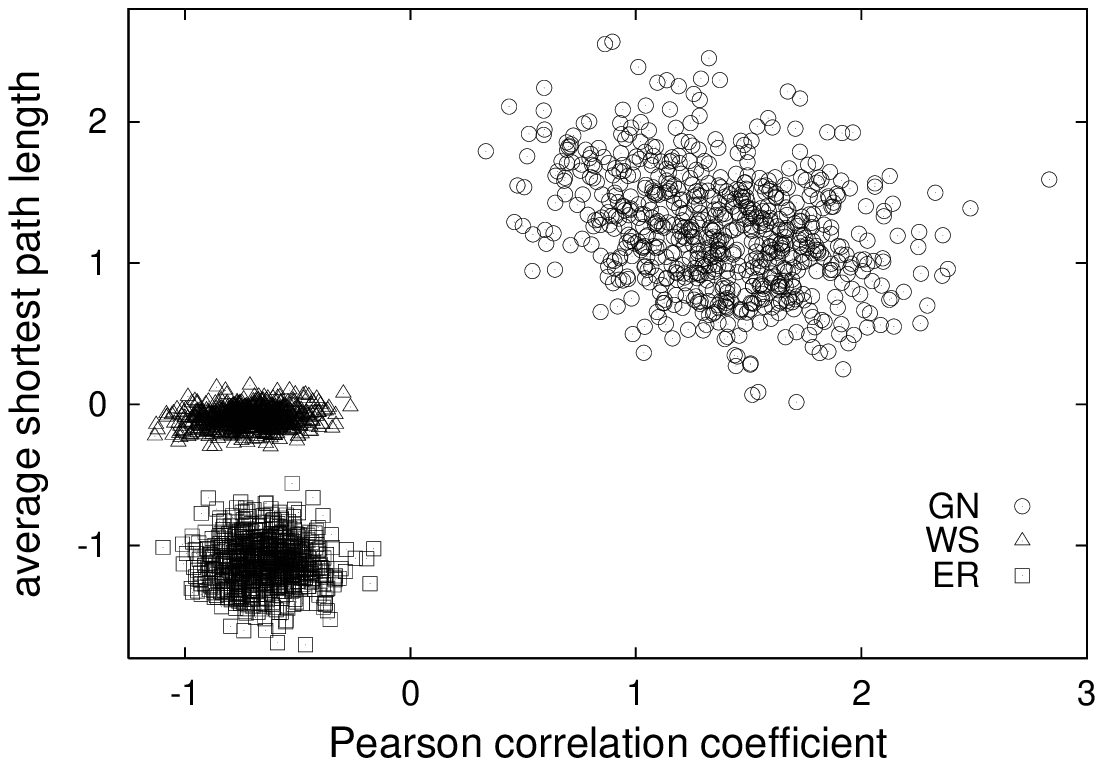}}
    \linebreak
    \subfigure[]{\includegraphics[width=0.42\linewidth]{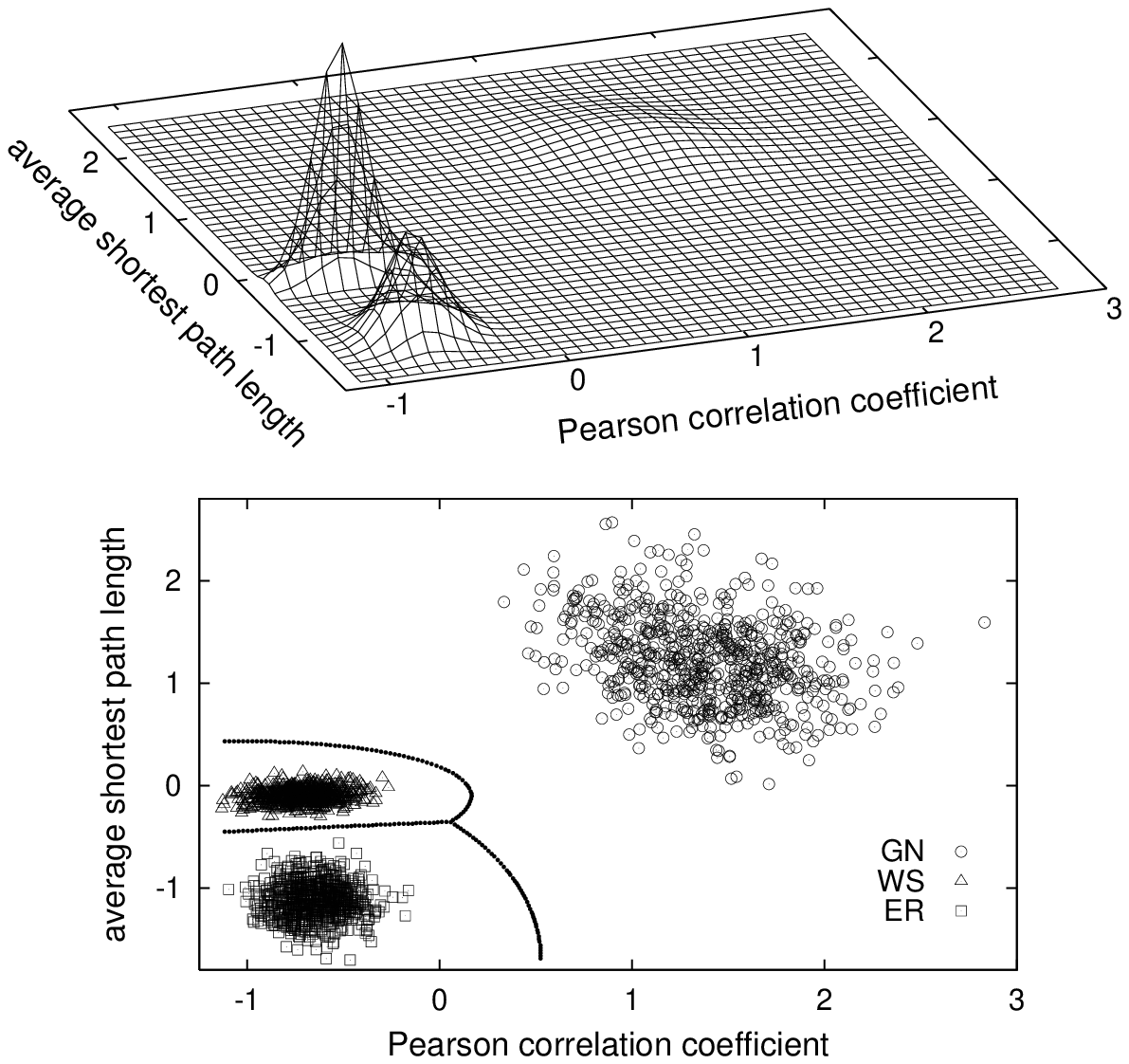}}
    \subfigure[]{\includegraphics[width=0.42\linewidth]{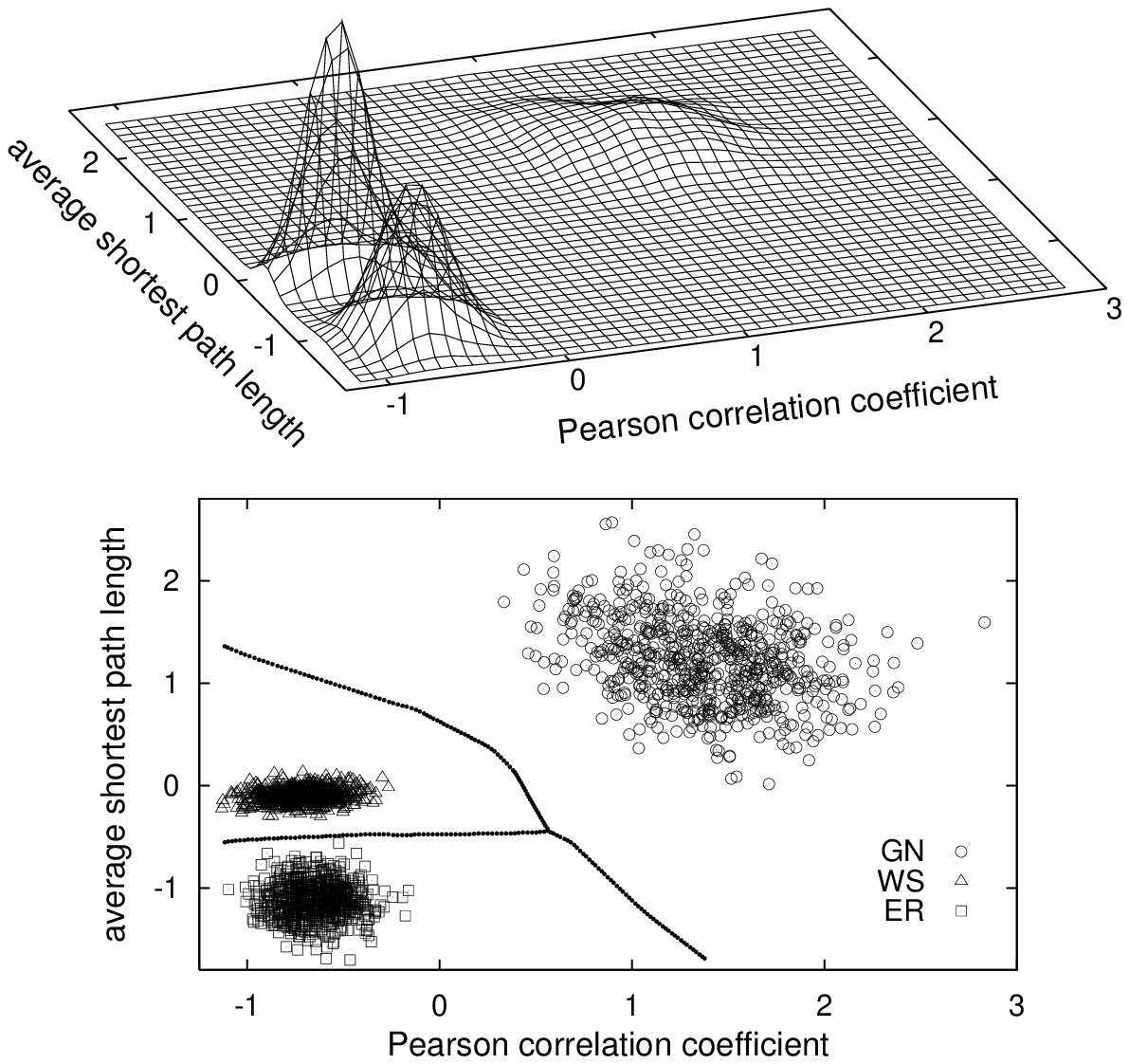}}
  \end{center}
  \caption{A scatterplot of normalized measurements containing several
    complex networks derived from three main categories, i.e.\
    Geographical Network (GN), Watts-Strogatz (WS) and Erd\H{o}s and
    R\'enyi random graph (ER) models (a), and respective Gaussians and
    decision regions obtained by using the Bayes method considering
    parametric (b) and non-parametric (c) estimation. Network
    parameters are $N=250, \langle k \rangle = 20$, with $1000$
    realizations for each model; rewiring probability in the WS model
    is $0.4$.}
  \label{fig:Bayes}
\end{figure}

Let us illustrate the above concepts and methodology in terms of a
situation involving three categories $C_1$, $C_2$ and $C_3$ of
complex networks, namely Geographical Network (GN), Watts-Strogatz
small-world network (WS) and Erd\H{o}s and R\'enyi random graph
(ER), characterized in terms of their normalized average shortest
path length $l$ and Pearson correlation coefficient of vertex
degrees $r$. The corresponding scatterplot is shown in
Figure~\ref{fig:Bayes}(a).

Usually we do not know the mass and conditional probabilities of
each type of networks, so they have to be estimated from the
available data.  This stage can be understood as the \emph{training
phase} of the Bayesian decision theory method.  There are two main
ways to estimate the probabilities required: parametric and
non-parametric. In the former, the mathematical form of the
probability functions is known (e.g., normal distribution) and the
respective parameters (mean and covariance matrix, in the case of
normal distributions) need to be estimated; in the latter, the
mathematical type of the densities is unknown , being estimated,
e.g., through some interpolation procedure such as the Parzen
windows methodology \cite{Duda_Hart:01}.  Once the training phase is
concluded, new objects whose classes are to be determined have their
measurements estimated and used to identify, among the probability
distributions of the trained models, which are the most likely
respective classes.  This categorization procedure corresponds to
the \emph{decision phase} of the Bayesian methodology.

Figure~\ref{fig:Bayes}(b) illustrates the parametric approach,
considering three normal density distributions, as applied to the
data in Figure~\ref{fig:Bayes}(a).  These distributions were defined
by having their parameters (namely average vector and covariance
matrix) estimated from the respective experimental measurements. The
separating frontiers are shown in the projection at the bottom of
the figure.  The decision regions obtained by using non-parametric
estimation through Parzen windows is shown in
Figure~\ref{fig:Bayes}(c).

Note that a very high dimensional feature space implies that a
substantially high number of individuals must be considered in order
to obtain properly estimated (i.e.\ not too sparse) densities.
Therefore, it is essential to limit the number of measurements to a
small set of more discriminative features.  An interesting alternative
involves the use of canonical projections in order to reduce the
dimensionality of the problem. A key open question which is briefly
addressed in this section regards which of the several topological
measurements available for complex networks characterization can yield
the best characterization and discrimination among the principal
network models.

\subsection{Combining Canonical Variable Analysis and Bayesian
Decision Theory}

\begin{figure}
  \begin{center}
    \subfigure[]{\includegraphics[width=0.4\linewidth]{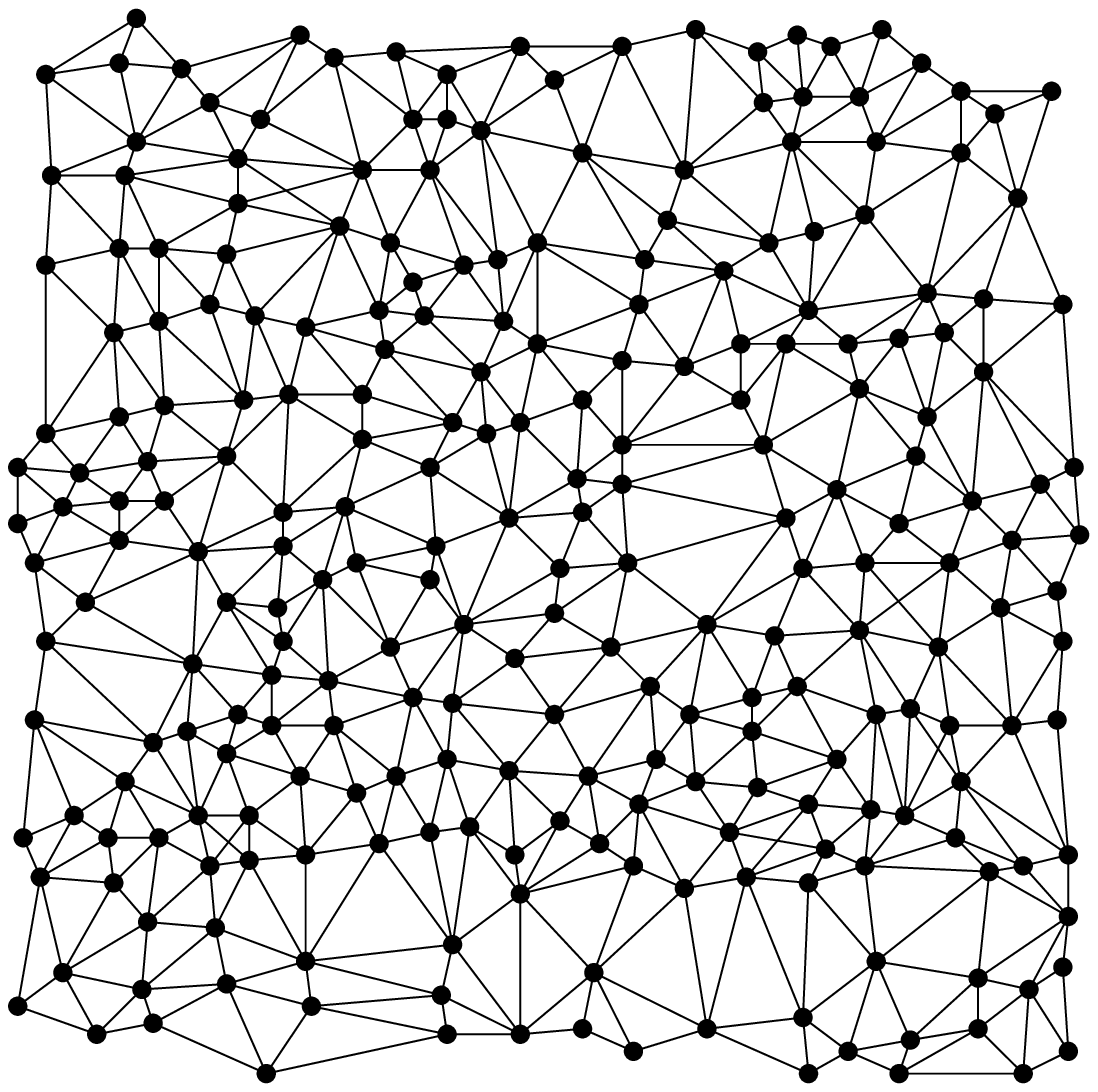}}
    \hspace{1cm}
    \subfigure[]{\includegraphics[width=0.4\linewidth]{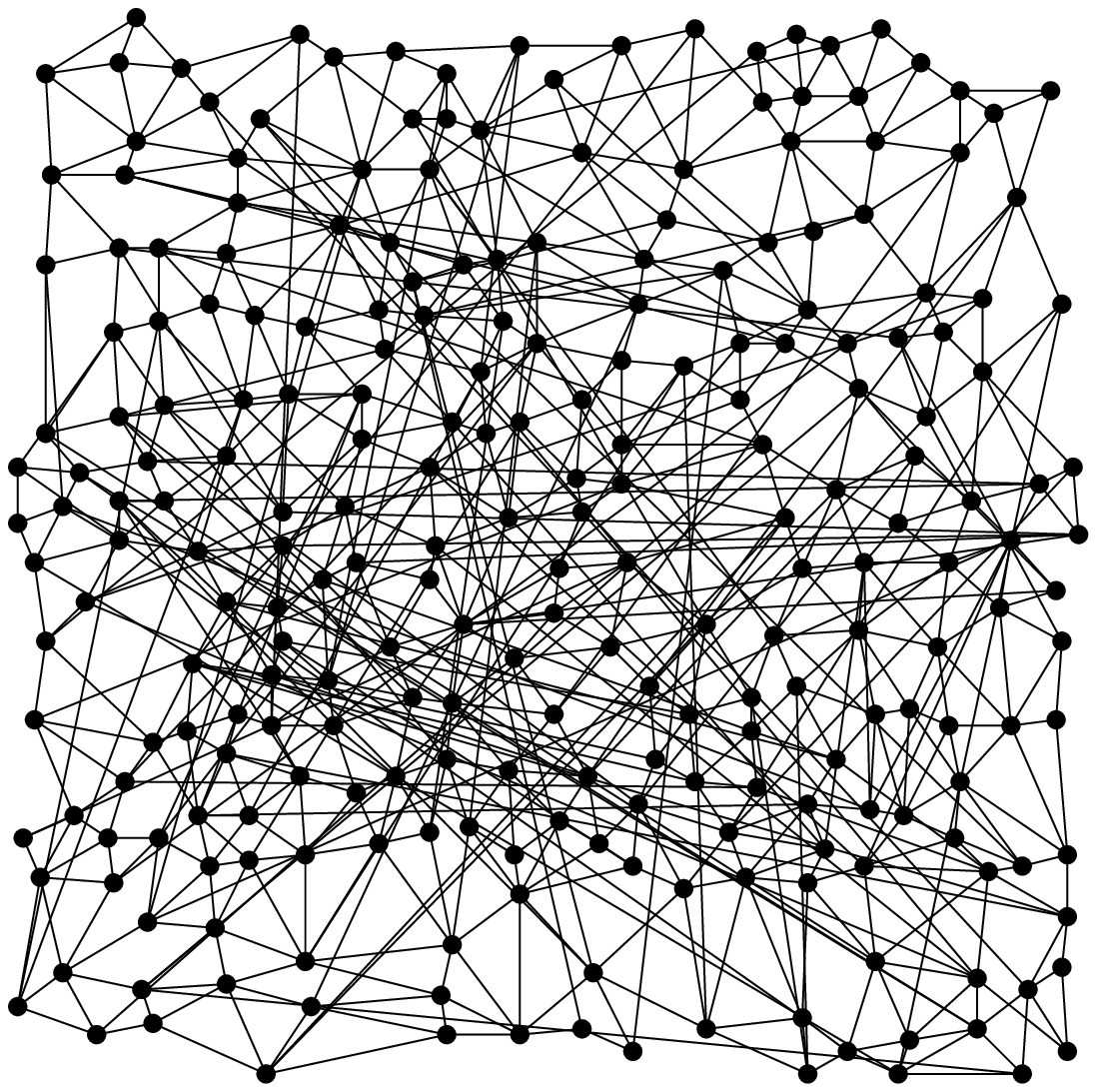}}
    \\ \vspace{0.25cm}
    \subfigure[]{\includegraphics[width=0.4\linewidth]{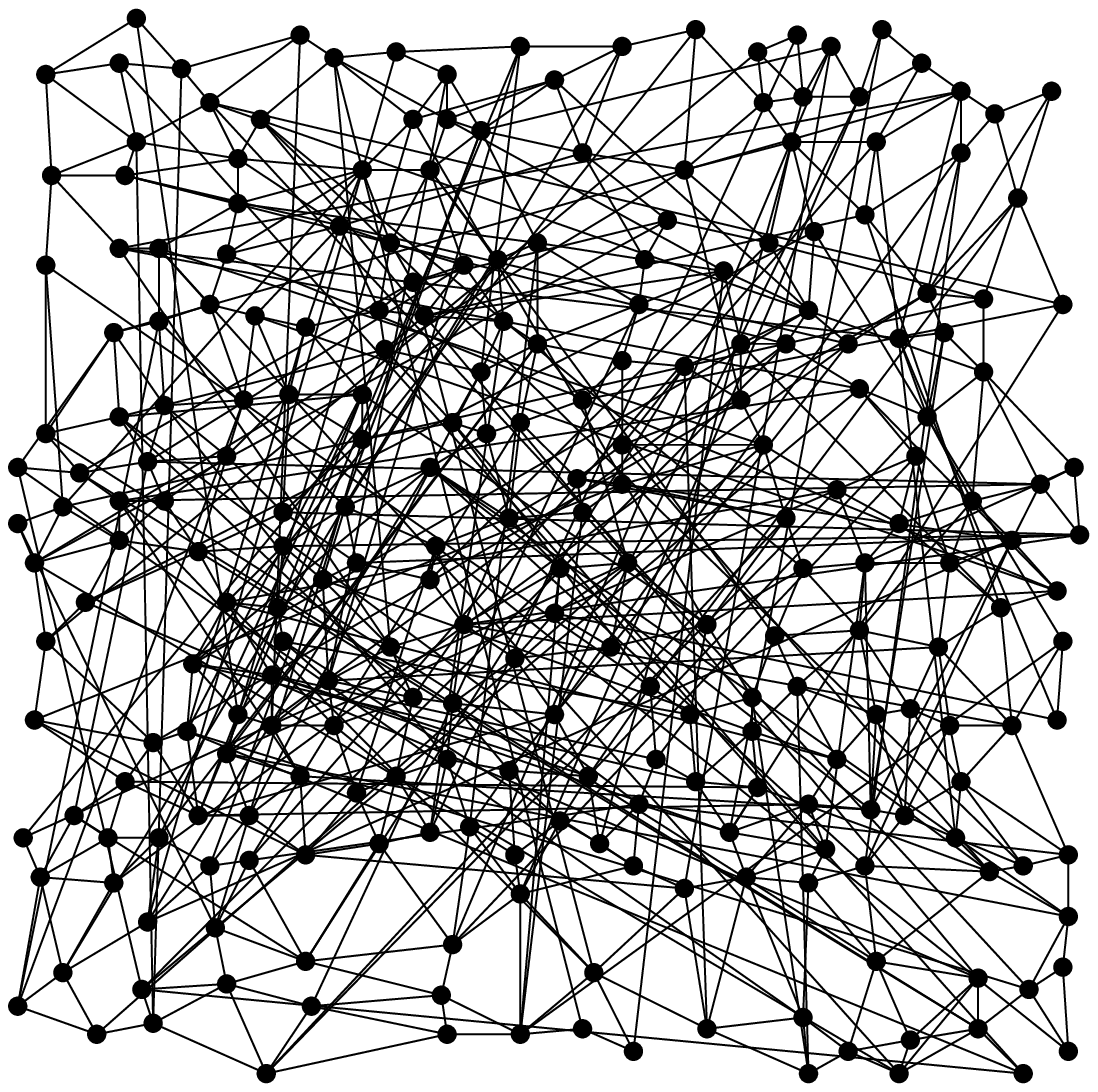}}
    \hspace{1cm}
    \subfigure[]{\includegraphics[width=0.4\linewidth]{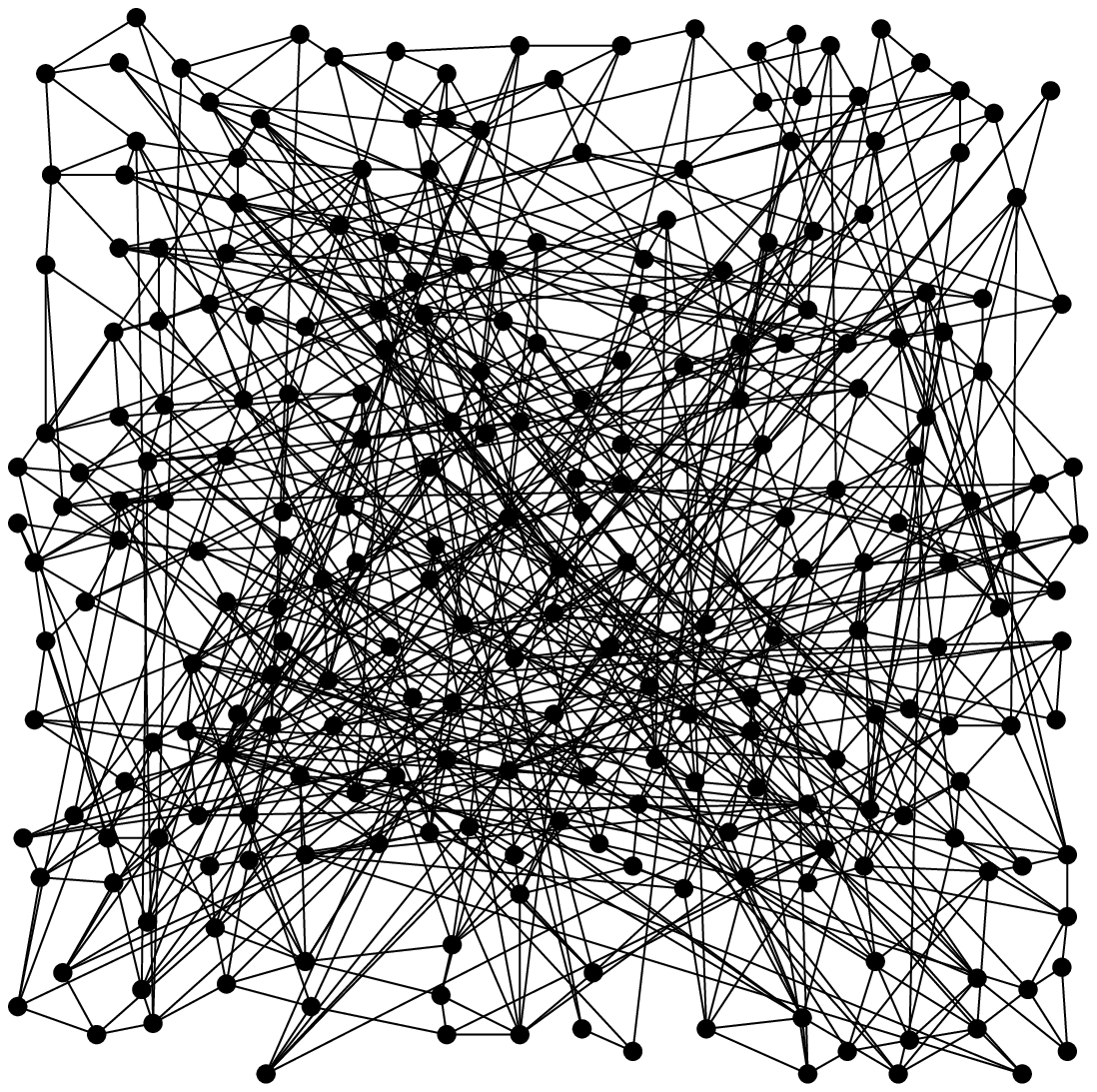}}
  \end{center}
  \caption{The Delaunay geographical network (DLN) for several numbers of
           rewirings: original (a) and after 60 (b), 120 (c) and
           200 (d) rewirings.}
  \label{fig:dln}
\end{figure}

An interesting possibility for classifying networks involves the
combination of canonical variable analysis and Bayesian decision
theory (e.g., \cite{Costa:01,McLachlan:04,Prado:04}).  More
specifically, the observations considered for the training stage are
projected into a reduced dimensional feature space by using
canonical analysis, so that the Bayesian decision method is applied
not over the larger original features space, but onto a more
manageable and representative features space. This possibility is
explored in this section to address the important issue of
classifying experimental complex networks into three main categories
defined by the similarity with the Barab\'asi-Albert (BA), Erd\H{o}s
and R\'enyi random graph (ER) and Geographical Network (GN) models.
The following experimental networks are considered in our
experiments:

\begin{trivlist}

\item \emph{US Airlines Transportation Network (USATN):} The USATN is
composed by $332$ airports, localized in the United States in 1997,
connected by flights. The data was collected from the Pajek datasets
\cite{pajek_data}. This kind of network exhibits a power law
behavior as described in \cite{Guimera04:EPJB,Guimera05}.  

\item \emph{Protein-Protein Interaction Network of the Saccharomyces
Cerevisiae (PPIN):} PPIN is formed by $1922$ proteins linked according
to identified direct physical interactions \cite{Jeong01}, a dataset
is available at the Center for Complex Network Research (The
University of Notre Dame). The vertex degree distributions of
protein-interaction networks tend to follow a power law
\cite{Jeong01}.

\item \emph{Autonomous System (AS):} In the Internet, an AS is a
collection of IP networks and routers under the control of one entity
that presents a common routing policy to the Internet.  Each AS is a
large domain of IP addresses that usually belong to one organization
such as a university, a business enterprise, or an Internet Service
Provider. In this type of networks, two vertices (AS) are connected if
there is at least one physical link between them. This kind of network
is usually described by the Barab\'{a}si-Albert model \cite{Yook02}
and the data considered in our work is available at the web site of
the National Laboratory of Applied Network
Research~(\texttt{http://www.nlanr.net}). We used the data collected
in Feb. $1998$, with the network containing $3522$ vertices and $6324$
edges.

\item \emph{Transcriptional Regulation Network of the E. coli (TRNE):}
In this network, the vertices represent operons (an operon is a group
of contiguous genes that are transcribed into a single mRNA molecule)
and each edge is directed from an operon that encodes a transcription
factor to another operon that is regulated by that transcription
factor. Hence, this kind of network, which is believed to be scale
free~\cite{ShenOrr:2002}, controls gene expression. We used the
undirected version of the network analyzed by Shen-Orr
\emph{et al.}\ \cite{ShenOrr:2002}, which is formed by $577$
interactions and $424$ operons. The original network was transformed
into the undirected form by the operation of \emph{symmetry} as
described in Section~\ref{basic_concepts}.

\item \emph{Delaunay Network (DLN):} This network was obtained by
distributing a set of points (the vertices) uniformly (but with an
exclusion radius in order to avoid points to become too close) along a
unit square and obtaining the edges from the respective Delaunay
triangulation (e.g.~\cite{Stoyan:book}).  Therefore, each point
defines a tile in the respective Voronoi diagram, and every pair of
adjacent vertices are connected (see Figure~\ref{fig:dln}). The
connectivity of this type of geometrical structure, henceforth called
\emph{Delaunay network}, is therefore completely determined by the
adjacency between the vertices, which is in turn defined by the
geographical distribution of the vertices.  As such, Voronoi networks
provide one interesting extreme case of geographical networks where
only the immediate spatial neighborhood is considered for
connection. The network considered here contains $251$ vertices and
$700$ edges.  Progressively rewired (degree preserving) versions of
this network are also considered in order to illustrate the evolution
of trajectories in decision spaces.  Figure~\ref{fig:dln} illustrates
four of these successive configurations.

\end{trivlist}

A total of three sets of $300$ realization of each reference model
(BA, ER and GN) were then generated. The networks for each set were
designed to have average vertex degrees near the experimental value.
The model and experimental networks were characterized in terms of
the following measurements: straightness $\mathit{st}$, average
vertex degree $\langle k \rangle$, Pearson correlation coefficient
of vertex degrees $r$, average clustering coefficient
$\widetilde{C}$, average shortest path length $\ell$, central point
dominance $\mathit{CPD}$, average hierarchical degree of second
level $\langle k_2(i) \rangle$, average hierarchical clustering
coefficient of second level $\langle C_2(i) \rangle$ and average
hierarchical divergence ratio of the third level $\langle
\mathit{dv}_3(i) \rangle$.

In order to provide a general and representative view of the effect
of these measurements in the classification of real networks by
using the methodology involving the canonical analysis followed by
Bayesian decision, we considered the following combinations of
measurements:

\renewcommand{\labelenumi}{\roman{enumi}.}
\begin{enumerate}
\item $\{\ell,st\}$,

\item $\{\langle k \rangle,\widetilde{C},\ell\}$,

\item $\{\langle k_2(i) \rangle,\langle C_2(i) \rangle,\langle \mathit{dv}_3(i)
\rangle\}$,

\item $\{st,r,\mathit{CPD}\}$,

\item $\{\langle k \rangle,\widetilde{C},\ell,st,r,\mathit{CPD}\}$,

\item $\{\langle k \rangle,\widetilde{C},\ell,\langle k_2(i) \rangle,\langle C_2(i) \rangle,\langle \mathit{dv}_3(i)
\rangle\}$,

\item $\{st,r,\mathit{CPD},\langle k_2(i) \rangle,\langle C_2(i)
\rangle,\langle \mathit{dv}_3(i) \rangle\}$,

\item all measurements.
\end{enumerate}

\noindent The combination (i) was the only one that did not require
canonical analysis. Table~\ref{tab:classification} shows the results,
i.e.\ the theoretical model and respective average vertex degree which
have been associated to each experimental network by the
classification procedure, obtained for each of these configurations.
More specifically, each experimental network was classified as having
the same category as the theoretical model defining the decision
region in the canonical projection space where the feature vector of
the experimental data was mapped.

\begin{sidewaystable}
  \caption{The classes assigned to the real networks by considering
  each combination of measurements. The classes in bold mean wrong
  identified model and, in italic style, wrong average vertex
  degree. The $*$ symbol represents an identified class well away from
  all theoretical models (see, for instance,
  Figure~\ref{fig:can_bayes}(c)).}  \centering
  \begin{footnotesize}
    \begin{tabular}{|c|c|c|c|c|c|c|c|c|c|}
      \hline \textbf{Experimental} & \textbf{Expected} &
      \multicolumn{8}{c|}{\textbf{Identified networks for the following
      combinations:}} \\ \cline{3-10} \textbf{Network} & \textbf{Network} &
      \textbf{(i)} & \textbf{(ii)} & \textbf{(iii)} & \textbf{(iv)} &
      \textbf{(v)} & \textbf{(vi)} & \textbf{(vii)} & \textbf{(viii)} \\
      \hline US Airlines Trans- & & & & & & & & & \\ portation Network &
      BA/GN & \emph{BA} & \emph{GN*} & GN* & \emph{BA} & \emph{BA*} &
      \emph{BA*} & \emph{GN*} & BA* \\ (USATN) & $\langle k \rangle = 12.8$
      & $\mathit{\langle k \rangle = 10.0}$ & $\mathit{\langle k \rangle =
    10.0}$ & $\langle k \rangle = 12.8$ & $\mathit{\langle k \rangle =
    10.0}$ & $\mathit{\langle k \rangle = 10.0}$ & $\mathit{\langle k
    \rangle = 10.0}$ & $\mathit{\langle k \rangle = 14.0}$ & $\langle k
      \rangle = 12.0$ \\ $\langle k \rangle = 12.8$ & & & & & & & & & \\
      \hline Autonomous & & & & & & & & & \\ System (AS) & BA & \emph{BA} &
      \textbf{GN} & \emph{BA} & BA & \textbf{GN} & \textbf{GN} & \emph{BA} &
      \textbf{GN} \\ $\langle k \rangle = 3.59$ & $\langle k \rangle = 3.59$
      & $\mathit{\langle k \rangle = 6.0}$ & \boldmath$\langle k \rangle =
      3.59$ & $\mathit{\langle k \rangle = 6.0}$ & $\langle k \rangle = 4.0$
      & \boldmath$\langle k \rangle = 3.59$ & \boldmath$\langle k \rangle =
      3.59$ & $\mathit{\langle k \rangle = 6.0}$ & \boldmath$\langle k
      \rangle = 3.59$ \\ & & & & & & & & & \\ \hline Transcriptional Regu- &
      & & & & & & & & \\ lation Network of & BA & BA & \textbf{GN} &
      \emph{\textbf{GN}} & \emph{BA} & \textbf{ER} & \textbf{ER} &
      \textbf{GN} & \textbf{ER} \\ the \emph{E. coli} (TRNE) & $\langle k
      \rangle = 2.45$ & $\langle k \rangle = 2.0$ & \boldmath$\langle k
      \rangle = 2.45$ & \boldmath$\mathit{\langle k \rangle = 4.0}$ &
      $\mathit{\langle k \rangle = 4.0}$ & \boldmath$\langle k \rangle =
      2.45$ & \boldmath$\langle k \rangle = 2.45$ & \boldmath$\langle k
      \rangle = 2.45$ & \boldmath$\langle k \rangle = 2.45$ \\ $\langle k
      \rangle = 2.45$ & & & & & & & & & \\ \hline Protein-Protein & & & & &
      & & & & \\ Interaction Network & & & & & & & & & \\ of the
      \emph{Saccharomyces} & BA & \emph{\textbf{ER}} & \textbf{GN} &
      \emph{\textbf{GN}} & \emph{\textbf{ER}} & \textbf{GN} & \textbf{GN} &
      \emph{\textbf{ER}} & \textbf{GN} \\ \emph{Cerevisiae} (PPIN) & $\langle k
      \rangle = 3.03$ & \boldmath$\mathit{\langle k \rangle = 2.0}$ &
      \boldmath$\langle k \rangle = 3.03$ & \boldmath$\mathit{\langle k
    \rangle = 2.0}$ & \boldmath$\mathit{\langle k \rangle = 2.0}$ &
      \boldmath$\langle k \rangle = 3.03$ & \boldmath$\langle k \rangle =
      3.03$ & \boldmath$\mathit{\langle k \rangle = 2.0}$ &
      \boldmath$\langle k \rangle = 3.03$ \\ $\langle k \rangle = 3.03$ & &
      & & & & & & & \\ \hline Delaunay  & & & & & & & & & \\ Network (DLN) &
      GN & \emph{GN} & \emph{GN} & GN & \emph{GN} & \emph{GN} & GN & GN & GN
      \\ $\langle k \rangle = 6.0$ & $\langle k \rangle = 6.0$ &
      $\mathit{\langle k \rangle = 4.0}$ & $\mathit{\langle k \rangle =
    4.0}$ & $\langle k \rangle = 6.0$ & $\mathit{\langle k \rangle = 4.0}$
      & $\mathit{\langle k \rangle = 4.0}$ & $\langle k \rangle = 6.0$ &
      $\langle k \rangle = 6.0$ & $\langle k \rangle = 6.0$ \\ & & & & & & &
      & & \\ \hline
    \end{tabular}
  \end{footnotesize}
  \label{tab:classification}
\end{sidewaystable}

\begin{figure}
  \begin{center}
    \subfigure[]{\includegraphics[width=0.75\linewidth]{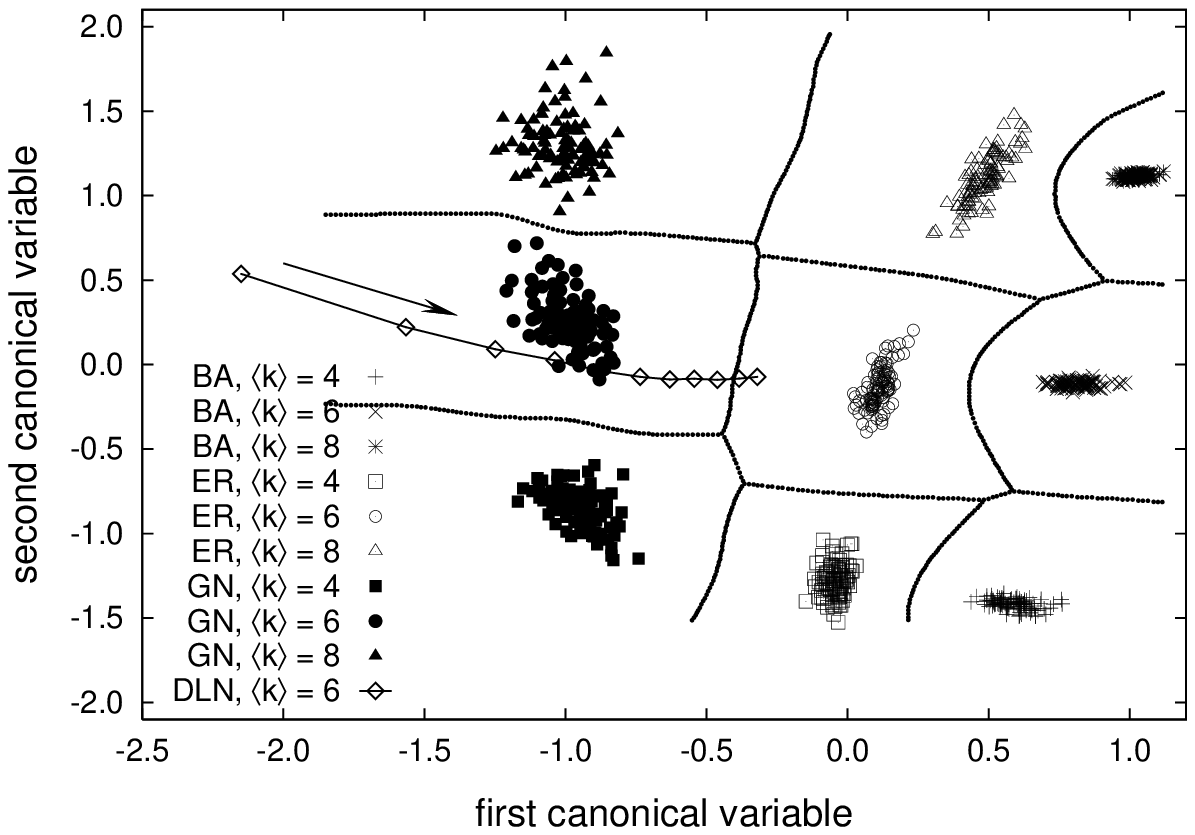}}
  \end{center}
  \caption{Separating frontiers between the decision regions in the
  scatterplots obtained by canonical analysis for the DLN. The
  separating frontiers were obtained by Bayesian decision theory. Note 
  the trajectory defined by the mapping of the progressively
  rewired versions of the original DLN network, extending from
  the GN towards the ER region with $\langle k \rangle = 6$.}
  \label{fig:DLN_bayes}
\end{figure}

\begin{sidewaysfigure}
  \begin{center}
    \subfigure[]{\includegraphics[width=0.4\linewidth]{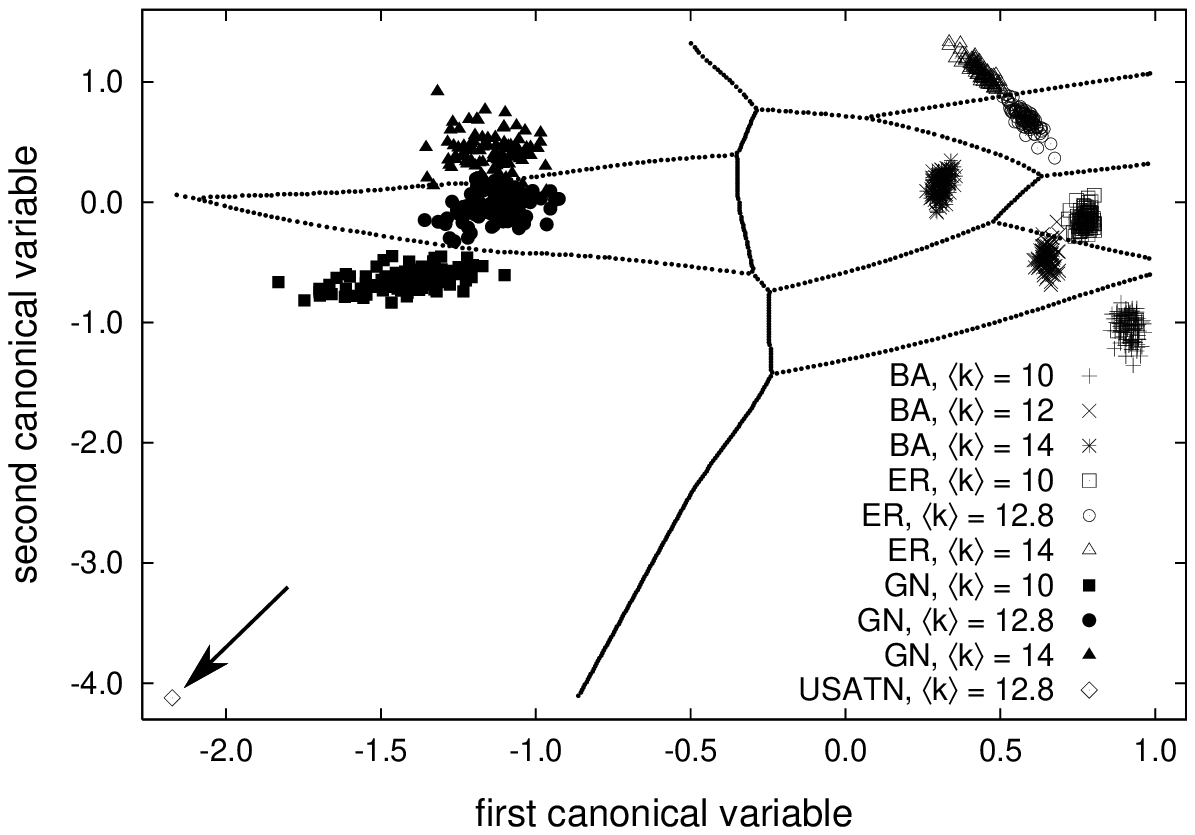}}
    \, \,
    \subfigure[]{\includegraphics[width=0.4\linewidth]{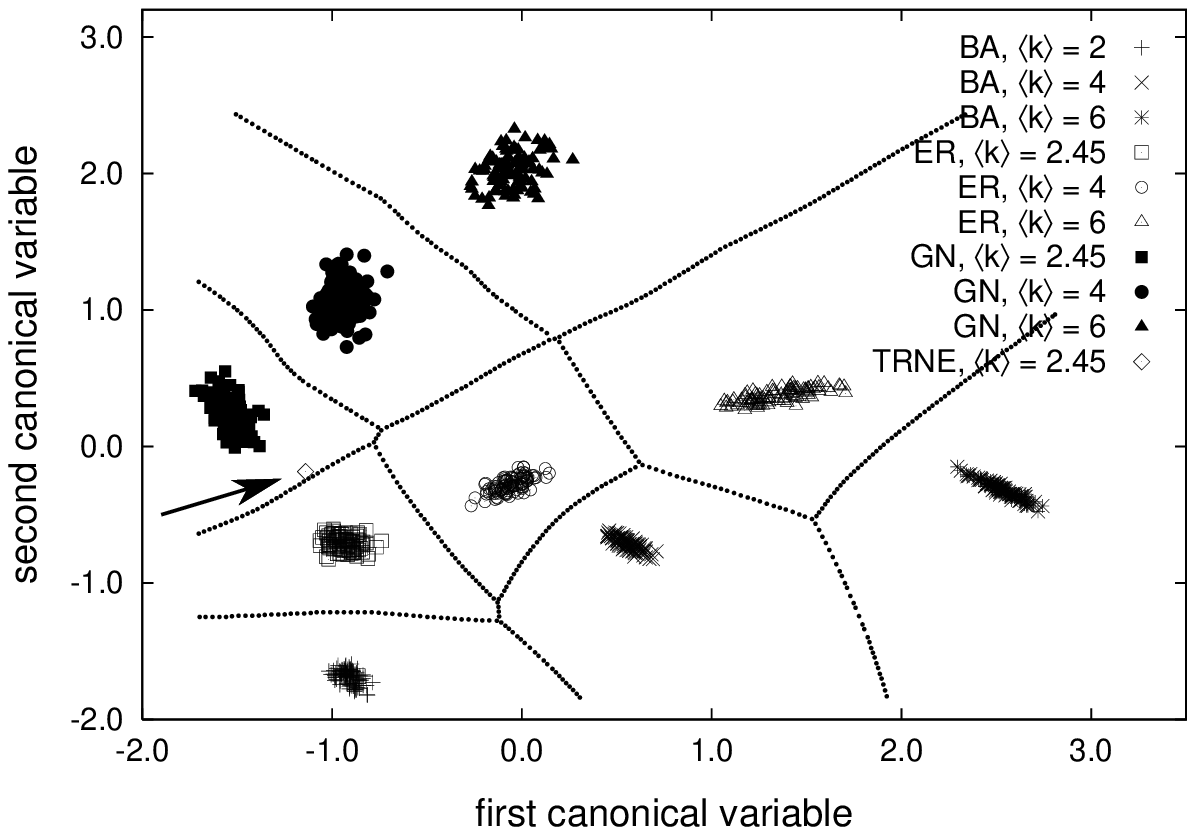}}
    \\
    \subfigure[]{\includegraphics[width=0.4\linewidth]{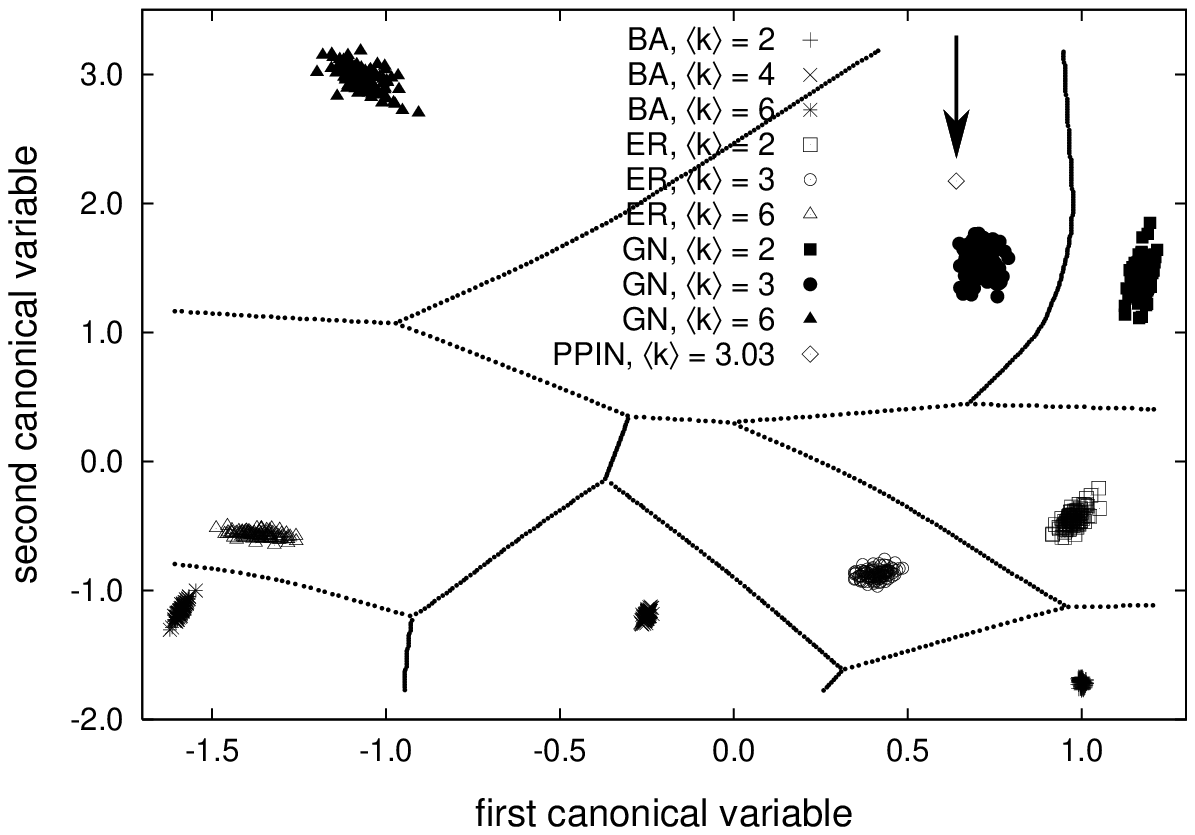}}
    \, \,
    \subfigure[]{\includegraphics[width=0.4\linewidth]{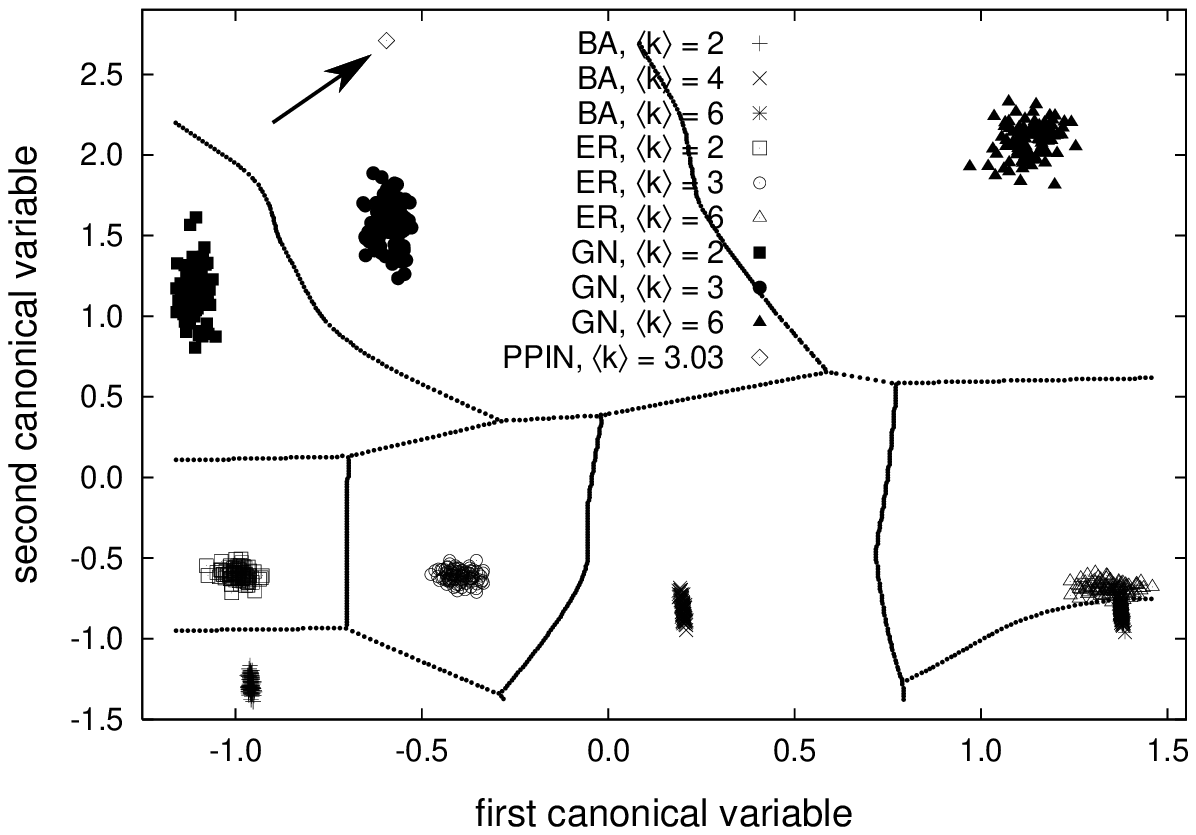}}
  \end{center}
  \caption{Examples of classification by canonical variable analysis
  and Bayesian decision theory: (a) US Airlines Transportation Network
  (USATN); (b) the Transcriptional Regulation Network of the
  \emph{E. coli} (TRNE); and (c) the Protein-Protein Interaction
  Network of the \emph{Saccharomyces Cerevisiae} (PPIN), considering
  all measurements; (d) the same protein network as in (c) but
  excluding the hierarchical measurements. Note the presence of the
  separating frontiers between the decision regions in the
  scatterplots.  The arrows indicate the experimental networks.}
  \label{fig:can_bayes}
\end{sidewaysfigure}

A number of interesting facts can be inferred from
Table~\ref{tab:classification}. To begin with, the compatibility
between the type of network model expected and obtained for each of
the experimental networks varies considerably for each case.  The best
compatibility was obtained for the DLN, i.e.\ the identified model was
compatible with the expected type (geographical) for all considered
combinations of measurements. Compatible average vertex degrees have
also been obtained for cases (iii), (vi)-(viii).
Figure~\ref{fig:DLN_bayes} illustrates the location of this network in
the scatterplot defined by the canonical projection of the combination
of all measurements.  In this figure, which also shows the separating
frontiers of the decision regions, the experimental network DLN
(represented as $\diamond$) resulted closer to GN with average vertex
degree of $6$. PPIN implied the highest number of incompatible
classifications which, instead of being identified as a scale-free
network (as expected \cite{Jeong01}), was understood as GN except for
the cases $\{\ell,st\}$ and $\{st,r,\mathit{CPD}\}$. A similar
situation was verified regarding the average vertex degrees.
Figure~\ref{fig:can_bayes}(c) and (d) show the resulting position of
this network within the scatterplots obtained by canonical projection
of the combination of all measurements (c) and all except those
hierarchical (d). Note the good agreement between the resulting
categories obtained for these two cases.  In both cases, the PPIN
resulted very close to the GN with average vertex degree of $3.03$.

A particularly interesting result has been obtained for the USATN,
which tended to appear well away from all theoretical groups in most
cases, as illustrated in the scatterplot shown in
Figure~\ref{fig:can_bayes}(a) with respect to the case $\{\langle k
\rangle,\widetilde{C},\ell\}$. Intermediate results were obtained
for the other networks. For instance, TRNE has been classified as
expected (i.e.\ as a BA network) in $2$ cases, identified as an ER
in only one case and as a GN in $5$ cases.
Figure~\ref{fig:can_bayes}(b) shows the position of this network in
the scatterplot defined for all measurements. Note that TRNE appears
almost in the middle of the ER and GN types for average vertex
degree of $2.45$.

\begin{figure}
  \begin{center}
    \includegraphics[width=0.75\linewidth]{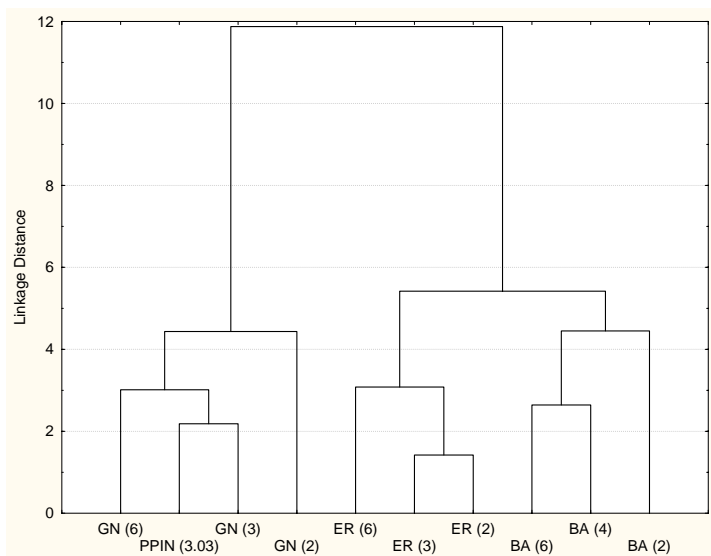}
  \end{center}
  \caption{Dendrogram obtained for the protein-protein interaction
  network considering all measurements except those hierarchical. Note
  that the BA, ER and GN networks resulted in well-separated branches,
  while the protein-protein network was included into the latter group.}
  \label{fig:dendro}
\end{figure}

It is also possible to use hierarchical clustering algorithms
(e.g.~\cite{Duda_Hart:01,Costa:01,Costa:2005a}) in order obtain
additional information about the relationship between the analyzed
networks.  Figure~\ref{fig:dendro} shows the dendrogram obtained for
the situation depicted in Figure~\ref{fig:can_bayes}(c) by using
Ward's aglomerative method.  In this method the networks, inititally
treated as individual clusters, are progressively merged in order to
guarantee minimal dispersion inside each cluster.  The linkage
distance is shown along the $y-$axis, indicating the point where the
clusters are merged (the sooner two clusters are merged, the most
similar they are).  The similarity between the cases belonging to each
of the three types of networks is reflected by the fact that three
respective main branches are obtained in the dendrogram in
Figure~\ref{fig:dendro}.  The GN cluster incorporates the experimental
protein-protein network, to which it is most closely related by the
measurements.  Note that the GN group, including the protein-protein
network, is significantly different from the ER and BA models at the
right-hand side of the figure, as indicated by the high linkage
distance at which these two groups (i.e. the GN and ER/BA) are merged.

The results discussed above illustrate the classification procedure
and its potential for identifying the category of networks of unknown
nature.  The fact that the assigned category sometimes varies
according to the choice of measurements suggests the presence of
specific topological features in some experimental networks which are
not fully compatible with any of the assumed theoretical models.
Indeed, the consideration of more measurements can, in principle,
provide a more comprehensive subclassification of the networks. Such
a possibility is particularly important in the case of scale-free
networks, which are known to involve subtypes \cite{Boccaletti05}.
For instance, TRNE has been identified in our experiments as having BA
type while considering two measurements (i.e.\ $\{st,\ell\}$), but was
understood as a GN model by considering three measurements (i.e.\
$\{\langle k \rangle,\widetilde{C},\ell\}$) and as ER when we
considered six measurements (i.e.\ $\{\langle k
\rangle,\widetilde{C},\ell,st,r,\mathit{CPD}\}$).

It should be always kept in mind that the consideration of an
excessive number of measurements may ultimately compromise the
quality of the classification.  Methodologies such as the canonical
analysis followed by Bayesian classification can be used to identify
the features which contribute the most for the correct
classifications.  This can be done by considering the measurements
which contribute more intensely for the canonical projections
providing the largest number of correct classifications.  A simpler
methodology involves the application of the principal component
analysis to remove the redundancies between the measurements. In the
case of a reduced number of measurements it is also possible to
consider all the respective combinations and evaluate which of them
yields the best classifications. Another interesting possibility for
investigating complex network connectivity is to consider outliers
analysis (e.g.~\cite{Costa0607272}). The reader interested in
additional information on multivariate statistics and feature
selection is referred to the specialized literature (e.g.,
\cite{Duda_Hart:01, Costa:01, McLachlan:04, Fukunaga:90}) for more
in-depth discussion and coverage.

\section{Concluding Remarks}

Measurements of the connectivity and topology of complex networks
are essential for the characterization, analysis, classification,
modeling and validation of complex networks.  Although initially
limited to simple features such as vertex degree, clustering
coefficient and shortest path length, several novel, powerful
measurements have been proposed.  We hope it has been made clear
that the several available measurements often provide complementary
characterization of distinct connectivity properties of the
structures under analysis.  It is only by becoming familiar with
such measurements that one can expect to identify proper sets of
features to be used for the characterization of complex networks.
The current survey has been organized to provide a comprehensive
coverage of not only the most traditional measurements but also
complementary alternatives which, though not so frequently used, can
provide valuable resources for characterizing specific topological
properties of complex networks. Special attention was also given to
the application of measurements in community finding algorithms, an
important issue in complex network research.

In addition to presenting such measurements according to coherent
categories, we also addressed issues such as visualization, in terms
of trajectories defined by measurements, of complex network growth.
As illustrated by the results presented, which considered several
important theoretical network models, such trajectories clearly
reflect, in graphical terms, important tendencies exhibited by
different network categories as their average degree is increased.
Another important point to be kept in mind in network measurements
is correlations.  While high correlation between a pair of
measurements indicates that they are largely redundant, our results
show that the own correlation values vary from one network model to
another, providing further useful information for network
characterization. Another important property of a specific
measurement is its sensitivity to small perturbations in the
network, such as the inclusion or removal of edges or vertices.  We
illustrated that different measurements can behave very differently
with respect to such induced changes.  Because one of the most
challenging issues related to network categorization regards the
choice of the features to be taken into account, we provided a
self-contained discussion about how multivariate statistics concepts
and methods can be applied for that aim. More specifically, we
showed how high dimensional measurement spaces can be effectively
projected, by using principal component analysis, into
lower-dimensional spaces favoring visualization and application of
computationally intensive measurements. We also described how two
useful methods, namely canonical analysis and Bayesian decision
theory, can be combined to provide the means for semi-automated
identification of the effective linear combinations of measurements,
in the sense of allowing good discrimination between network
categories.  The potential of such multivariate methodologies was
illustrated for theoretical models and experimental networks.  The
results clearly suggested that considering a comprehensive set of
measurements can provide more complete characterization of the
topological properties of the networks to the point of requiring a
revision of the traditional classification of experimental networks
into subclasses or new models.

All in all, this survey provides for the first time an integrated
presentation and discussion of a comprehensive set of measurements
previously covered in separated works.  In addition, it addresses
important issues related to application of these measurements for
characterization and classification of networks, including dynamical
representations in terms of trajectories, redundancy between
measurements as quantified by correlations, perturbation effects and
a powerful multivariate framework for classification of networks of
unknown category.  The systematic application of such concepts and
tools is poised to yield a wealthy of new results in the study of
complex networks.

\section*{Acknowledgments}

We are grateful to Lucas Antiqueira, Carlos A.-A.  Castillo-Ocaranza,
Ernesto Es\-tra\-da, A. D\'{\i}az-Guilera, Shalev Itzkovitz, Marcus
Kaiser, Xiang Lee, Jon Machta, Adilson E. Motter, Osvaldo
N. Oliveira-Jr, Andrea Scharnhorst, Matheus Viana, and Duncan Watts
for comments and suggestions.  Luciano da F. Costa is grateful to
FAPESP (proc.\ 99/12765-2), CNPq (proc.\ 308231/03-1) and the Human
Frontier Science Program (RGP39/2002) for financial support.
Francisco A. Rodrigues is grateful to FAPESP (proc.\ 04/00492-1) and
Paulino R. Villas Boas is grateful to CNPq (proc.\ 141390/2004-2).

\begin{table}
  \caption{Summary of discussed measurements.}
  \label{Measurements}
  \begin{center}
    \begin{tabular}{lcc}
      \hline
      Measurement & Symbol & Equation \\
      \hline
      Mean geodesic distance & $\ell$ & (\ref{meandist}) \\
      Global efficiency & $E$ & (\ref{netefficiency}) \\
      Harmonic mean distance & $h$ & (\ref{hmeandist}) \\
      Vulnerability & $V$ & (\ref{netvulnerability}) \\
      Network clustering coefficient& $C$ and $\widetilde{C}$ &
(\ref{clustcoeff}) and (\ref{clustercoeff2}) \\
      Weighted clustering coefficient & $C^w$ &
      (\ref{weightedcluster}) \\
      Cyclic coefficient & $\Theta$& (\ref{eq:cyclicnet}) \\
      Maximum degree& $k_{\mathrm{max}}$ & (\ref{maxdeg}) \\
      Mean degree of the neighbors& $k_{\mathrm{nn}}(k)$ & (\ref{knn}) \\
      Degree-degree correlation coefficient & $r$ & (\ref{pearson})\\
      Assortativity coefficient& $\widetilde{\mathbb{Q}}$, $\mathbb{Q}$ & (\ref{eq:guptaassor}) and
      (\ref{eq:newmanassor}) \\
      Bipartivity degree & $b$ and $\beta$ & (\ref{eq:bipartholme})
      and (\ref{eq:bipartestrada})\\
      Degree Distribution  entropy & $H(i)$ & (\ref{eq:entropy})\\
      Average search information & $\mathcal{S}$ &
      (\ref{searchinfo})\\
      Access information & $\mathcal{A}_i$ & (\ref{accessinfo})\\
      Hide information & $\mathcal{H}_i$ & (\ref{hideinfo})\\
      Target entropy& $\mathcal{T}$ & (\ref{targetentropy}) \\
      Road entropy & $\mathcal{R}$& (\ref{roadentropy}) \\
      Betweenness centrality & $B_i$ & (\ref{betweenness})\\
      Central point dominance & $\mathit{CPD}$ &(\ref{eq:cpd})\\
      $l$th moment & $M_l$& (\ref{moment}) \\
      Modularity & $Q$& (\ref{eq:modularity}) \\
      Participation coefficient & $P_i$& (\ref{pcoeff}) \\
      $z$-score & $z_i$& (\ref{eq:zscore}) \\
      Significance profile & $\mathit{SP}_i$ &(\ref{SP}) \\
      Subgraph centrality & $\mathit{SC}$ &(\ref{eq:subcentnet})\\
      Hierarchical clustering coefficient &$C_{rs}$ & (\ref{eq:hcc})\\
      Convergence ratio & $\mathit{cv}_d(i)$ & (\ref{cratio})\\
      Divergence ratio & $\mathit{dv}_d(i)$ & (\ref{dratio})\\
      Edge reciprocity & $\varrho$ and $\rho$ & (\ref{eq:recipnormal})
      and (\ref{eq:reccorr})\\
      Matching index of edge $(i,j)$ & $\mu_{ij}$ & (\ref{eq:matching})\\
      \hline
    \end{tabular}
  \end{center}
\end{table}

\pagebreak

\bibliographystyle{unsrt}
\bibliography{meassurv}

\end{document}